\documentclass[prd,aps,twocolumn,a4paper,floatfix,nofootinbib]{revtex4-2}

\usepackage[utf8]{inputenc}
\usepackage{graphicx,psfrag}
\usepackage{mathrsfs}
\usepackage{amsmath,amsfonts,amssymb}
\usepackage{multirow}
\usepackage{diagbox}
\usepackage{comment}
\usepackage{xcolor}
\usepackage{enumerate}
\usepackage{booktabs}
\usepackage[normalem]{ulem}
\usepackage{hyperref}
\usepackage{mathtools}
\usepackage{siunitx}
\usepackage{textgreek}
\hypersetup{
    colorlinks = true,
    linkcolor = {blue},
    citecolor = {blue},
    urlcolor = {blue},
    linkbordercolor = {white},
    }

\usepackage{color}
\definecolor{cyan}{rgb}{0,0.9,0.9}
\definecolor{orange}{rgb}{0.9,0.5,0}
\definecolor{magenta}{rgb}{1,0,1}
\definecolor{purple}{rgb}{0.8,0.4,0.8}
\definecolor{gray}{rgb}{0.8242,0.8242,0.8242}
\definecolor{green}{rgb}{0.,0.8,0.}

\DeclareMathAlphabet\mathbfcal{OMS}{cmsy}{b}{n}

\usepackage[normalem]{ulem}

\begin{document}

\title{M1 neutrino transport within the numerical-relativistic code BAM with application to low mass binary neutron star mergers}

\author{Federico \surname{Schianchi}$^{1}$}
\author{Henrique \surname{Gieg}$^{2}$}
\author{Vsevolod \surname{Nedora}$^{3}$}
\author{Anna \surname{Neuweiler}$^{1}$}
\author{Maximiliano \surname{Ujevic}$^{2}$}
\author{Mattia \surname{Bulla}$^{4,5,6}$}
\author{Tim \surname{Dietrich}$^{1,3}$}

\address{
${}^1$Institut f\"ur Physik und Astronomie, Universit\"at Potsdam, Haus 28, Karl-Liebknecht-Str. 24/25, 14476, Potsdam, Germany \\
${}^2$Centro de Ci\^encias Naturais e Humanas, Universidade Federal do ABC, 09210-170, Santo Andr\'e, S\~ao Paulo, Brazil \\
${}^3$Max Planck Institute for Gravitational Physics (Albert Einstein Institute), Am M\"uhlenberg 1, Potsdam 14476, Germany\\
$^{4}$Department of Physics and Earth Science, University of Ferrara, via Saragat 1, I-44122 Ferrara, Italy\\
$^{5}$INFN, Sezione di Ferrara, via Saragat 1, I-44122 Ferrara, Italy\\
$^{6}$INAF, Osservatorio Astronomico d’Abruzzo, via Mentore Maggini snc, 64100 Teramo, Italy
}

\date{\today}

\begin{abstract}
Neutrino interactions are essential for an accurate understanding of the binary neutron star merger process. In this article, we extend the code infrastructure of the well-established numerical-relativity code BAM that until recently neglected neutrino-driven interactions. In fact, while previous work allowed already the usage of nuclear-tabulated equations of state and employing a neutrino leakage scheme, we are moving forward by implementing a first-order multipolar radiation transport scheme (M1) for the advection of neutrinos. 
After testing our implementation on a set of standard scenarios, we apply it to the evolution of four low-mass binary systems, and we perform an analysis of ejecta properties. We also show that our new ejecta analysis infrastructure is able to provide numerical relativity-informed inputs for the codes \texttt{POSSIS} and \texttt{Skynet}, for the computation of kilonova lightcurves and nucleosynthesis yields, respectively. 
\end{abstract}

\maketitle

\date{\today}


\section{Introduction}
\label{sec:intro}
 Simulations of binary neutron star (BNS) mergers are a fundamental tool to support interpretations of multimessenger observations combining gravitational waves (GWs) and electromagnetic (EM) signals produced by the same transient event, allowing, among others, the study of matter at supranuclear densities e.g.,\cite{Radice:2016rys,Margalit:2017dij,Bauswein:2017vtn,Radice:2017lry,Dietrich:2018upm}, the expansion rate of the Universe~\cite{LIGOScientific:2017adf}, and the production of heavy elements, e.g. \cite{Lattimer:1974slx,Symbalisty:1982a,Rosswog:1998hy,Freiburghaus:1999,Rosswog:2005su}.

The strong interest in BNS mergers is 
partially caused by the myriad of observational data that has recently become available with the detection of the GW signal GW170817 \cite{TheLIGOScientific:2017qsa} by advanced LIGO \cite{LIGOScientific:2014pky} and advanced Virgo \cite{VIRGO:2014yos} and its associated EM counterparts: the kilonova AT2017gfo \cite{LIGOScientific:2017ync, Arcavi:2017xiz, Coulter:2017wya, Lipunov:2017dwd, Tanvir:2017pws, Valenti:2017ngx} and the short \textgamma-ray burst GRB170817A \cite{Savchenko:2017ffs, Abbott:GRB}, with long-lived signatures of its afterglow \cite{Hajela:2019mjy,Hajela:2021faz,Balasubramanian:2022sie}. 
With the start of the O4 observation run of the LIGO-Virgo-Kagra collaboration in May 2023, more events of this kind are expected to be detected, e.g.,~\cite{KAGRA:2013rdx,Petrov:2021bqm,Colombo:2022zzp,Kiendrebeogo:2023hzf}. 

In the neutron-rich matter outflow, $r$-process nucleosynthesis can set in, which can power transient EM phenomena due to heating caused by radioactive decay of newly synthesized nuclei in a wide range of atomic numbers \cite{Lippuner:2015gwa}, corroborating the hypothesis that kilonovae are connected to the production of heavy nuclei~\cite{Watson:2019xjv,Domoto:2022cqp}. 
Analysis of AT2017gfo has shown that kilonovae can consist of multiple components, each one generated by ejecta with different electron fractions and entropy \cite{Arcavi:2017xiz,Coulter:2017wya,Drout:2017ijr,Evans:2017mmy,Hallinan:2017woc,Kasliwal:2017ngb,Nicholl:2017ahq,Smartt:2017fuw,Soares-santos:2017lru,Tanvir:2017pws,Troja:2017nqp,Mooley:2018dlz,Ruan:2017bha,Lyman:2018qjg}. 
Studies of ejecta based on numerical-relativity (NR) simulations of BNS mergers suggest that the properties of the ejecta depend on the different ejection mechanisms during and after the merger, e.g., Refs.~\cite{Cowperthwaite:2017dyu,Villar:2017wcc,Tanvir:2017pws,Tanaka:2017qxj,Perego:2017wtu,Kawaguchi:2018ptg}. 

Most NR simulations of BNS mergers are relatively short ($\leq100\,$ms after the merger) and thus provide information on the early time,  dynamical ejecta, which is generally divided into a tidal component (driven by tidal torques) and a shocked component (driven by shocks launched during NS core bounces) 
\citep{Hotokezaka:2013iia,Bauswein:2013yna,Wanajo:2014wha,Sekiguchi:2015dma,Radice:2016dwd,Sekiguchi:2016bjd,Radice:2018pdn,Vincent:2019kor}. In equal-mass mergers, the shocked component is found to be up to a factor
${\sim}10$ more massive than the tidal one \cite{Bernuzzi:2020tgt}. 
However, the dynamical ejecta found in NR simulations cannot account alone for the
bright blue and late red components of the observed kilonova in AT2017gfo \citep{Cowperthwaite:2017dyu,Villar:2017wcc,Siegel:2019mlp}.

Winds powered by neutrino absorption and angular momentum transport can unbind $\mathcal{O}(0.1~M_{\odot})$ from the disk surrounding the remnant on timescales of $\mathcal{O}(0.1-1\,{\rm s})$ and could (if present) give the largest contribution to the kilonova signal \cite{Dessart:2008zd,Fernandez:2014bra,Just:2014fka,Lippuner:2017bfm,Siegel:2017nub,Fujibayashi:2017puw,Radice:2018xqa,Fernandez:2018kax,Janiuk:2019rrt,Miller:2019dpt,Fujibayashi:2020qda,Mosta:2020hlh}. Until recent years, these winds have been mostly studied by means of long-term simulations of neutrino-cooled disks  \cite{Metzger:2008av,Beloborodov:2008nx,Lee:2009uc,Fernandez:2012kh}.
Ab-initio NR simulations of the merger with advanced neutrino-transport and
magnetohydrodynamics were not yet fully developed at sufficiently long timescales \citep{Sekiguchi:2011zd,Wanajo:2014wha,Sekiguchi:2015dma,Palenzuela:2015dqa,Radice:2016dwd,Lehner:2016lxy,Sekiguchi:2016bjd,Foucart:2016vxd,Bovard:2017mvn,Fujibayashi:2017puw,Fujibayashi:2017xsz,Radice:2018xqa,Nedora:2019jhl,Vincent:2019kor,Bernuzzi:2020txg}, but large progress has been made recently, e.g., \cite{Fujibayashi:2022ftg}. 
Additionally, shorter (up to $100$~ms post-merger) NR simulations 
pointed out the existence of moderately neutron-rich spiral-wave wind 
that is sufficiently massive and fast to contribute to the early blue kilonova emission \citep{Nedora:2019jhl}.
Another contribution to post-merger ejecta can come from neutrino-driven winds that can lead to ${\sim} 10^{-4}-10^{-3}M_{\odot}$ ejecta with high electron fraction \citep{Dessart:2008zd,Perego:2014fma,Just:2014fka}.

To perform multimessenger analyses of future GW and EM detections associated with BNS mergers, NR simulations, including microphysical modeling, are essential. In particular, for the estimation of nucleosynthetic yields and kilonova light curves, it is important to account for the interaction of nuclear matter with neutrinos. This is because neutrino emission and absorption are responsible for determining the electron fraction of the ejecta, which influences kilonova light curves and nucleosynthesis strongly.

 In the past, several attempts were made to map ejecta properties to binary parameters like deformability and mass ratio, e.g., \cite{Dietrich:2016fpt,Coughlin:2018miv,Coughlin:2018fis,Radice:2020ddv,Kruger:2020gig,Nedora:2020qtd,Dietrich:2020efo,Henkel:2022naw}, with the aim of building phenomenological fits for Bayesian analysis of kilonova light curves and GW signal simultaneously. These studies showed that neutrino radiation treatment plays an important role in determining the mass, composition, and geometry of the ejecta. The extension and improvements of such fits with new data require the use of an advanced scheme to include neutrino radiation.
 
The first attempt to include neutrino interactions in a BNS merger simulation was made more than 20 years ago in \cite{Ruffert:1995fs, Rosswog:2003rv} by means of a neutrino leakage scheme (NLS). NLS employs an effective neutrino emissivity assigned to each fluid element according to its thermodynamical configuration and the optical depth of the path from it to infinity. This effective emission represents the rate of neutrino energy/number that escapes a fluid element. Hence, NLS is limited to model neutrino cooling. Unfortunately, this quantity is only known in the diffusive and free-streaming regimes, and phenomenological interpolation is used for gray zones. The main issue of NLS is the fact that, by neglecting the neutrino heating and pressure on the nuclear matter, it leads to a significant underestimation of the ejecta's electron fraction \cite{Radice:2018pdn} and affects the matter dynamics. The more recent development of an advanced spectral leakage (ASL) scheme tried to solve this issue by phenomenological modeling of neutrino flux anisotropies \cite{Perego:2015agy,Gizzi:2019awu,Gizzi:2021ssk}.

A more accurate theoretical approach to incorporate neutrino effects would require evolving the neutrinos distribution function according to the General Relativistic Boltzmann equation \cite{Lindquist:1966}. In principle, it is possible to follow this approach in a conservative 3+1 formulation, e.g., \cite{Cardall:2013kwa}. However, since the distribution function is defined in the 6+1-dimensional one-particle phase space, the computational cost of such an approach is prohibitive. Therefore, in recent years, more computationally efficient neutrino radiation transport approaches have become increasingly popular. Amongst them is the so-called moment scheme, which is based on a multipolar expansion of the moments of the radiation distribution function~\cite{Thorne:1981}. The 3+1 decomposition of such a formalism has been first studied in \cite{Shibata:2011kx}. The basic idea of this framework is to dynamically evolve the distribution function of neutrino intensity in a base of multipoles up to a certain rank and evolve them as field variables. Most of the radiation transport codes used in NR consider the transport of the zeroth and first-rank moments, thus referred to as M0~scheme \cite{Radice:2018pdn} or M1 moments scheme~\cite{Shibata:2011kx, Shibata:2012ty,Foucart:2015vpa,Radice:2021jtw,Musolino:2023pao}. It is worth noting that the aforementioned M1 implementations rely on the grey approximation, i.e., the considered moments are frequency integrated. This description makes the computation significantly less expensive but less accurate regarding the matter-neutrino interaction rates, which are strongly dependent on the neutrino energy \cite{Burrows:2004vq}. We want to point out that not only in BNS simulations but also in core-collapse supernovae and disk simulations, multipolar radiation transport schemes are regularly used. In most cases, even in more sophisticated versions, like non-gray, energy-dependent schemes, e.g., \cite{OConnor:2014sgn,Just:2015fda, Kuroda:2016, OConnor:2018, Rahman:2019, Skinner:2019, Laiu:2021}.

One important artifact of multipolar radiation transport schemes is the well-known unphysical interaction of crossing beams \cite{Foucart:2015vpa}, which is due to the inability of the M1 scheme to treat higher-order moments of the distribution. The crossing beams and energy-dependent interaction rates issues are both cured by Monte-Carlo radiation transport. In the latter, radiation is modeled by an arbitrary number of neutrino packets, each one with its own energy and momentum. This scheme has recently been adapted to NR simulation \cite{Foucart:2020qjb,Foucart:2021mcb,Kawaguchi:2022tae}.

Another scheme that can, in principle, solve these issues is the relativistic Lattice-Boltzmann \cite{Weih:2020qyh}, where the momenta component of the Boltzmann equation is solved in every space point on a discretized spherical grid in order to model the transport even in presence of higher order momenta. Overall, we refer to \cite{Foucart:2022bth} for a detailed review of neutrino transport methods.

Among all the mentioned schemes, we decided to implement M1 transport because of its ability to treat neutrino heating and pressure with a reasonable computational cost and for being well-tested in NR simulations of BNS mergers.

This article is structured as follows: In Sec. \ref{Governing Equations}, we recap the governing equations of General Relativistic Radiation Hydrodynamics (GRRHD) and M1 transport. In Sec.~\ref{Numerical Scheme}, we discuss the numerical methods used to integrate M1 transport equations, paying particular attention to the stiff source terms and the advection of radiation in the trapped regime. In Sec.~\ref{Numerical testes}, we show the results of the tests we performed to validate the code in different regimes. In Sec.~\ref{Binary Neutron Star Mergers}, we present the application of our newly developed code to the merger of binary neutron stars with two different EoSs and mass ratios, with a description of ejecta geometry, neutrino luminosity, GW signal, nucleosynthesis  yields, and kilonova light curves.

Throughout this article, we will use the Einstein notation for index summation with the $(-,+,+,+)$ signature of the metric and (unless differently specified) geometric units, i.e., $G=c=M_{\odot}=1$. Also, the Boltzmann constant is $\kappa_B = 1$.


\section{Governing Equations}
\label{Governing Equations}

\subsection{3+1-Decomposition and spacetime evolution}

The spacetime dynamics is considered by numerically solving Einstein's field equations in 3+1 formulation, for which the line element reads
\begin{equation}\label{eq:4d-lin-el}
    ds^2 = -\alpha^2dt^2 + (dx^i + \beta^i dt)(dx^j + \beta^j dt) \gamma_{ij},
\end{equation}
where $\alpha$ is the lapse function, $\beta^i$ is the shift vector, and $\gamma_{ij}$ is the 3-dimensional spatial metric (or 3-metric) induced on the 3-dimensional slices of the 4-dimensional spacetime, identified by $t = {\rm constant}$. 
The 3-metric is given by
\begin{equation}
    \gamma_{\alpha \beta} = g_{\alpha\beta} + n_{\alpha}n_{\beta},
\end{equation}
where $g_{\alpha \beta}$ is the 4-dimensional spacetime metric and $n^{\alpha}$ is the timelike, normal vector field. By construction, $n^\alpha$ is future-directed, normal to each point of a given $t = {\rm constant}$ slice and normalized to $n^\mu n_\mu = -1$.
In the coordinate system given by the line element of Eq. ~\eqref{eq:4d-lin-el}, the normal vector field has components
\begin{equation}
    n^{\alpha} = \left ( \frac{1}{\alpha}, -\frac{\beta^k}{\alpha} \right), \quad \quad n_{\alpha} = \left (  -\alpha, 0  \right ).
\end{equation}

In this framework, the BAM code ~\cite{Bruegmann:2006at,Thierfelder:2011yi,Dietrich:2015iva,Bernuzzi:2016pie,Gieg:2022} can solve for the Einstein field equations. In this work, we do so using the Z4c formulation with constraint damping terms \cite{Bona:2003fj, Bona:2005pp, Gundlach:2005eh} as implemented in \cite{Bernuzzi:2009ex}.

\subsection{General Relativistic Hydrodynamics}
As in the previous version of BAM that included a neutrino leakage scheme \cite{Gieg:2022}, we solve general-relativistic radiation hydrodynamics equations arising from the conservation of stress-energy tensor of matter with source terms representing neutrino interactions. Furthermore, the conservation of baryon number and transport of electron fraction lead to:
\begin{align}
    \nabla_{\mu} (\rho u^{\mu}) &= 0,                                             \label{eq:mass_cons} \\
    \nabla_{\mu} T_{\rm matter}^{\mu \nu} & = -S^{\nu},                           \label{eq:stress_cons} \\
    \nabla_{\mu} (\rho Y_{\rm e} u^{\mu}) &= m_{\rm b} \mathcal{R},               \label{eq:Ye_cons}
\end{align}
with $\rho$ being the rest mass density, $T^{\rm matter}_{\mu\nu}$ the stress energy tensor of matter, $u^{\mu}$ its four-velocity, $m_{\rm b}$ the baryon mass, and $Y_{\rm e} = n_{\rm p}/n_{\rm b} = (n_{\rm e^-} - n_{\rm e^+})/n_{\rm b}$  the electron fraction. $n_{\rm e^-}$, $n_{\rm e^+}$, $n_{\rm p}$, and $n_{\rm b}$ are the number densities of electrons, positrons, protons, and baryons, respectively. The source terms $S^{\mu}$ and $\mathcal{R}$ represent the interaction of the fluid with neutrinos, i.e., neutrino cooling and heating, and the lepton number deposition rate, respectively.

We employ the usual decomposition of the fluid's 4-velocity as follows:
\begin{equation}
    u^{\alpha} = W(n^{\alpha} + v^{\alpha}), \quad \quad n^{\alpha}v_{\alpha}=0,
\end{equation}
with $W = -n^{\alpha}u_{\alpha}=1/\sqrt{1-v^iv_i}$, being the Lorentz factor.

Assuming matter to be an ideal fluid, its stress-energy tensor can be decomposed as
\begin{equation}
    T_{\rm matter}^{\mu\nu} = \rho h u^{\mu}u^{\nu} + pg^{\mu\nu},
\end{equation}
where $h$ and $p$ are the specific enthalpy and the pressure of the fluid, respectively. Equations \eqref{eq:mass_cons}, \eqref{eq:stress_cons}, and \eqref{eq:Ye_cons} are expressed as conservative transport equations following the standard Valencia formulation \cite{Banyuls:1997} as already implemented in previous versions of BAM, e.g., \cite{Thierfelder:2011yi,Gieg:2022}, which is based on the evolution of the conservative variables
\begin{align}
    D    & = \sqrt{\gamma} W \rho, \label{D} \\  
    \tau & = \sqrt{\gamma} (W^2 h \rho - p) - D,\\ 
    \mathcal{S}_i  & = \sqrt{\gamma} W^2 h \rho v_i, \\
    D_Y  & = \sqrt{\gamma} W \rho Y_e.
\end{align}
As an upgrade, in comparison to our previous implementation, we modified the source terms $S^{\mu}$ and $\mathcal{R}$ to adapt them to the new neutrino scheme, as we will discuss in the following. 

\subsection{Multipolar formulation for radiation transport}

In this article, we implement a first-order multipolar radiation transport scheme following the formulation of Ref.~\cite{Thorne:1981,Shibata:2011kx}. The multipolar formulation was originally developed to reduce the dimensionality of the general-relativistic Boltzmann equation for the neutrino distribution function in the phase space
\begin{equation}
\frac{\text{d}f(x^{\mu},p^{\mu})}{\text{d}l} = S_{\rm coll}(x^{\mu},p^{\mu},f),
    \label{eq:Boltzmann}
\end{equation}
with $f$ being the distribution of neutrinos, $l$ being the proper length traveled by neutrinos in a fiducial observer frame, and $S_{\rm coll}$ being a collisional term that takes into account the interaction of neutrinos with matter, i.e., emission, absorption, and scattering. The derivative $\text{d}/\text{d}l$ is along the trajectory in the phase space of the neutrinos, so it will have a component in physical spacetime and one in momentum space. Since neutrinos are assumed to travel on light-like geodesics, their momentum has to satisfy the constraint $p^{\alpha}p_{\alpha}=0$. This reduces the dimensionality of the problem by one and allows us to describe the 4-momentum via the variables $\Omega$ and $\nu$, representing the space direction of particles on a solid angle and their frequency in the fiducial observer frame. To make the evaluation of collisional sources easier, we chose the fluid frame as the fiducial frame. In the following text, $\nu$ will always be the frequency of neutrinos as measured in the fluid frame. 

At this point, we still have to handle a 6+1 dimensional problem that, if we want to ensure a sufficient resolution and accuracy for proper modeling, would computationally be too expensive. 
Hence, we need to work out a partial differential equation on the physical 3D space that can capture the main features of radiation even without fully solving for $f$ in the momentum space. 
In this regard, Thorne \cite{Thorne:1981} showed that it is convenient to decompose the intensity of radiation $I=\nu^3 f$ and the source $S_{\rm coll}$ in multipoles of the radiation momentum $p^{\alpha}$
\begin{align}
    M^{\alpha_1 ... \alpha_k} (x^\beta) &:= \int_0 ^{\infty} \text{d} \nu\ \nu^3 \int \text{d}\Omega \frac{f(\nu,\Omega,x^{\mu})}{\nu^k}p^{\alpha_1}...p^{\alpha_k} ,
    \label{eq:M} \\
    S^{\alpha_1 ... \alpha_k} (x^\beta) &:= \int_0 ^{\infty} \text{d} \nu\ \nu^3 \int \text{d}\Omega \frac{S_{\rm coll}(\nu,\Omega,x^{\mu},f)}{\nu^k}p^{\alpha_1}...p^{\alpha_k} .
    \label{eq:S}
\end{align}
Plugging this ansatz into the Boltzmann Equation, Eq.~\eqref{eq:Boltzmann}, one can derive the following evolution equations for every radiation moment $M^{A_k}$:
\begin{equation}
    \nabla_{\beta} M^{A_k \beta} - (k-1)M^{A_k \beta \gamma} \nabla_{\gamma}u_{\beta} = S^{A_k},
    \label{eq:multipolar_master}
\end{equation}
where $A_k$ is a multi-index of order $k$; cf.~Ref.~\cite{Thorne:1981}.

In this work, we will focus on the second-order multipole $M^{\alpha \beta}$, which is known to be equal to the stress-energy tensor of radiation. It can be decomposed employing the laboratory frame or employing the fluid frame by choosing two different decompositions of the radiation momentum $p^{\alpha}$: 
\begin{eqnarray}
   & \text{fluid frame:} & \quad p^{\alpha} = \nu  (u^{\alpha} + \ell^{\alpha}), \\
   & \text{laboratory frame:}  & \quad  p^{\alpha} = \nu' (n^{\alpha} + l^{\alpha}),
\end{eqnarray} 

where $\nu$ and $\nu'$ are neutrino frequency in the fluid and lab frame, respectively\footnote{Note that $\nu$ appearing in the integrals of Eqs.~\eqref{eq:M} and \eqref{eq:S} is always the frequency in the fluid's frame.}, with the constraints $u^{\alpha}\ell_{\alpha} = n^{\alpha}l_{\alpha} = 0$ and $l^{\alpha} l_{\alpha} = \ell^{\alpha}\ell_{\alpha} = 1$. \\
In such a way, we can express the radiation stress-energy tensor in the fluid frame as 
\begin{equation}
    T_{\rm rad}^{\alpha\beta} = M^{\alpha\beta} = Ju^{\alpha}u^{\beta} + u^{\alpha}H^{\beta} +  H^{\alpha}u^{\beta} + \mathcal{K}^{\alpha\beta},
\end{equation}
with $u^{\alpha}H_{\alpha}=S^{\alpha\beta}u_{\alpha}=0$ and
\begin{align}
    J &:= \int_o^{\infty} \text{d}\nu\ \nu^3 \int \text{d}\Omega f(\nu,\Omega, x^{\mu}), \\
    H^{\alpha} &:= \int_o^{\infty} \text{d}\nu\ \nu^3 \int \text{d}\Omega f(\nu,\Omega, x^{\mu}) \ell^{\alpha},\\
    \mathcal{K}^{\alpha \beta} &:= \int_o^{\infty} \text{d}\nu\  \nu^3 \int \text{d}\Omega f(\nu,\Omega, x^{\mu}) \ell^{\alpha} \ell^{\beta},
    \label{eq:fluid_vars_3}
\end{align}
representing respectively the energy, the momentum, and the stress-energy tensor measured in the fluid frame. We will use these variables to express source terms since interaction rates of radiation with matter are usually evaluated in the fluid frame, in particular, we will write the 1st source multipole following Shibata et.~al.~\cite{Shibata:2011kx} as 
\begin{equation}
    S^{\alpha} = \eta u^{\alpha} -\kappa_a J u^{\alpha} -(\kappa_a + \kappa_s)H^{\alpha},
 \label{eq:collisional_sources}
\end{equation}
with $\kappa_a$ being the absorption opacity, $\kappa_s$ being the scattering opacity, and $\eta$ being the emissivity. 
In our work, we incorporate neutrino emission, absorption, and elastic scattering into the source term, but we neglect inelastic scattering.

In general, the emissivity $\eta$ and the opacities $\kappa_a, \kappa_s$ depend both on the fluid properties, namely on the density of the matter $\rho$, the temperature $T$, and the electron fraction $Y_e$, but also on the neutrino spectrum. 
Unfortunately, the latter information is not available in our formalism since we only evolve averaged quantities. This represents one of the weaknesses of the employed scheme. Hence, to enable dynamical simulations, we have to employ additional assumptions that we will outline in the following. 

\subsection{M1 evolution equations}
Ref.~\cite{Shibata:2011kx} showed that fluid-frame variables are not suitable for obtaining a well-posed system of partial differential equations in conservative form. For such a purpose, we need to perform a decomposition in the laboratory frame as
\begin{equation}
    T_{\rm rad}^{\alpha \beta} = M^{\alpha \beta} = E n^{\alpha} n^{\beta} + F^{\alpha}n^{\beta} + F^{\beta}n^{\alpha} + P^{\alpha \beta},
    \label{eq:lab_frame_dec}
\end{equation}
with $n^{\alpha}F_{\alpha} = P^{\alpha \beta}n_{\alpha} = F^t = P^{t\alpha} = 0$, in this case, $E$, $F^{i}$, and $P^{ij}$ represent, respectively, the energy density, momentum density, and stress tensor as measured in the lab frame and are defined in an analogous way as their fluid frame equivalents.

We can work out laboratory frame variables starting from the fluid frame ones and vice versa performing different projections of $T_{\textrm{rad}}^{\alpha\beta}$. For our work, we will use:
\begin{align}
    J &= W^2 E - 2 W F^iu_i + P^{ij}u_i u_j,
    \label{eq:J} \\
H^{\alpha} &= (EW - F^iu_i)h^{\alpha}_{\ \beta}n^{\beta} + Wh^{\alpha}_{\beta}F^{\beta} - h^{\alpha}_iu_jP^{ij},
\label{eq:H} 
\end{align}
where we have defined the 3-metric in the fluid frame as
\begin{equation}
        h^{\alpha \beta} = g^{\alpha \beta} + u^{\alpha}u^{\beta}.
\end{equation}
Ref.~\cite{Shibata:2011kx} showed that by decomposing $M^{\alpha \beta}$ as in Eq.~\eqref{eq:lab_frame_dec} and plugging it into Eq.~\eqref{eq:multipolar_master} with $k=1$, we can get the following conservative evolution equations for the energy and momentum of neutrinos:
\begin{align}
  \phantom{i + j + k}
  &\begin{aligned}
    \partial_t \tilde{E} &+ \partial_j (\alpha \tilde{F}^j - \beta^j \tilde{E})\\
      &= \alpha (\tilde{P}^{ij} K_{ij} - \tilde{F}^j\partial_j \ln(\alpha) - \tilde{S}^{\alpha}n_{\alpha}),
      \label{eq:evolution_E}
  \end{aligned}\\
  &\begin{aligned}
    \partial_t\tilde{F_i} &+ \partial_j(\alpha \tilde{P}^j_i - \beta^j\tilde{F_i})\\
       &= \left(-\tilde{E} \partial_i \alpha + \tilde{F}_k\partial_i\beta^k + \frac{\alpha}{2}\tilde{P}^{jk} \partial_i \gamma_{jk} + \alpha \tilde{S}^{\alpha}\gamma_{i\alpha}\right),
       \label{eq:evolution_F}
  \end{aligned}
\end{align}
where $K_{ij}$ is the extrinsic curvature of the spatial hypersurface, and we defined the densitized variables $\tilde{E}=\sqrt{\gamma}E$, $\tilde{F}^i=\sqrt{\gamma}F^i$, $\tilde{P}^{ij}=\sqrt{\gamma}P^{ij}$ and $\tilde{S}^\alpha = \sqrt{\gamma}S^{\alpha}$. Here, we can clearly see that the source terms of this equation can be divided into two categories, the gravitational ones, proportional to the first derivatives of the metric and the gauge, and the collisional ones, proportional to $\tilde{S}^{\alpha}$. While the former is responsible for effects like neutrino path bending and gravitational blueshift/redshift, the latter describes neutrino emission, absorption, and scattering by the fluid. 

\subsection{Closure relation}
Since we do not have an evolution equation for $P_{ij}$, we can only estimate it from $E$ and $F^i$. We follow the prescription discussed in \cite{Shibata:2011kx}:
\begin{equation}
    P^{ij} = \frac{1}{2}(3\chi(\zeta)-1)P^{ij}_{\rm thin} + \frac{3}{2}(1-\chi(\zeta))P^{ij}_{\rm thick},
    \label{eq:closure}
\end{equation}
with $P_{\rm thin}$ and $P_{\rm thick}$ being the closures in the optically thin and thick regimes, respectively, and the quantity $\chi$ is called Eddington factor. The Eddington factor was introduced to model the transition from a trapped radiation regime ($\chi=1/3$) to a free streaming radiation regime ($\chi=1$). In this work, we use the so-called Minerbo closure \cite{MINERBO1978541}
\begin{equation}
    \chi(\zeta) = \frac{1}{3} + \zeta^2 \frac{6 - 2\zeta + 6\zeta^2}{15},
    \label{eq:minerbo closure}
\end{equation}
where
\begin{equation}
    \zeta^2 = \frac{H^{\alpha}H_{\alpha}}{J^2},
    \label{eq:closure_parameter}
\end{equation}
is the closure parameter. We expect $\zeta \rightarrow 1$ for free streaming radiation and $\zeta \rightarrow 0$ for trapped radiation.

The free streaming closure can be expressed as \cite{Shibata:2011kx}:
\begin{eqnarray}
    P^{ij}_{\rm thin} = E \frac{F^i F^j}{F^2}. 
    \label{eq:closure_thin}
\end{eqnarray}
The computation of $P^{ij}_{\rm thick}$ is more elaborated since the thick closure must be defined in such a way to be isotropic in the fluid frame, i.e, we want
\begin{equation}
    \mathcal{K}^{\alpha \beta}_{\rm thick} = \frac{1}{3}J h^{\alpha \beta}.
    \label{eq:closure_thick}
\end{equation}
Refs.~\cite{Radice:2021jtw,Foucart:2015vpa} showed that $\mathcal{K}^{\alpha \beta}_{\rm thick}$ of Eq. \eqref{eq:closure_thick} leads to 
\begin{equation}
    P^{ij}_{\rm thick} = \frac{4}{3} J_{\rm thick} W^2 v^i v^j + 2 W v^{(i} \gamma^{j)}_{\alpha}H^{\alpha}_{\rm thick} + \frac{1}{3}J_{\rm thick} \gamma^{ij},
\end{equation}
with 
\begin{align}
    J_{\rm thick} &= \frac{3}{2W^2+1} \left [ E(2W^2-1) - 2W^2F^iv_i \right ], \\
    \gamma ^i_{\alpha} H_{\rm thick}^{\alpha} &= \frac{F^i}{W} - \frac{4}{3}J_{\rm thick}Wv^i + W[F^i v_i -E + J_{\rm thick}]v^i.
\end{align}

The system composed of Eqs.~\eqref{eq:evolution_E} and \eqref{eq:evolution_F} is proven to be strongly hyperbolic using the closure \eqref{eq:closure} as long as the causality constraint $E \leq |F|$ is satisfied. Equation~\eqref{eq:closure} also guarantees that characteristic velocities of the system (\ref{eq:evolution_E}, \ref{eq:evolution_F}) are not superluminal. \\

Similar to other implementations of this scheme, e.g., \cite{Shibata:2012zz,Foucart:2015vpa,Radice:2021jtw,Musolino:2023pao}, we divide neutrinos into three species: $\nu_e$, $\bar{\nu}_e$, and $\nu_x$, with this last species collecting all heavy neutrinos and respective anti-neutrinos together. In this way, we are solving three M1 systems (\ref{eq:evolution_E}, \ref{eq:evolution_F}) coupled to each other only through the fluid. \\

\subsection{Neutrino number density}
The previously described scheme still misses any information about the neutrinos energy spectrum, which will be important for having an accurate estimate of the fluid's neutrino opacities $\kappa_a$ and $\kappa_s$ (since cross sections of the involved processes are strongly dependent on neutrino energy). The simplest way to improve the previous scheme in this sense is adding the evolution of neutrino number density in such a way to be able to get the neutrino average energy $\langle \epsilon_{\nu_i} \rangle$ in every point.

In our implementation, we set up the neutrino number evolution following \cite{Radice:2021jtw} and \cite{Foucart:2016rxm}, 
i.e., through the transport equation 
\begin{equation}
    \nabla_{\alpha} (n f^{\alpha}) = \eta_n - \kappa_n n,
    \label{eq:number_transp_cov}
\end{equation}
where $n$ is the neutrino number density in the fluid frame and $\eta_n$ and $\kappa_n$ are the neutrino number emissivity and opacity respectively and $nf^{\alpha}$ is the 4 dimensional number flux. According to Ref.~\cite{Radice:2021jtw} we chose 
\begin{equation}
    f^{\alpha} = u^{\alpha} + \frac{H^{\alpha}}{J},
\end{equation}
in such a way that the projection of $nf^{\alpha}$ along $u^{\alpha}$ gives the neutrino number density in the fluid frame:
\begin{equation}
    n = -nf^{\alpha} u_{\alpha}.
\end{equation}

Expressed in slice adapted coordinates, Eq.~\eqref{eq:number_transp_cov} reads
\begin{equation}
    \partial_t (\alpha \sqrt{\gamma} n f^0) + \partial_i(\alpha\sqrt{\gamma}nf^i) = \alpha \sqrt{\gamma}(\eta_n - \kappa_n n),
    \label{eq:number_transp_coo}
\end{equation}
which is a transport equation for the conservative variable
\begin{equation}
    N := \alpha \sqrt{\gamma}nf^0.
\end{equation}

\noindent Finally, we can find $f^0$ and $f^i$ using the definition of slice adapted coordinates:
\begin{align}
    \alpha f^0 &= -f^{\alpha}n_{\alpha} = W - \frac{H^{\alpha}n_{\alpha}}{J}, \\
    f^i &= Wv^i + \frac{\gamma^i_{\alpha}H^{\alpha}}{J} - \beta^i f^0.
\end{align}
We solve Eq.~\eqref{eq:number_transp_coo} together with Eq.~\eqref{eq:evolution_E} and Eq.~\eqref{eq:evolution_F} to get a complete and closed system of hyperbolic transport equations in conservative form.
Note that in this formulation, the average energy of neutrinos in the fluid frame can be simply obtained by:
\begin{equation}
    \langle \epsilon_{\nu} \rangle = \frac{J}{n}.
\end{equation}

\subsection{Coupling to hydrodynamics}
To model the exchange of energy and momentum between neutrinos and the fluid, we modify the conservation of the matter's stress-energy tensor into
\begin{equation}
    \nabla_{\beta}T_{\rm matter}^{\beta\alpha} = - \sum_{\nu_i} S^{\alpha}_{\nu_i},
\end{equation}
where the sum runs over all three neutrino species. This means 
\begin{align}
    \partial_t \tau &= \text{standard hydro rhs} +  \sum_{\nu_i}  \alpha n^{\alpha}\tilde{S}_{\alpha,\nu_i}  \label{eq:tau_corr}, \\
    \partial_t \mathcal{S}_i &= \text{standard hydro rhs} - \sum_{\nu_i}  \alpha \gamma_{i}^{\alpha} \tilde{S}_{\alpha,\nu_i},
\end{align}
where $\tau$ and $\mathcal{S}_i$ are the conservative internal energy and momentum in the standard Valencia formulation of GRHD. 

We also take the variation of the electron fraction of the fluid into account and solve the transport equation
\begin{equation}
    \nabla_{\alpha}(\rho Y_e u^{\alpha}) = m_b \mathcal{R},
\end{equation}
with a source term given by interaction with neutrinos
\begin{equation}
    \mathcal{R} = -\sum_{\nu_i} \text{sign}(\nu_i) (\eta_{n,\nu_i} - \kappa_{n,\nu_i} n_{\nu_i}),
    \label{eq:Ye_source}
\end{equation}
where 
\begin{equation}
\text{sign}(\nu_i)=\begin{cases}
1, &  \text{ if } \nu_i = \nu_e,\\
-1, & \text{ if } \nu_i = \bar{\nu}_e, \\
0, &  \text{ if } \nu_i = \nu_x,
\end{cases}
\end{equation}
is a function that accounts for different signs of contributions given by different neutrinos species.

\subsection{Opacities and emissivities}

Within our gray scheme, we evolve energy-integrated variables and lose information about the neutrino spectrum. This means opacities contained in Eqs.~\eqref{eq:collisional_sources} and \eqref{eq:Ye_source} represent effective frequency-averaged quantities. 
In the case of neutrinos in thermal equilibrium with the fluid, we can define the equilibrium opacities as
\begin{align}
    \kappa^{\rm eq}_{a,s} &= \frac{\int_0^{+\infty} \kappa_{a,s}(\epsilon) I^{\rm eq}(\epsilon, T, \mu) \text{d}\epsilon}{\int_0^{+\infty} I^{\rm eq}(\epsilon, T, \mu) \text{d}\epsilon}, 
    \label{eq:ka_gray} \\
    \kappa^{\rm eq}_{n}   &= \frac{\int_0^{+\infty} \kappa_{a}(\epsilon)  n^{\rm eq}(\epsilon,  T, \mu) \text{d}\epsilon}{\int_0^{+\infty} n^{\rm eq}(\epsilon, T, \mu) \text{d}\epsilon},
    \label{eq:kn_gray}
\end{align}
where $\epsilon$ is the neutrino energy, $I^{\rm eq}$ and $n^{\rm eq}$ are the spectral energy density and number density at equilibrium, respectively. $T$ is the fluid's temperature and $\mu$ is the neutrino chemical potential at equilibrium. We assume $I^{\rm eq} \sim \epsilon^3 f_{\rm FD}(\epsilon, T, \mu)$ and $n^{\rm eq} \sim \epsilon^2 f_{\rm FD}(\epsilon, T, \mu)$ with $f_{\rm FD}$ being the ultrarelativistic Fermi-Dirac distribution function.

The fluid's temperature $T$ is one of the primitive variables provided by the hydrodynamic sector in our new BAM implementation~\cite{Gieg:2022} while the chemical potential at equilibrium $\mu$ is obtained by the nuclear EoS table. The latter actually provides the chemical potential for $e^-$, $n$, and $p$. Based on these, we can compute the potentials for neutrinos assuming $\beta$-equilibrium, i.e.
\begin{eqnarray}
    \mu_{\nu_e} = \mu_{e^-} + \mu_{p} - \mu_{n}, \quad \mu_{\overline{\nu}_e} = - \mu_{\nu_e}, \quad  \mu_{\nu_x} = 0.
\end{eqnarray}

The frequency-dependent opacities $\kappa_{a,s}(\epsilon)$ are obtained from the open source code \texttt{NuLib} \cite{OConnor:2014sgn} available at \url{http://www.nulib.org}. For a given EoS, they are evaluated as functions of the fluid's rest mass density $\rho$, temperature $T_{f}$, and electron fraction $Y_e$ and given in the form of a 4D table.
For every value of $\epsilon$ in the opacity table, we perform a 3D interpolation with respect to the other three variables ($\rho$, $T$, $Y_e$) to get $\kappa_{a,s}(\epsilon)$. Finally, we use those values to discretize and evaluate the integrals in Eqs.~(\ref{eq:ka_gray}-\ref{eq:kn_gray}) to obtain the desired opacities.
For our work, we use 400 points for $\rho$, 180 for $T$, 60 for $Y_e$, and 24 for $\epsilon$. 
Table~\ref{tab:reactions} lists all reactions taken into account for the calculation of the spectral opacities and the emissivities with related references about the calculation method. We note that, in principle, \texttt{NuLib} could include more reactions.

As also reported in Ref.~\cite{Endrizzi:2019trv}, \texttt{NuLib} tables give an unphysically high opacity in regions with $\rho<10^{11}$~g/cm$^3$ and $T \lesssim 0.35$~MeV. This is because of blocking factors that are applied to the absorption opacities for $\rho>10^{11}$~g/cm$^3$. Unfortunately, the application of blocking factors in lower-density regions leads to numerical issues for 1~MeV $\lesssim T \lesssim$ 30~MeV. 
Therefore, we modified the original \texttt{NuLib} code to extend the domain of application of absorption blocking factors to the regions where $T<0.35$~MeV and $Y_e \lessgtr 0.4$ ($>$ for $\nu_e$, $<$ for $\overline{\nu}_e$) independently on $\rho$, in addition to the region $\rho>10^{11}$~g/cm$^3$. This ensures that we obtain a smooth table that is free of unphysical absorption opacities that were previously affecting the low-density and low-temperature regions. \\

\begin{table}[t]
\begin{tabular}{l||l}
                          & References         \\ \hline \hline
Charged Current Processes & \multirow{4}{*}{} \\ \hline
$\nu_e + n \leftrightarrow p + e^-$                   &    \cite{Burrows:2004vq} \cite{Horowitz:2002weak}\\
$\overline{\nu}_e + p \leftrightarrow n + e^+$        &    \cite{Burrows:2004vq} \cite{Horowitz:2002weak}\\
 $\nu_e + (A,Z) \leftrightarrow (A,Z+1) + e^-$        &    \cite{Burrows:2004vq} \cite{Bruenn:1985stellar}\\ \hline \hline
Thermal Processes         & \multirow{3}{*}{} \\ \hline
$e^- + e^+ \leftrightarrow \nu_x + \overline{\nu}_x$ &   \cite{Burrows:2004vq} \cite{Bruenn:1985} \cite{Yueh:1976neutrino}\\
$N+N \leftrightarrow N+N+\nu_x + \overline{\nu}_x$    & \cite{Burrows:2004vq}    \\ \hline \hline
Elastic Scattering        &                   \\  \hline
$\nu + \alpha \rightarrow \nu + \alpha$  &    \cite{Burrows:2004vq} \cite{Bruenn:1985stellar}              \\
$\nu + p \rightarrow \nu + p$        &    \cite{Burrows:2004vq} \cite{Bruenn:1985stellar}   \cite{Horowitz:2002weak}            \\
$\nu + n \rightarrow \nu + n$            &    \cite{Burrows:2004vq} \cite{Bruenn:1985stellar}   \cite{Horowitz:2002weak}      \\
$\nu + (A,Z) \rightarrow \nu + (A,Z)$    &     \cite{Burrows:2004vq} \cite{Bruenn:1985stellar}  \cite{Horowitz:1997neutrino}   
\end{tabular}
    \caption{Weak interaction processes taken into account in our work using \texttt{NuLib}. All processes involving charged particles include Weak-Magnetism and recoil corrections from \cite{Horowitz:2002weak}. Charged current and thermal processes sections only include absorption and emission processes. Their inverse processes are also taken into account through the Kirkhhoff law. In the last section, the neutrino family is not specified since all families are involved in these processes, even though, eventually, with different cross sections.}
    \label{tab:reactions}
\end{table}

So far, we assumed neutrinos to be in equilibrium with the fluid. However, this is, in general, not the case. Since the cross sections of neutrinos scale with $\epsilon^2$, the assumption of neutrinos at equilibrium with the fluid would lead to an underestimate of opacities when hot neutrinos out of equilibrium cross a region of cooler fluid. To take this energy dependence into account, we apply the correction
\begin{equation}
    \kappa_{a,s,n} = \kappa_{a,s,n}^{eq} \left( \frac{T^{\nu}_{\rm eff}}{T} \right)^2,
    \label{eq:ka_corr}
\end{equation}
which is also used in most other grey M1 implementations \cite{Foucart:2016rxm,Radice:2021jtw,Musolino:2023pao}. Where $T^{\nu}_{\rm eff}$ is the effective temperature of neutrinos.
To obtain $T^{\nu}_{\rm eff}$, we assume neutrinos spectrum to be Planckian with temperature $T^{\nu}_{\rm eff}$ and reduced chemical potential $\eta_{\nu} = \mu/T_f$. We can then evaluate the average neutrino energy as
\begin{equation}
    \langle \epsilon_{\nu} \rangle = \frac{F_3(\eta_{\nu})}{F_2(\eta_{\nu})} T^{\nu}_{\rm eff},
    \label{eq:average_en_nu}
\end{equation}
with $F_k$ being the Fermi integral of order $k$. Since we know $\langle \epsilon_{\nu} \rangle = J/n$, we can solve Eq.~\eqref{eq:average_en_nu} for
\begin{equation}
    T^{\nu}_{\rm eff} = \frac{F_2(\eta_{\nu})}{F_3(\eta_{\nu})} \frac{J}{n},
    \label{eq:T_eff}
\end{equation}
and plug $T^{\nu}_{\rm eff}$ into Eq.~\eqref{eq:ka_corr} to obtain the corrected opacity.

EoS tables only provide chemical potentials of neutrinos at thermal equilibrium with the fluid. When neutrinos get decoupled from the latter, we expect to approach a distribution with zero chemical potential, i.e., the distribution describing a fixed number of particles. As in \cite{Foucart:2015vpa}, to qualitatively account for this transition, we are evaluating the reduced chemical potentials of Eq. \eqref{eq:T_eff} in the following way:
\begin{equation}
    \eta_{\nu} =\frac{\mu}{T}(1 - e^{-\tau}),
    \label{eq:nu_red}
\end{equation}
where $\tau$ is the optical depth provided by the NLS of \cite{Gieg:2022}.

Finally, once we have set $\kappa_{a,s,n}$, we remain with $\eta$ and $\eta_n$ to be set. For that, we assume the Kirchhoff law
\begin{align}
    \eta   &= \kappa^{eq}_a \frac{4\pi}{(hc)^3} F_3(\eta_{\nu}) T^4, \\
    \eta_n &= \kappa^{eq}_n \frac{4\pi}{(hc)^3} F_2(\eta_{\nu}) T^3.
    \label{eq:emissivity}
\end{align}
Such a choice ensures that neutrinos thermalize with the fluid and reach the expected thermal equilibrium state when trapped.

\section{Numerical Scheme}
\label{Numerical Scheme}

We implemented the above multipolar formalism for radiation transport as a new module in the BAM code~\cite{Bruegmann:2006at,Thierfelder:2011yi,Dietrich:2015iva,Bernuzzi:2016pie,Gieg:2022}, and will provide implementation details below.

\subsection{Closure factor}
Equation~\eqref{eq:closure_parameter} cannot be evaluated directly since $H^{\alpha}$ and $J$ are functions of $P^{ij}$ (and so of $\zeta$); cf.~Eqs.~\eqref{eq:J} and \eqref{eq:H}. 
Hence, the closure factor must be found by solving the following implicit equation
\begin{equation}
    \frac{\zeta^2 J^2(\zeta) - H_{\alpha}(\zeta)H^{\alpha}(\zeta)}{E^2} = 0
    \label{eq:implicit_root}
\end{equation}
using a root finder algorithm.
 
In our implementation, we solve Eq.~\eqref{eq:implicit_root} for $\zeta$ using a Dekker algorithm \cite{dekker1969finding}, which improves the convergence speed compared to the bisection scheme.

\subsection{Fluxes}

To evaluate the numerical fluxes at cell interfaces, we follow Ref.~\cite{Radice:2021jtw}. Given a field variable $u$ and its flux $\mathcal{F}(u)$, we employ a linear combination of a low-order diffusive flux $\mathcal{F}_{\rm LO}$ and a second-order non-diffusive flux $\mathcal{F}_{\rm HO}$ so that:
\begin{equation}
    \mathcal{F}_{i+1/2} = \mathcal{F}_{\rm HO}(u_{i+1/2}) - A \left [ \mathcal{F}_{\rm HO}(u_{i+1/2}) - \mathcal{F}_{\rm LO}(u_{i+1/2})  \right ] , 
    \label{eq:fluxes}
\end{equation}
where $A = {\rm min}(1, \frac{1}{\overline{\kappa} \Delta x})$ and $\overline{\kappa} = (\kappa_{s,i} +  \kappa_{s,i+1} + \kappa_{a,i} +  \kappa_{a,i+1}$)/2.
This ansatz leads to $\mathcal{F}_{i+1/2} = \mathcal{F}_{\rm LO}$ in the free streaming regime and to $\mathcal{F}_{i+1/2} \simeq \mathcal{F}_{\rm HO}$ in the scattering/absorption regime. 
The low-order diffusive flux is computed using fluxes at the cell center, and a local Lax-Friedrichs (LLF) Riemann solver~\cite{Nessyahu:1990, Kurganov:2000}:
\begin{equation}
    \mathcal{F}_{\rm LO}(u_{i+1/2}) = \frac{\mathcal{F}(u_i) + \mathcal{F}(u_{i+1})}{2} - \lambda_{\rm amax} \frac{u_{i+1} - u_i}{2},
    \label{eq:llf}
\end{equation}
with
\begin{equation}
\lambda_{\rm amax} = \max _{a \in\{i, i+1\} b \in [1,2] } \left\{ |\lambda^b_a| \right\}, 
\end{equation}
where $\lambda^b$ are the characteristic velocities of the system. This choice ensures the monotonicity preservation of the solution in case of shocks and leads to better stability in the free streaming regime due to an increased numerical dissipation. 
Certainly, in some cases, the latter can also be a disadvantage, e.g., in the case of radiation in an optically thick medium, it would introduce an unphysical diffusion, leading to a wrong estimation of the neutrinos diffusion rate. To avoid this effect, we employ the following non-diffusive scheme in optically thick regions:
\begin{eqnarray}
    \mathcal{F}_{\rm HO}(u_{i+1/2}) = \frac{1}{2} \left [ \mathcal{F}(u_i) + \mathcal{F}(u_{i+1}) \right ].
    \label{eq:non_diff_flux}
\end{eqnarray}
Our choice of $\mathcal{F}_{\rm HO}$ cures the unphysical diffusion of $\mathcal{F}_{\rm LO}$, but in case of shocks, it can violate monotonicity preservation.

To make the scheme described by Eq.~\eqref{eq:fluxes} able to handle shocks in a thick regime without adding unphysical diffusion in smooth regions, we first compute $\mathcal{F}$ in every point using Eq.~\eqref{eq:fluxes} and then set $\mathcal{F} = \mathcal{F}_{\rm LO}$ if one of the following conditions is satisfied:
\begin{itemize}
    \item $\Delta^n_{i-1}     \Delta^n_i     < 0$    or   $\Delta^n_i     \Delta^n_{i+1}     < 0$, i.e., if the solution at the current time step shows an extremum.
    \item $\tilde{E}^{n+1}_{i} \leq 0$ or $\tilde{E}^{n+1}_{i+1} \leq 0$, i.e., if the energy solution at the next time step would be overshoot to a negative value.
    \item $\frac{\Delta^{n+1}_{i-1}}{\Delta^{n+1}_i} < \frac{1}{4}$ or $\frac{\Delta^{n+1}_{i}}{\Delta^{n+1}_{i+1}} < \frac{1}{4}$, i.e., if the solution at next time step would develop an extremum or if the change in the slope happens too quickly,
    \label{eq:LO_lux_conditions}
\end{itemize}
where
\begin{align*}
    \Delta_i^n & = u^n_{i+1} - u^n_i, \\
    u^{n+1}_i  & = u_i^n - \frac{\Delta t}{\Delta x} \left ( \mathcal{F}^n_{i} - \mathcal{F}^n_{i-1} \right ). \\
\end{align*}
Characteristic velocities are constructed as a linear combination of thin and thick velocities with the same coefficients used to compose the closure in Eq.~\eqref{eq:closure}. We chose to employ the following velocities in a generic $i^\textrm{th}$-direction:
\begin{align}
    \lambda_{\rm thin} ^ {1,2} =& -\beta^i \pm \alpha |F^i|/|F|, \\
    \lambda_{\rm thick} ^{1,2} =& -\beta^i \pm \alpha \sqrt{\gamma^{ii}/3}.
\end{align}
In our tests, this particular choice increases the stability of our scheme without affecting the accuracy. Thin velocities are the same as employed in \cite{Foucart:2015vpa} while the thick ones are taken from \cite{Radice:2021jtw}. For a complete list and discussion of the characteristic velocity, we refer to \cite{Shibata:2011kx}.

We note that, in principle, Eq.~\eqref{eq:fluxes} is second-order accurate in the diffusive region far away from shocks or solution's extrema and first-order accurate in the free streaming regime.

\subsection{Implicit-explicit time step}
Scattering and absorption opacities, even in the geometrized units handled by BAM, can reach large values up to $10^{3} \Delta t$ in very thick regions, e.g., in the neutron star interior. For such values, the collisional source terms $S^{\alpha}$ of Eqs.~(\ref{eq:evolution_E}, \ref{eq:evolution_F}) become stiff, i.e., we have to treat the terms through an implicit scheme. Fluxes and gravitational sources are instead handled via a second-order explicit-implicit method.
A full time step of our radiation evolution algorithm is given by:
\begin{align}
    \frac{\textbf{q}^* - \textbf{q}^n}{\Delta t} &= - \partial_i \mathbfcal{F}^i(\textbf{q}^n) + \textbf{G}(\textbf{q}^n) + \textbf{S}_{\textrm{coll}}(\textbf{q}^*),
    \label{eq:evolution_substep1} \\
    \frac{\textbf{q}^{n+1} - \textbf{q}^n}{\Delta t} &= - \partial_i \mathbfcal{F}^i(\textbf{q}^*) + \textbf{G}(\textbf{q}^*) + \textbf{S}_{\textrm{coll}}(\textbf{q}^{n+1}),
    \label{eq:impl_sys} 
\end{align}
where $\textbf{q} = (\tilde{E}, \tilde{F}_i, N)$, and $\mathbfcal{F}^i$, $\textbf{G}$, and $\textbf{S}_{\textrm{coll}}$ represent fluxes, gravitational-source terms, and collisional source terms of Eq.~\eqref{eq:evolution_E}, \eqref{eq:evolution_F}, and \eqref{eq:number_transp_coo}, respectively. This method is 2nd-order accurate in fluxes and gravitational terms but only 1st-order accurate in the implicit terms. \\
Hydrodynamics variables are kept constant during the radiation substep of Eq.~\eqref{eq:evolution_substep1}, and are updated after the second step using $\textbf{S}_{\textrm{coll}}(\textbf{q}^{n+1})$. Analogously radiation variables are kept constant during the hydrodynamics and spacetime evolution substeps, which are performed ignoring radiation-fluid interactions.

The application of a partially implicit method requires the solution of a system in $\tilde{E}^{n+1}$, $\tilde{F}^{n+1}_i$, and $N^{n+1}$.
The last variable is decoupled from the rest of the system since its implicit time step can be written as
\begin{equation}
    \frac{N^{n+1} - N^n}{\Delta t} = -\partial_i \mathcal{F}^i_N(N^{n}) + \alpha\sqrt{\gamma}\eta_n - \kappa_n \frac{N^{n+1}}{f^0},
\end{equation}
which can be solved straightforwardly for $N^{n+1}$ as:
\begin{equation}
    N^{n+1} = \frac{N^n -\partial_i \mathcal{F}^i_N(N^{n}) \Delta t + \alpha\sqrt{\gamma}\eta_n \Delta t}{1 + \frac{k_n}{f^0} \Delta t }.
    \label{eq:impl_N}
\end{equation}

Solving for the neutrino momenta, unfortunately, requires more effort. We follow the linearized scheme of \cite{Foucart:2015vpa}. Plugging the expression of $J$ and $H^{\alpha}$ in Eqs.~\eqref{eq:J} and \eqref{eq:H}, into the definition of $S^{\alpha}$ in Eq.~\eqref{eq:collisional_sources}, and linearizing assuming $\zeta$ and $F^i/|F|$ to be constant, we can write
\begin{equation}
\label{eq:sources_linear}
    \tilde{S}_{\alpha}^{n+1} = \tilde{E}^{n+1} A_{\alpha} + \tilde{F}^{n+1}_iB^{i}_{\alpha} + \sqrt{\gamma} \eta u_{\alpha},
\end{equation}
where $A^{\alpha}$ and $B^{i\alpha}$ are tensor functions of $\kappa_a$, $\kappa_s$, $\zeta$, $F^i/|F|$, and $u^{\alpha}$ only. In our implementation, we solve for the implicit time step using the value of $\zeta$ and $F^i/|F|$ at time step $n$. This is necessary for obtaining a linear system in ($\tilde{E}^{n+1},\tilde{F}_i^{n+1}$). Plugging this expression into Eq.~\eqref{eq:impl_sys}, we get a system of four linear equations for four variables, whose analytical solution can be found in Appendix \ref{sec:appendixA}.

We have pointed out that the scheme used for collisional terms is not fully implicit since $A^{\alpha}$ and $B^{i\alpha}$ are also dependent on neutrino variables, and the values of the fluid's opacities are not updated according to the radiation-fluid interaction at each substep. 
However, the solution of a fully coupled implicit system would require a non-linear root finder and an update of fluid variables at each substep with a significantly higher computational cost. For this reason, we limit ourselves to this linearized implicit scheme, which we found to be enough to ensure the stability of the code.

Finally, it is worth mentioning that other M1 implementations \cite{Anninos_2020,Radice:2021jtw,Musolino:2023pao} treat the nonlinear terms of $S_{\textrm{coll}}$ implicitly. This is equivalent to treating the ratio $f^i = F^i/|F|$ as a variable at time $n+1$ and solving for a four-dimensional nonlinear root-finding problem. This scheme is believed to be more accurate in describing the interaction of radiation with a fast-moving fluid since it handles better the terms proportional to $v \cdot f$. However, in the next section, we will show that our linearized scheme properly captures the advection of trapped radiation by a moving fluid, which is a stringent test that must be satisfied by a radiation transport code oriented to the simulation of BNS mergers. \\

\subsection{Neutrino right-hand side routine}
In the following, we summarize the steps followed to evaluate the full right-hand side (RHS) of the neutrino sector from $E$, $F^i$, and $N$, i.e., Eqs.~(\ref{eq:evolution_E}, \ref{eq:evolution_F}, \ref{eq:number_transp_coo}): \\
\begin{itemize}
    \item We check for causality constraint $E \leq |F|$, if it is not satisfied we set $E=|F|$ by rescaling $F^i$. This step is necessary to ensure the causality and hyperbolicity of the scheme.
    \item We evaluate the closure factor $\zeta$ solving Eq.~\eqref{eq:implicit_root}. We use $\zeta$ to evaluate the fluid frame energy $J$ and the neutrino number density $n$.
    \item We set opacities using the scheme described in the previous section.
    \item We compute fluxes at the cell's interfaces using Eq.~\eqref{eq:fluxes}.
    \item We check whether the reconstructed fluxes satisfy one of the conditions listed above. If they do, we recompute them using Eq.~\eqref{eq:llf}.
    \item We add the fluxes divergence to the RHS.
    \item We evaluate the gravitational source terms of Eq.~\eqref{eq:evolution_E} and \eqref{eq:evolution_F} and add them to RHS.
    \item We solve Eq.~\eqref{eq:impl_sys} for $E^{n+1}$, $F_i^{n+1}$, and $N^{n+1}$ using the explicit part of RHS we have evaluated in the previous points.
\end{itemize}

\bigskip

All the steps listed above are repeated for the three neutrino species.

\section{Numerical Tests}
\label{Numerical testes}

\subsection{Geodesics}

\begin{figure}
    \centering 
    \includegraphics[width=0.45\textwidth]{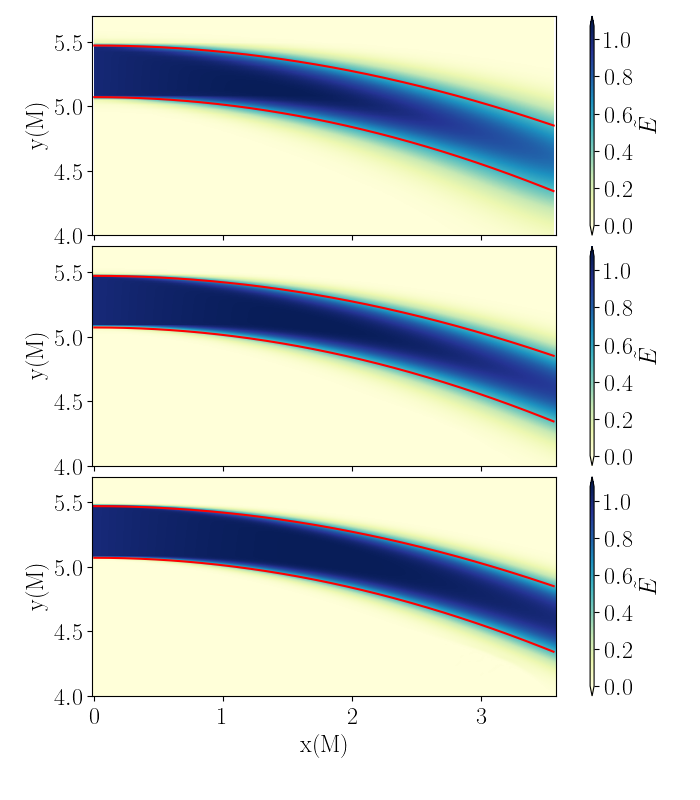}
    \caption{A neutrino beam traveling in a Kerr-Schild spacetime with three different resolutions. In the top panel, we use a grid spacing of $\Delta x = 0.075$. In the central and bottom panels, we have $\Delta x /2$ and $\Delta x /4$, respectively. Coordinates with respect to the singularity are expressed in mass units. The color map represents the energy density of the neutrinos $\tilde{E}$. Red lines represent the expected geodesics that should contain the beam. } 
    \label{fig:test_bh2d}
\end{figure}

To test the fluxes and gravitational sources, we set up a test employing Kerr-Schild spacetime with zero angular momentum, where we shoot a beam of free streaming neutrinos with $\tilde{E} = |\tilde{F}| = 1$ from left to the right of our numerical domain. 
In Fig.~\ref{fig:test_bh2d}, the neutrino beam is injected in the simulation tangentially to the black hole (BH) horizon at a coordinate distance $5M$ to $5.5M$ from the singularity (which is located in the origin of the coordinate system). The red lines in the figure show light-like geodesics that neutrinos at the top and bottom of the beam are supposed to follow. The whole beam should then be contained between these two lines.
We observe that most of the neutrino energy remains confined between the two geodesics with a small part that is dispersed outside, mostly because of the low-order reconstruction scheme employed to handle the free streaming region, which introduces numerical dispersion. This interpretation is strengthened by the fact the dispersion happens on both sides of the beam and decreases with increasing grid resolution. However, we expect that this is not an issue in BNS simulations since we do not expect to have sharp variations of the energy density in free streaming regions as in this test case.

\subsection{Absorption}

To test the collisional source terms that model the neutrino-fluid interaction in the pure absorption regime, we set up two tests on a flat spacetime, one with a static and one with a stationary moving fluid. In both tests, we shoot a beam of neutrinos similar to the previous case.
Since, in these conditions, neutrinos should be only absorbed and not scattered, we expect their regime to remain purely thin and the momentum vectors to remain parallel to each other.

\begin{figure}
    \centering 
    \includegraphics[width=0.49\textwidth]{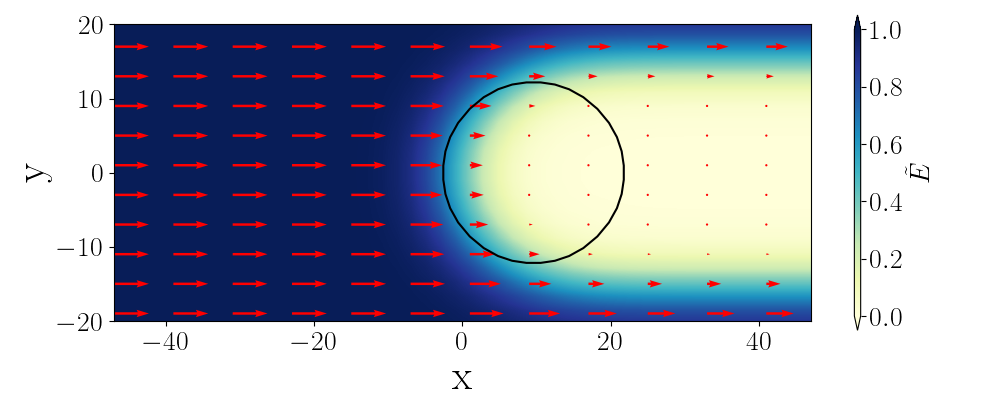}
    \caption{
        Neutrinos traveling from left to right, encountering a sphere of matter at rest with pure absorption opacity. The center of the sphere is located at $(10,0)$ and has an opacity of $\kappa_a=0.5$ that decreases radially as a Gaussian. Red arrows represent neutrino momentum $\tilde{F}^i$. The black line shows the contour level of $\kappa_a=0.1$ for the sphere of matter.
    } 
    \label{fig:test_shadow}
\end{figure}

In Fig.~\ref{fig:test_shadow}, we show a wide beam of neutrinos moving from left to right encountering a sphere of matter with $\kappa_a=0.5$ in the center and decreasing radially as a Gaussian. As expected, neutrino momenta remain parallel to each other, and as a consequence, the region behind the sphere receives a much smaller amount of radiation when compared to regions on the sides, projecting a very clear shadow on the right edge of the simulation domain.

\begin{figure}
   \centering 
    \includegraphics[width=0.49\textwidth]{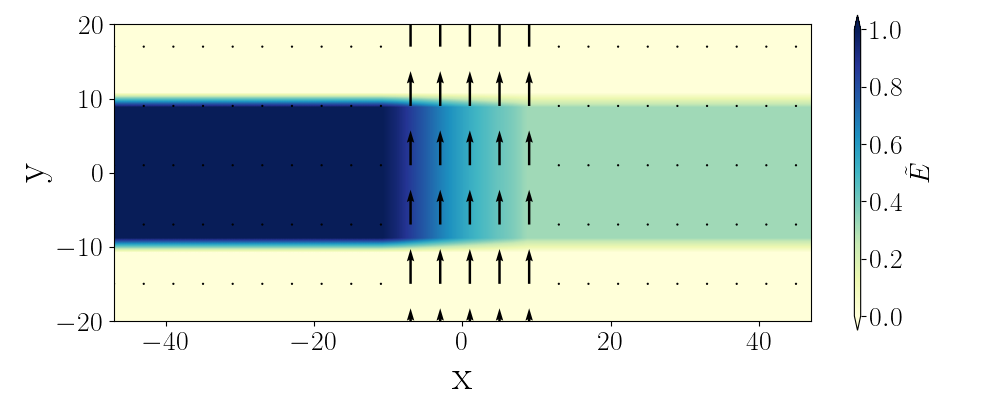}
    \caption{
        A Neutrino beam traveling from left to right passing through a tube of fluid with density $\rho=10^{-5}$ and absorption opacity $\kappa_a = 0.05$ traveling upward with velocity $v_y=0.5$ (represented by black arrows).
    } 
    \label{fig:test_ka_tube_E}
\end{figure}

In the second absorption test, shown in Fig.~\ref{fig:test_ka_tube_E}, we distribute matter on a vertical tube with homogeneous properties, $\kappa_a=0.05$ and $v_y = 0.5$. As expected, we observe that a part of the radiation is absorbed by the fluid, and another part passes through it without being scattered. In contrast to the previous test, the fluid is not at rest. Hence, the test is well suited to probe the conversion between the fluid and the laboratory frame, which were equal in our previous test, i.e., $\zeta=1$ was trivially satisfied along the beam. 
In this new test, we still expect to find $\zeta=1$. However, since now $H^{\alpha}H_{\alpha} \neq F_i F^i$ and $E \neq J$, this is not trivial anymore. 
Based on the success of the test, we can conclude that the root finder algorithm used to evaluate $\zeta$ is converging to the correct solution.

\subsection{Advection}
\begin{figure}[t]
    \centering 
    \includegraphics[width=0.45\textwidth]{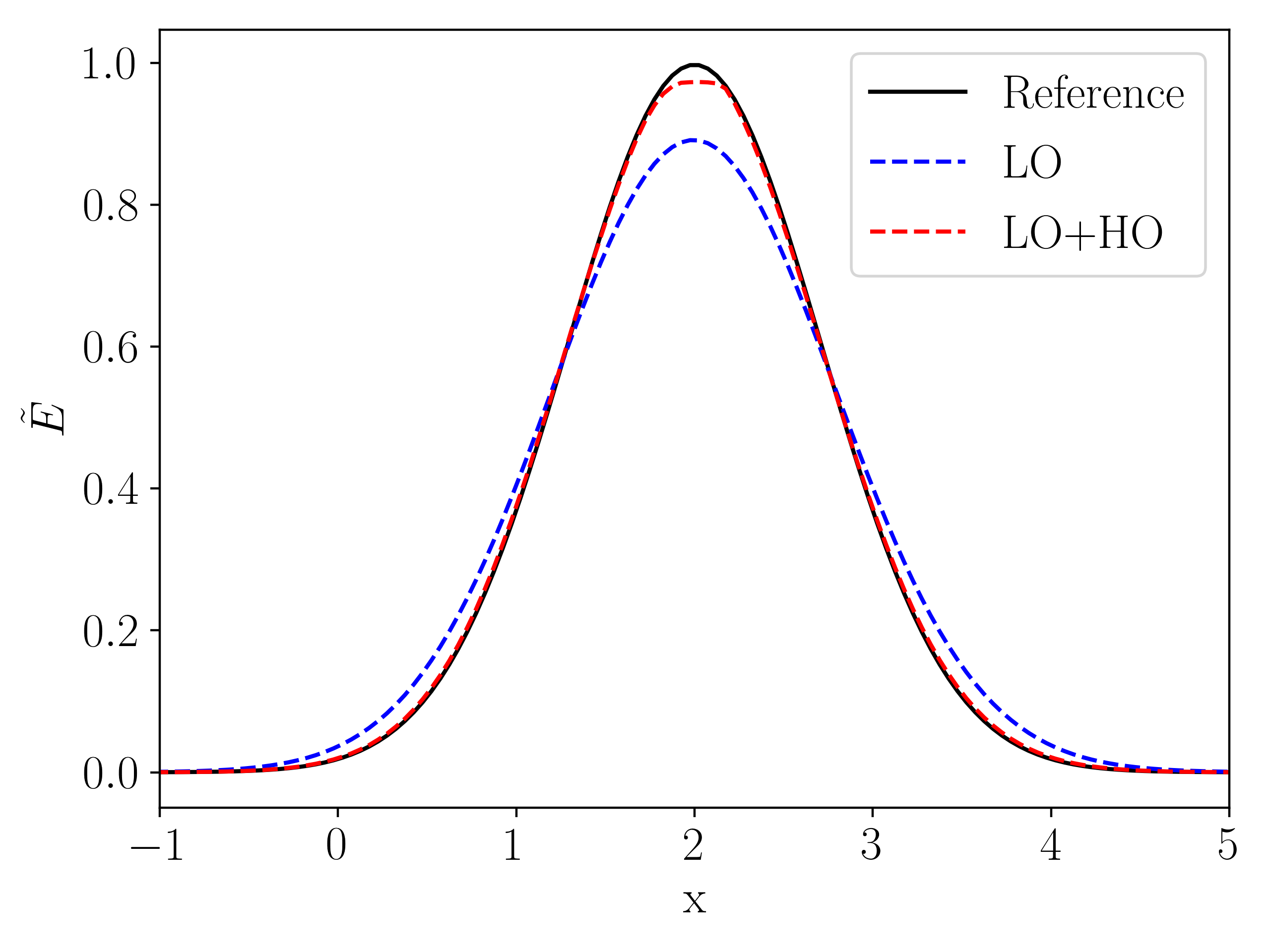}
    \includegraphics[width=0.45\textwidth]{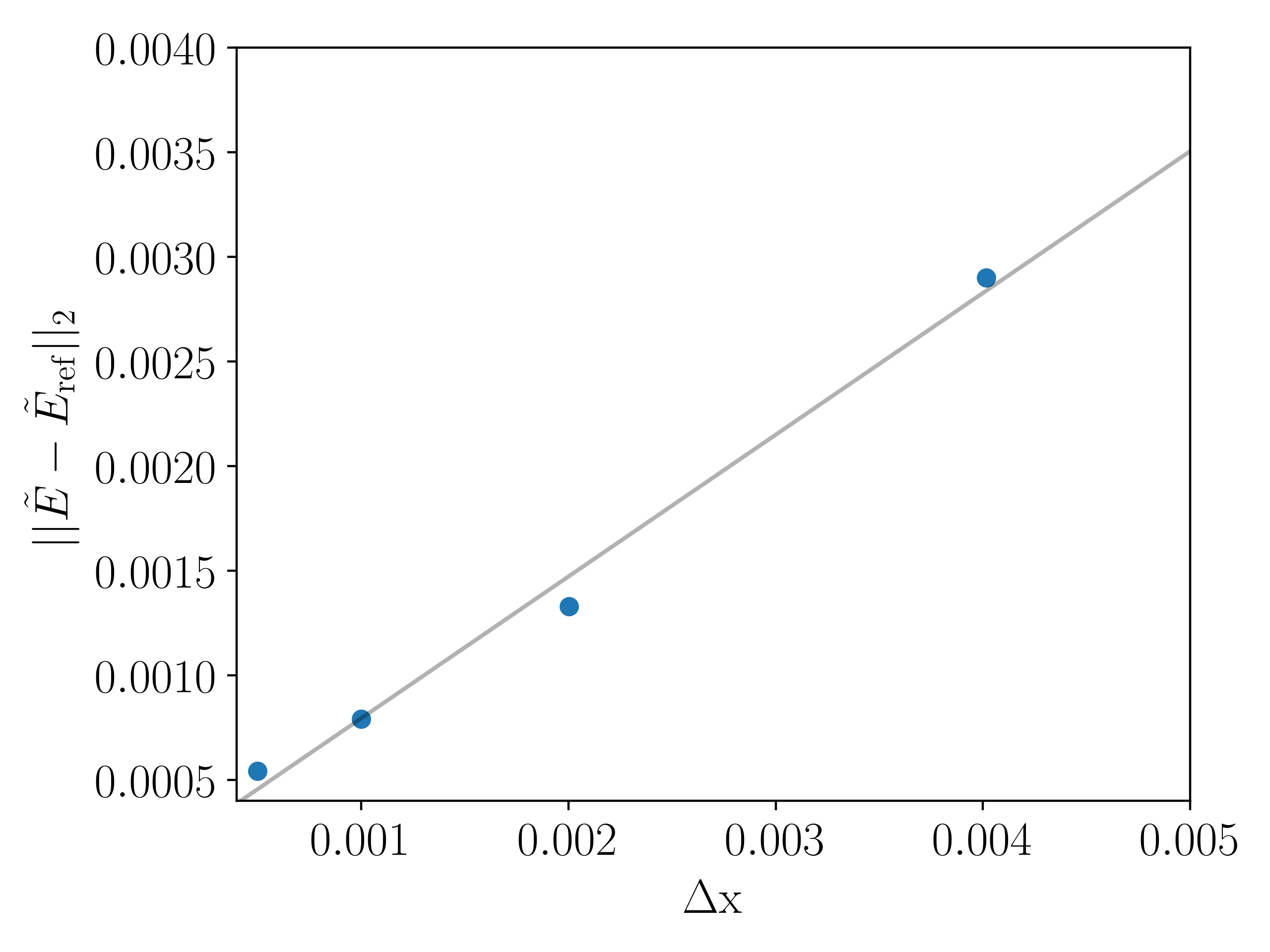}
    \caption{\textbf{Top panel}: Gaussian neutrino packet being advected and diffused in high scattering regime by a fluid with $\kappa_s=10^3$ and moving to the right with a velocity of $v=0.5$. The snapshot is taken at time $t=4$. Different dashed lines represent different schemes for fluxes reconstruction, while the black line represents the analytical solution of the translated heat equation. \textbf{Bottom panel:} L2-norm of the error with respect to resolution for the test shown above with HO+LO fluxes. The black line shows first-order convergence.}
    \label{fig:test_advection} 
\end{figure}
Advection of trapped radiation in a moving fluid is one of the most challenging situations that our code has to handle. To test such a scenario, we set up a test similar to the one shown in Sec.~4 of Ref.~\cite{Radice:2021jtw}, i.e., we evolve a one-dimensional Gaussian neutrino packet trapped in a homogeneous fluid moving at mildly relativistic velocity with stiff, pure-scattering opacity. \\
As initial conditions, we chose 
\begin{equation}
    \tilde{E}(t=0,x) = e^{-x^2}, \quad J = \frac{3E}{4W^2-1}, \quad F_i = \frac{4}{3}JW^2 v_i.
\end{equation}
As shown in \cite{Radice:2021jtw}, with this condition for $F_{\alpha}$, it is ensured that $H^{\alpha}=0$, i.e., we model a fully thick regime. For the fluid, we chose $\kappa_s=10^3$ and $|v|=v_x=0.5$. We use a single uniform grid with $\Delta x = 0.05$ and employ a Courant-Friedrich-Levi (CFL) factor of $0.25$. We test two different flux reconstruction schemes to check whether they can capture the correct diffusion rate in the regime $k_s \Delta x \gg 1$. In this article, we test two different schemes: a constant reconstruction ($u_{i+1/2}=u_i$) with an LLF Riemann solver (Eq.~\eqref{eq:llf}) and the composed flux of Eq.~\eqref{eq:fluxes} proposed in \cite{Radice:2021jtw}.

Results are shown in Fig.~\ref{fig:test_advection} together with the reference solution, which we assume to be the advected solution of the diffusion equation
\begin{equation}
    \tilde{E}(x,t) = \frac{1}{\sqrt{1+4Dt}} \exp\left[-\frac{(x-v_xt)^2}{1+4Dt}\right],
\end{equation}
with $D=1/(3\kappa_s)$ being the diffusivity. We see that the lowest order reconstruction scheme [Eq.~\eqref{eq:llf}] fails in reproducing the correct diffusion rate because of its intrinsic numerical dispersion. The scheme used in \cite{Radice:2021jtw}, instead, performs better except near the maximum, where it reduces again to the lower order one. Moreover, we observe no unphysical amplification of the package, contrary to the test performed using $\texttt{ZelmaniM1}$ library \cite{Roberts:2016lzn} in \cite{Radice:2021jtw}. In our implementation, we find that both the neutrino energy and the neutrino number are advected with the correct velocity.

\begin{figure}
    \centering 
    \includegraphics[width=0.45\textwidth]{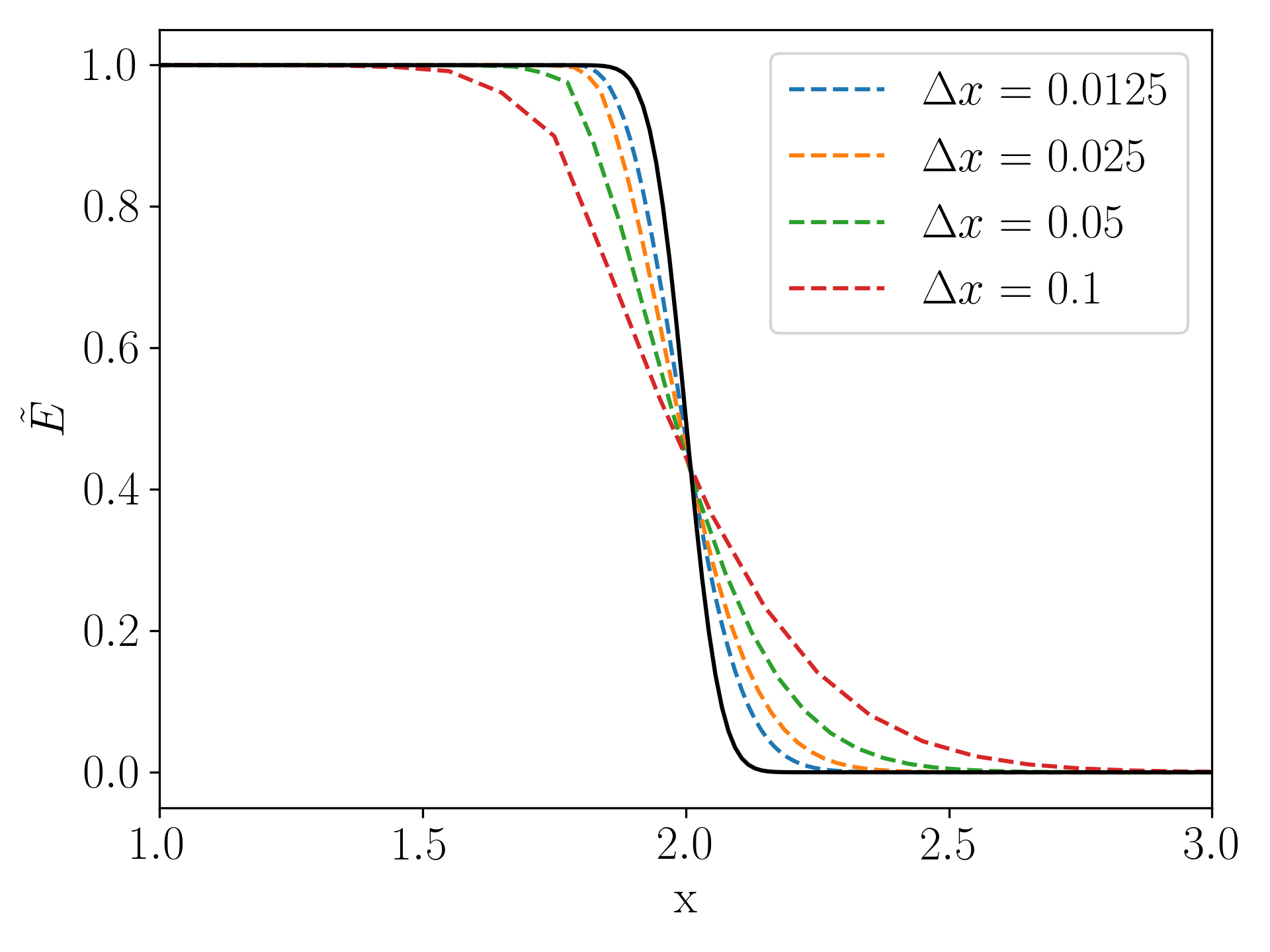}
    \caption{A similar test as performed in Fig.~\ref{fig:test_advection} but using a step function as initial condition. The exact solution is shown by a black line, while colored lines show numerical solutions at different resolutions. The evolution scheme and grid configuration are the same as in Fig.~\ref{fig:test_advection}.}
    \label{fig:test_advection_step} 
\end{figure}

To test the robustness of the scheme, we performed an additional test with identical fluid configuration but neutrinos' initial data given by a step function. Results at time $t=4$ are shown in Fig.~\ref{fig:test_advection_step} for different resolutions together with the reference solution
\begin{equation}
    \tilde{E}(x,t) = \frac{1}{2} \left [ 1 - \mathrm{erf} \left (\frac{x-v_x t}{2\sqrt{Dt}} \right ) \right ],
\end{equation}
with $\mathrm{erf}$ being the error function. 
This test shows that the flux reconstruction, Eq.~\eqref{eq:fluxes}, together with the linearized collisional sources, Eq.~\eqref{eq:sources_linear}, can handle shocks even in the presence of stiff source terms preserving the monotonicity of the solution and with a numerical dispersion that decreases with the increase of resolution.

\subsection{Uniform Sphere}
The uniform sphere test is the closest configuration to an idealized star for which we have an analytical solution of the Boltzmann equations \cite{smit1997hyperbolicity}. For this reason, several groups have shown such simulations to test their implementations, e.g., Refs.~\cite{Radice:2021jtw, Foucart:2015vpa, smit1997hyperbolicity,Musolino:2023pao}. It consists of a sphere of radius $r_s=1$. In its interior, we set $\kappa_a=\eta={\rm constant}$ and $\kappa_s=0$. We set up this test on a 3-dimensional Cartesian grid with $\Delta x = \Delta y = \Delta z = 0.05$ imposing reflection symmetry with respect to $x-y$, $x-z$ and $y-z$ planes. Evolution is performed using a RK3 algorithm with a CFL factor of $0.25$. We perform this test with two different opacities $\kappa_1 = 5$ and $\kappa_2=10^{10}$ to test different regimes; cf.~\cite{Foucart:2015vpa}.

\begin{figure}
    \centering 
    \includegraphics[width=0.45\textwidth]{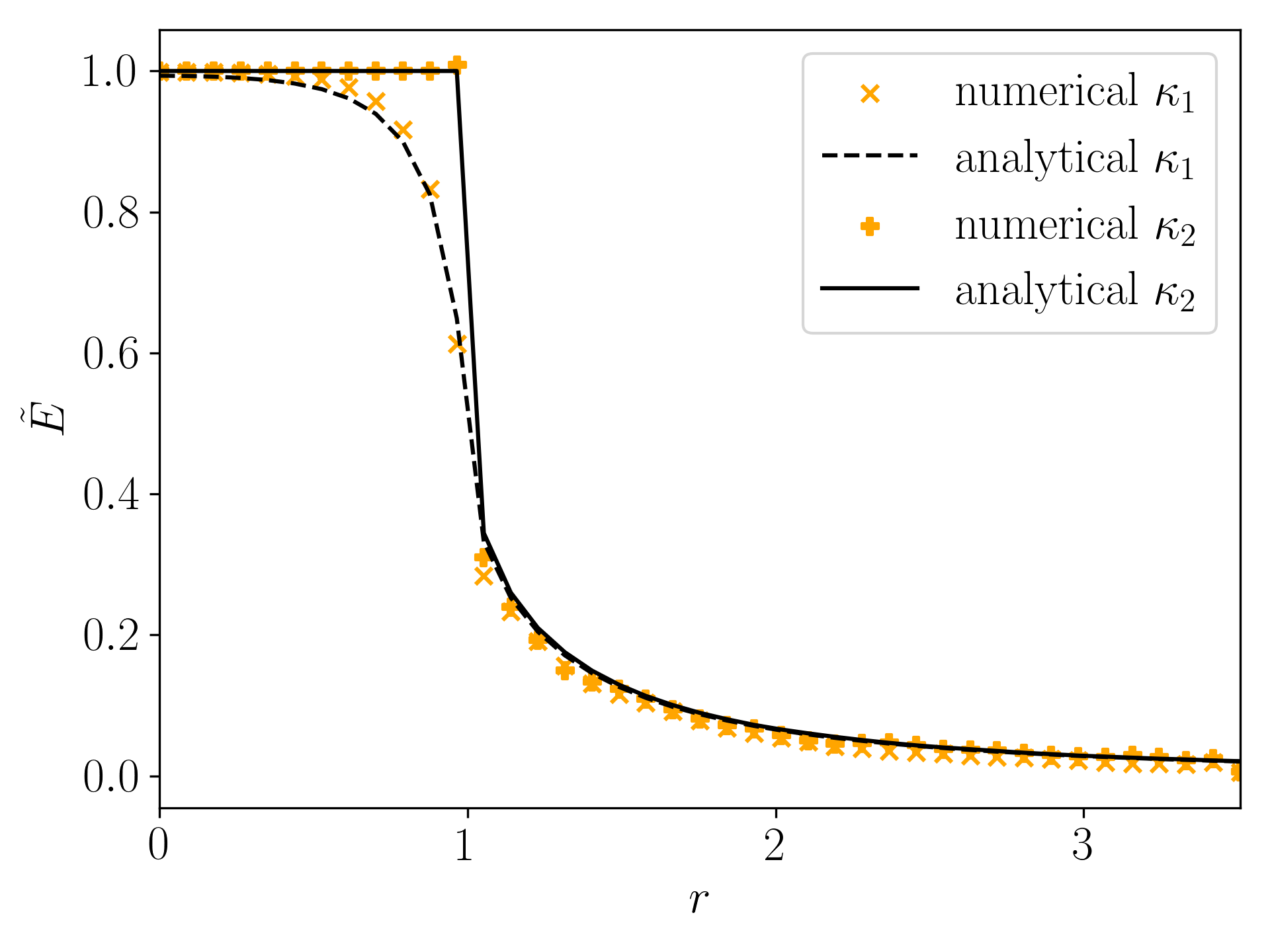}
    \caption{Radial dependence of the neutrino energy density for the uniform sphere test at time $t=12$ (profiles are extracted along the diagonal direction). We show the numerical solution of the M1 scheme and the analytical solution of the Boltzmann equation. We use two different values of opacity $\kappa_1=5$ and $\kappa_2=10^{10}$.}
    \label{fig:test_uniform_sphere} 
\end{figure}

Figure~\ref{fig:test_uniform_sphere} shows both the numerical and analytical $\tilde{E}$ as a function of the radius for the opacities. The numerical solution is taken at $t=12$ along the diagonal $x=y=z$. Overall, we find a good agreement between our numerical result and the analytical solution of the Boltzmann equation, comparable to the results obtained in other works. However, we point out that one cannot expect to converge to the exact solution since the M1 scheme is only an approximation to the Boltzmann equation and is only  exact in the  fully trapped or free streaming regimes (without crossing beams).

\subsection{Single isolated hot star}
We evolve a single isolated hot neutron star employing the SFHo EoS~\cite{Steiner:2012rk}. Initial data are constructed by solving the TOV equations with the assumption of constant entropy and beta-equilibrium as in \cite{Gieg:2022}. For the integration of the TOV equation, we choose $\rho_c = 8.65 \times 10^{14}$~g/cm$^3$ and an entropy per baryon $s = 1k_B$. This leads to a total baryonic mass $M_{\rm bar} = 1.64~M_{\odot}$, which corresponds to a gravitational mass of 1.52~$M_{\odot}$, a coordinate radius $R=9.8$~km, and a central temperature of 27.8~MeV. We evolve the system on a grid with a grid spacing of $\Delta x = \Delta y = \Delta z = 182~{\rm m}$ using a CFL factor of $0.25$. In this test, we evolve the hydrodynamics with the module of \cite{Gieg:2022} using 4th order Runge-Kutta (RK4) integration algorithm and WENOZ \cite{Borges:2008} primitive reconstruction with LLF Riemann solver for the fluxes at cell interfaces.

\begin{figure}
    \centering 
    \includegraphics[width=0.45\textwidth]{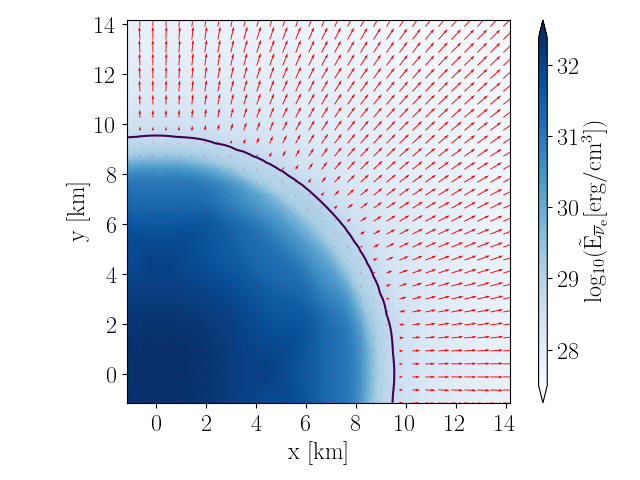}
    \caption{2D snapshot at $t = 4.5$~ms. The colormap shows the energy density of anti-electron neutrinos. Red arrows represent the normalized neutrino momentum $F^i/E$ and the black contour line shows the neutrino photosphere defined by $\tau = 2/3$.}
    \label{fig:TOV_E_F} 
\end{figure}

Fig.~\ref{fig:TOV_E_F} shows the transition of neutrinos from trapped to the free streaming regime on the surface of the star. As expected, the neutrino energy density reaches its peak in the star's core due to the higher density and temperature of the fluid in this region. Moreover, we can observe that neutrinos inside the star have a zero average momentum since they constitute a particle gas in thermal equilibrium with the fluid, and the transport phenomena are negligible. When the optical depth $\tau$ drops below $2/3$, interactions with the fluid start becoming subdominant, and neutrinos start traveling freely, developing an average momentum in the radial direction.

\begin{figure}
    \centering 
    \includegraphics[width=0.45\textwidth]{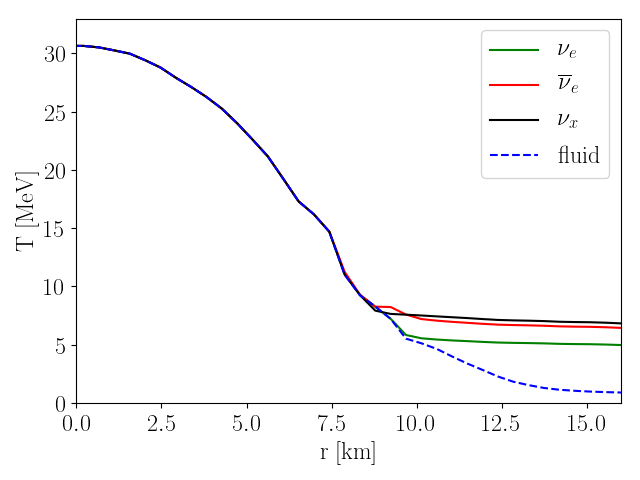}
    \caption{Temperature of fluid and neutrinos at $t=4.5$~ms as a function of radius. We can see the three neutrino species decoupling from the fluid at three different temperatures near $r=10$~km.}
    \label{fig:TOV_thermalization} 
\end{figure}

Another important consequence of the neutrino-baryon decoupling can be seen in Fig.~\ref{fig:TOV_thermalization}, where we can observe all three species of neutrinos being thermalized with the fluid in the inner part of the star and decoupling next to the relative photosphere at three different temperatures. After decoupling, the neutrino temperature remains constant due to the lack of interactions with the fluid. The average energy hierarchy is, as reported in the literature, $\langle \epsilon_{\nu_e} \rangle < \langle \epsilon_{\overline{\nu}_e} \rangle < \langle \epsilon_{\nu_x} \rangle$ \cite{Endrizzi:2019trv, Radice:2021jtw}.

\section{Binary Neutron Star Mergers}
\label{Binary Neutron Star Mergers}

\subsection{Configurations and Setup}

We run 10 different BNS configurations employing two different EoSs (SFHo \cite{Steiner:2012rk} and DD2 \cite{Hempel:2009mc}) with the same total baryonic mass of 2.6~$M_{\odot}$, and two different mass ratios of $q=M_1/M_2=1$ and $q=1.2$, where $M_{i}$ is the gravitational mass of the $i$-th star. All binary systems are considered to be irrotational, i.e., the stars are non-spinning. Further details about the setups are given in Table~\ref{tab:BNS_config_table}. We run the simulations with SFHo EOS and neutrino transport at two different resolutions: R1 with 96 points per dimension in each of the two finest boxes covering the stars. This corresponds to a grid spacing in the finest level of $\Delta x_{\rm min} = 248$~m and $\Delta x_{\rm max} = 31.8$~km in the coarsest one. R2 with 128 points on each finest box for $\Delta x_{\rm min} = 186$~m and $\Delta x_{\rm max} = 23.8$~km on the coarsest level. Initial data was produced using the pseudo-spectral code SGRID \cite{Tichy:2009yr, Tichy:2012rp, Dietrich:2015pxa, Tichy:2019ouu} under the assumption that matter is in beta-equilibrium with a constant initial temperature of $T=0.1$~MeV; cf.~\cite{Gieg:2022}. 
The proper initial distance between the stars' centers is set to $38$~km. This corresponds to about three orbits before the merger of the stars. Given that we will primarily focus on the post-merger evolution, we did not perform any eccentricity reduction procedure. Both spacetime and hydrodynamics variables are evolved using a method of lines with RK4 algorithm with a CFL factor of $0.25$.
Time evolution is performed using a Berger-Oliger algorithm with eight refinement levels. The two finest refinement levels are composed of two moving boxes centered around the stars.
Spacetime is evolved employing the Z4c formulation~\cite{Bernuzzi:2009ex,Hilditch:2012fp}. It is discretized using a finite difference scheme with a fourth-order centered stencil for numerical derivatives. Lapse and shift are evolved using $1+\log$ slicing ~\cite{Bona:1994dr} and gamma-driver conditions \cite{Alcubierre:2002kk} respectively.
For hydrodynamic variables we use a finite volume scheme with WENOZ \cite{Borges:2008} reconstruction of primitives at cell interfaces and HLL Riemann solver \cite{Harten:1983} for computing numerical fluxes. We apply the flux corrections of the conservative adaptive mesh refinement \cite{Dietrich:2015iva} to the conservative hydrodynamics variables but not to the radiation fields. \\

\begin{table*}
\begin{tabular}{l|l|l|l|l|l|l|l|l|l|l|l}
Model name     & EoS  & $M^1_b [M_{\odot}]$ & $M^2_b [M_\odot]$ & $M^1_G [M_{\odot}]$ & $M^2_G [M_{\odot}]$ & $C_1$ & $C_2$ & $q$ & $\tilde{\Lambda}$ & $M_{\rm ADM} [M_{\odot}]$ & $J_{\rm ADM} [M_{\odot}^2]$ \\ \hline \hline
SFHo\_q1  & SFHo & 1.301  & 1.301  & 1.200  & 1.200  & 0.148  & 0.148  & 1   & 843 & 2.376 & 5.673     \\ \hline
SFHo\_q12 & SFHo & 1.432  & 1.172  & 1.309  & 1.091  & 0.162  & 0.134  & 1.2 & 875 & 2.377 & 5.617     \\ \hline  
DD2\_q1   & DD2  & 1.292  & 1.292  & 1.183  & 1.183  & 0.134  & 0.134  & 1   & 1585 & 2.377 & 5.749     \\ \hline
DD2\_q12  & DD2  & 1.420  & 1.165  & 1.309  & 1.091  & 0.146  & 0.123  & 1.2 & 1618 & 2.379 & 5.866     \\ 
\end{tabular}
\caption{Table of BNS parameters. From left to right: EoS, baryonic masses, gravitational masses, compactnesses, mass ratio, reduced tidal deformability parameter \cite{Damour:2009vw, Favata:2013rwa, Flanagan:2007ix}, ADM mass, and angular momentum, respectively.}
\label{tab:BNS_config_table}
\end{table*}

\begin{table*}
\begin{tabular}{l|l|l|l|l|l|l|l|l}
Simulation name     & $M^{20~{\rm ms}}_{\rm ej} [10^{-2}~M_{\odot}]$  & $\langle v^{20~{\rm ms}}_{\infty} \rangle$ & $\langle Y_e^{20~{\rm ms}} \rangle$ & $M_{\rm disk} [M_{\odot}]$ & $\langle Y_e^{\rm disk} \rangle$   & $M^{\rm tot}_{\rm ej} [10^{-2} \rm M_{\odot}]$ & $\langle v_{\infty}^{\rm tot} \rangle$ & $\langle Y_e^{\rm tot} \rangle$ \\ \hline \hline
SFHo\_q1\_M1\_R1   & 0.22 & 0.15 & 0.29 & 0.21 & 0.13 & 0.37 & 0.13 & 0.37     \\ \hline
SFHo\_q1\_M1\_R2   & 0.20 & 0.16 & 0.28 & 0.21 & 0.15 & 0.30 & 0.14 & 0.37     \\ \hline
SFHo\_q12\_M1\_R1  & 0.36 & 0.17 & 0.20 & 0.28 & 0.13 & 0.50 & 0.15 & 0.29     \\ \hline 
SFHo\_q12\_M1\_R2  & 0.31 & 0.16 & 0.17 & 0.24 & 0.15 & 0.48 & 0.14 & 0.28     \\ \hline 
SFHo\_q1\_R2       &   -   & -  & -    &   0.20   & -    & 0.20 & 0.15 & -     \\ \hline
SFHo\_q12\_R2      &  -    & -  & -    &   0.27   & -    &  0.32 & 0.17  & -    \\ \hline
DD2\_q1\_M1\_R2    & 0.13 & 0.14 & 0.21 & 0.24  & 0.12 & 0.17 & 0.13 & 0.28     \\ \hline
DD2\_q12\_M1\_R2   & 0.26 & 0.16 & 0.20 & 0.20  & 0.15 & 0.37 & 0.14 & 0.22     \\ \hline
DD2\_q1\_R2        & - & - & - & N.A. & - & 0.12 & 0.14 & -       \\ \hline
DD2\_q12\_R2       & - & - & - & N.A & - & 0.24 & 0.17 &  -       \\
\end{tabular}
\caption{Summary of ejecta and disk properties for all simulations. The first three columns include the mass, the average velocity, and the average $Y_e$ of the ejecta extracted up to $20$~ms after merger. The fourth and fifth columns contain the mass and the average $Y_e$ of the disk, respectively. Finally, in the last three columns are the mass, the average velocity, and the average $Y_e$ of all the ejecta, i.e., also including the component identified as neutrino wind. For simulations without neutrino transport, we do not show $\langle Y_e \rangle$, and since we do not have neutrino wind ejecta, we only show the total ejecta. Hence, we mark them as non-available data (N.A.).}
\label{tab:BNS_results_table}
\end{table*}

\subsection{Ejecta}
We compute ejecta properties using a series of concentric spheres centered around the coordinate origin with radii varying from $300$~km to $1000$~km. On each sphere, the total flux of mass, energy, and momentum of outgoing, unbound matter is computed. On such extraction spheres, the matter is assumed to be unbound according to the geodesic criterion \cite{Hotokezaka:2013b}, i.e., if 
\begin{equation}
u_t < -1 \qquad \text{and} \quad u_r>0.
\end{equation}
From now on, we will always refer to the unbound mass as the one that satisfies this criterion unless stated otherwise. Differently from previous BAM versions, the spheres with radius $450$~km and $600$~km also save the angular coordinates $(\theta,\phi)$ of the matter flux together with $u_t$, $\rho$, $T$, and $Y_e$. This allows a more detailed analysis of the ejecta that includes its geometry and thermodynamical properties, e.g., the use of the Bernoulli criterion \cite{kastaun2015properties, Nedora:2019jhl} for determining unbound mass, i.e., 
\begin{equation}
h u_t < -1 \qquad \text{and} \quad u_r>0.
\end{equation}
Since $u_t$ [or $hu_t$ for Bernoulli] is assumed to be conserved and at infinity $u_t=-W$ [or $hu_t=-W$], it is also possible to compute the asymptotic velocity $v_{\infty}$ of each fluid element as $v_{\infty} = \sqrt{1 - 1/u_t^2}$ [or $\sqrt{1-1/(hu_t)^2}$].

\subsubsection{Ejecta Mass}

Mass ejection from BNS mergers within a dynamical timescale $\mathcal{O}(10~{\rm ms})$ has already been the subject of several detailed studies, e.g., \cite{Bauswein:2013yna, Hotokezaka:2013b,Sekiguchi:2015dma,Radice:2016dwd,Sekiguchi:2016bjd,Radice:2018ghv,Vincent:2019kor,Nedora:2020pak}. There is a general consensus on dividing dynamical ejecta into two components: tidal tail and shocked ejecta. The former is composed of matter shed from the star's surface right before the merger due to tidal forces. Since this matter does not undergo any shock heating or weak interaction, it has a low $Y_e$ comparable to the one of neutron stars' outer layers and low entropy $\lesssim 10~\kappa_B$. Shocked ejecta, on the opposite, is launched by the high pressure developed in the shock formed at the star's surface during the plunge. It has significantly higher entropy and $Y_e$ with respect to the tidal tails. It is produced later but with higher velocity, rapidly reaching the tidal tails and interacting with them \cite{Radice:2018ghv}.

\begin{figure}
    \centering 
    \includegraphics[width=0.5\textwidth]{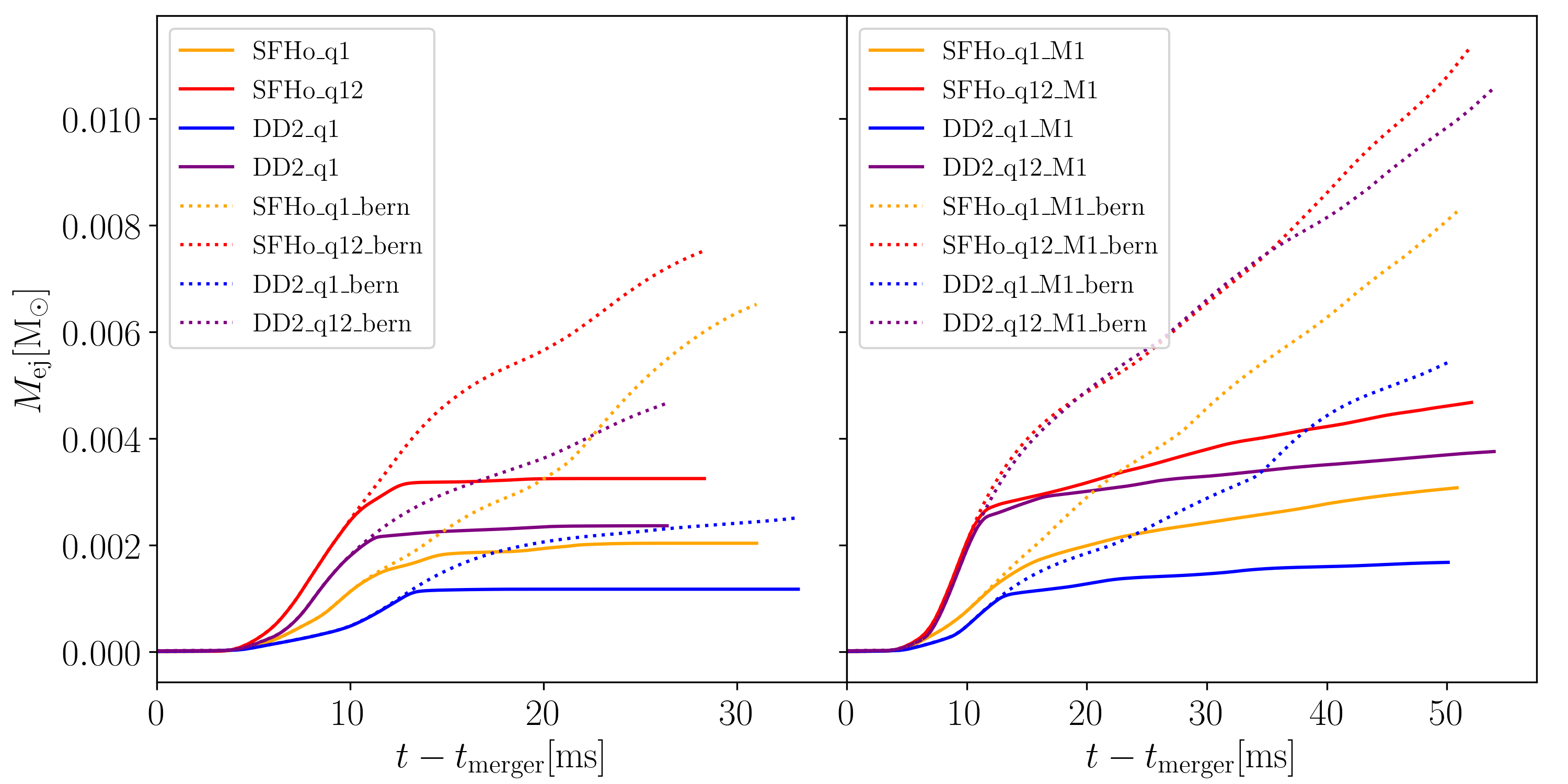}
    \caption{Mass of the ejecta passed through the detection sphere at $r \simeq 450$~km as a function of time for the SFHo simulations. The left and right panels show simulations without and with neutrino radiation, respectively. Solid lines represent the mass unbound according to the geodesic criterion, while dashed lines refer to the Bernoulli criterion.
    }
    \label{fig:Mej_vs_time} 
\end{figure}

\begin{figure*}
    \centering 
    \includegraphics[width=.9\textwidth]{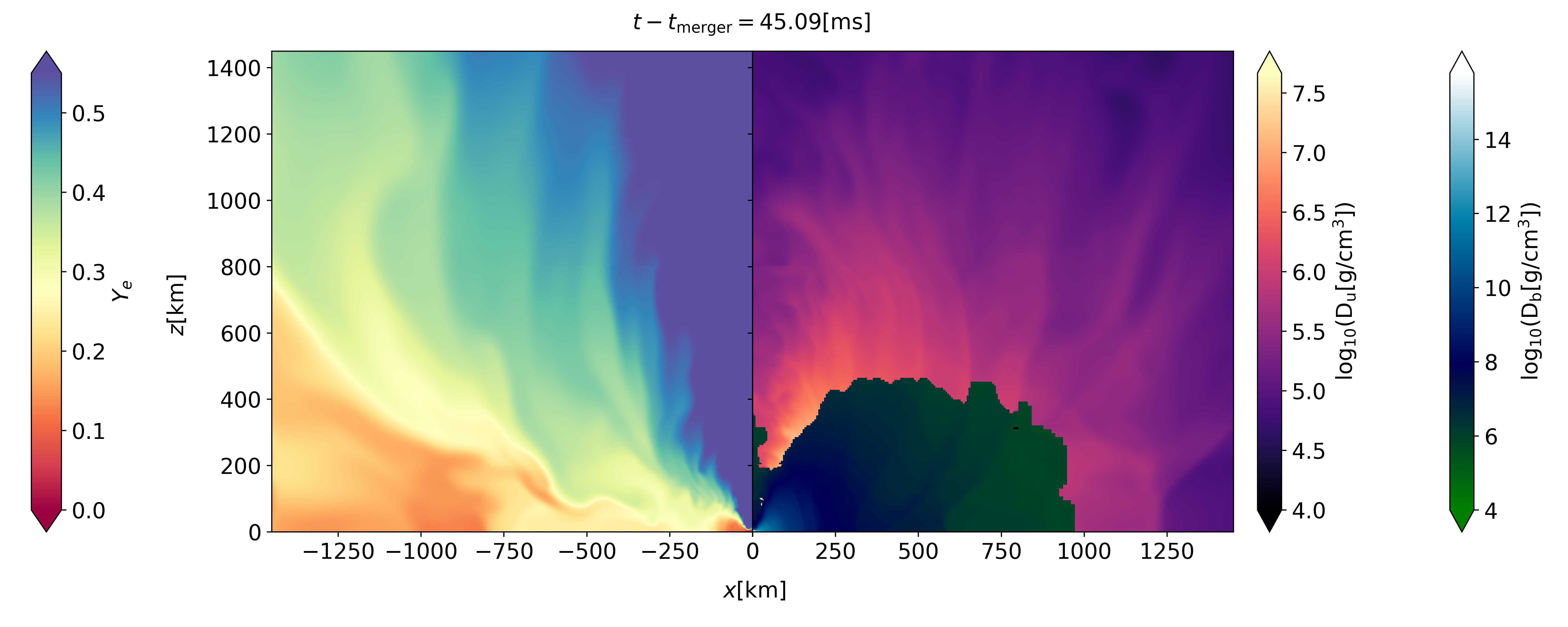}
    \caption{{\it Right}: $xz$-plane snapshot of bound $D_b$ and unbound $D_u$ matter for the simulation SFHo\_q1\_M1\_R2 represented by two different colormaps. Bound matter is identified using the ballistic criterion. {\it Left}: $xz$-plane snapshot of $Y_e$ for the same simulation.}
    \label{fig:Du_xz} 
\end{figure*}

In Fig.~\ref{fig:Mej_vs_time}, we show the mass of the unbound matter moving through the detection sphere at $r\simeq450$~km as a function of time for both geodesic and Bernoulli criteria. There is an important qualitative difference between simulations where neutrinos are neglected and the ones including neutrino transport. While in the former case, the ejecta mass saturates within 20~ms after the merger, in the latter one, we observe a non-negligible matter outflow continuing for the whole duration of the simulation, although with decreasing intensity. Such a phenomenon has been observed in other BNS simulations with M1 transport in \cite{Foucart:2015gaa,Zappa:2022rpd}, where a very similar numerical implementation of M1 is used, and in much smaller amount also in \cite{Sekiguchi:2016bjd}. We attribute it to the neutrinos emitted from the remnant. Through scattering/absorption processes in the upper parts of the disk, they can indeed accelerate material, making it gravitationally unbound. This hypothesis is consistent with what we see in Fig.~\ref{fig:Du_xz}, where we show the conserved mass density for bound and unbound matter on the $xz$-plane roughly 45~ms after the merger. We denote by $D_u$ the conserved mass density of unbound matter, i.e., $D_u=D$ where matter is unbound and $D_u=0$ otherwise. Most of the unbound matter is concentrated in the inner part of the upper edge of the disk, as we would expect from a neutrino wind mechanism powered by the remnant emission.
In particular, in \cite{Foucart:2015gaa}, equal mass simulations using the SFHo and DD2 EoSs are performed, and an early neutrino wind mechanism is also observed. However, such simulations only show results up to $\simeq 10$~ms after the merger. 

For both EoSs, the ejecta mass is higher and more rapidly growing for asymmetric configurations. This is in agreement with the higher amount of tidal tails ejecta that asymmetric binaries are known to produce. In the same figure, the amount of ejecta according to the Bernoulli criterion is also shown. 
Bernoulli-criterion ejecta corresponds to the geodesic one in the very initial phase of the matter outflow but predicts a significantly higher mass after the dynamical phase. More importantly, Bernoulli ejecta is not close to saturation at the end of the simulation time. These features are comparable with the results of other works, e.g.,\cite{Kastaun:2014fna,Bovard:2017mvn,Foucart:2021ikp,Nedora:2019jhl, Nedora:2020pak, Zappa:2022rpd}. This continuous matter outflow is attributed to the so-called spiral wave wind, i.e., the outward transport of angular momentum through the disk due to the shocks.

\subsubsection{Electron fraction and velocity}

\begin{figure}
    \centering 
    \includegraphics[width=0.45\textwidth]{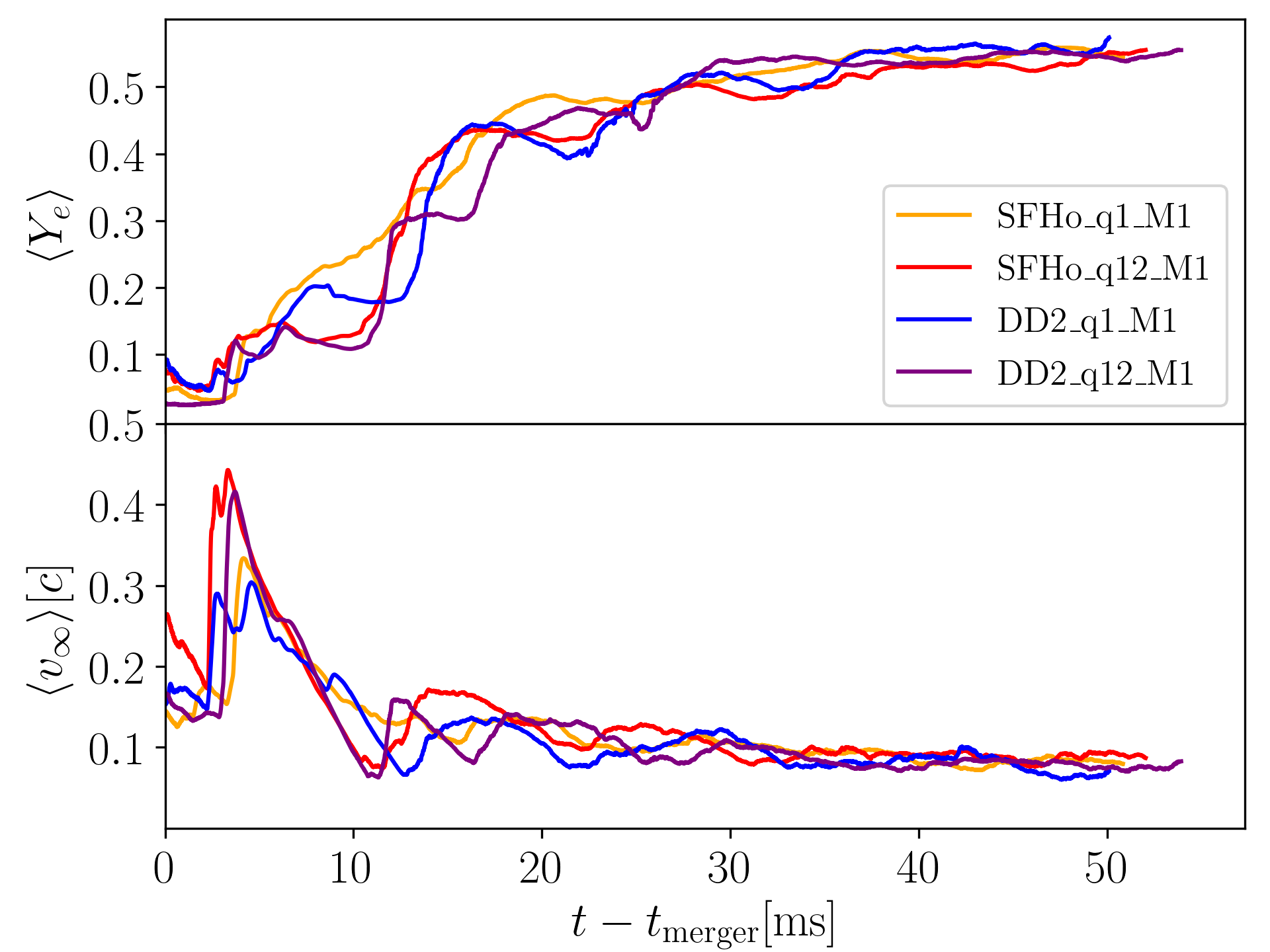}
    \caption{Average $\langle Y_e \rangle$ and $\langle v_{\infty} \rangle$  of matter passing through the detection sphere at radius $r \simeq 450$~km.}
    \label{fig:dMdt} 
\end{figure}

Figure~\ref{fig:dMdt} shows the average of $Y_e$ and $v_{\infty}$ ($\langle Y_e \rangle$ and $\langle v_{\infty} \rangle$ respectively), of matter flowing through the detection sphere, located at $r\simeq 450$~km, as a function of time. These quantities are defined as:
\begin{align}
    \langle Y_e \rangle (t) &= \frac{\int d\Omega F_{D_u} (t, \Omega) Y_e(t,\Omega)}{\int d\Omega F_{D_u} (t, \Omega)}, \\
    \langle v_{\infty} \rangle (t) &= \frac{\int d\Omega F_{D_u} (t, \Omega) v_\infty(t,\Omega)}{\int d\Omega F_{D_u} (t, \Omega)},
\end{align}
where $r$ is the radius of the extraction sphere and $F_{D_u} = D_u (\alpha v^r - \beta^r)$ is the local radial flux of unbound matter through the detection sphere.

\begin{figure}
    \centering 
    \includegraphics[width=0.45\textwidth]{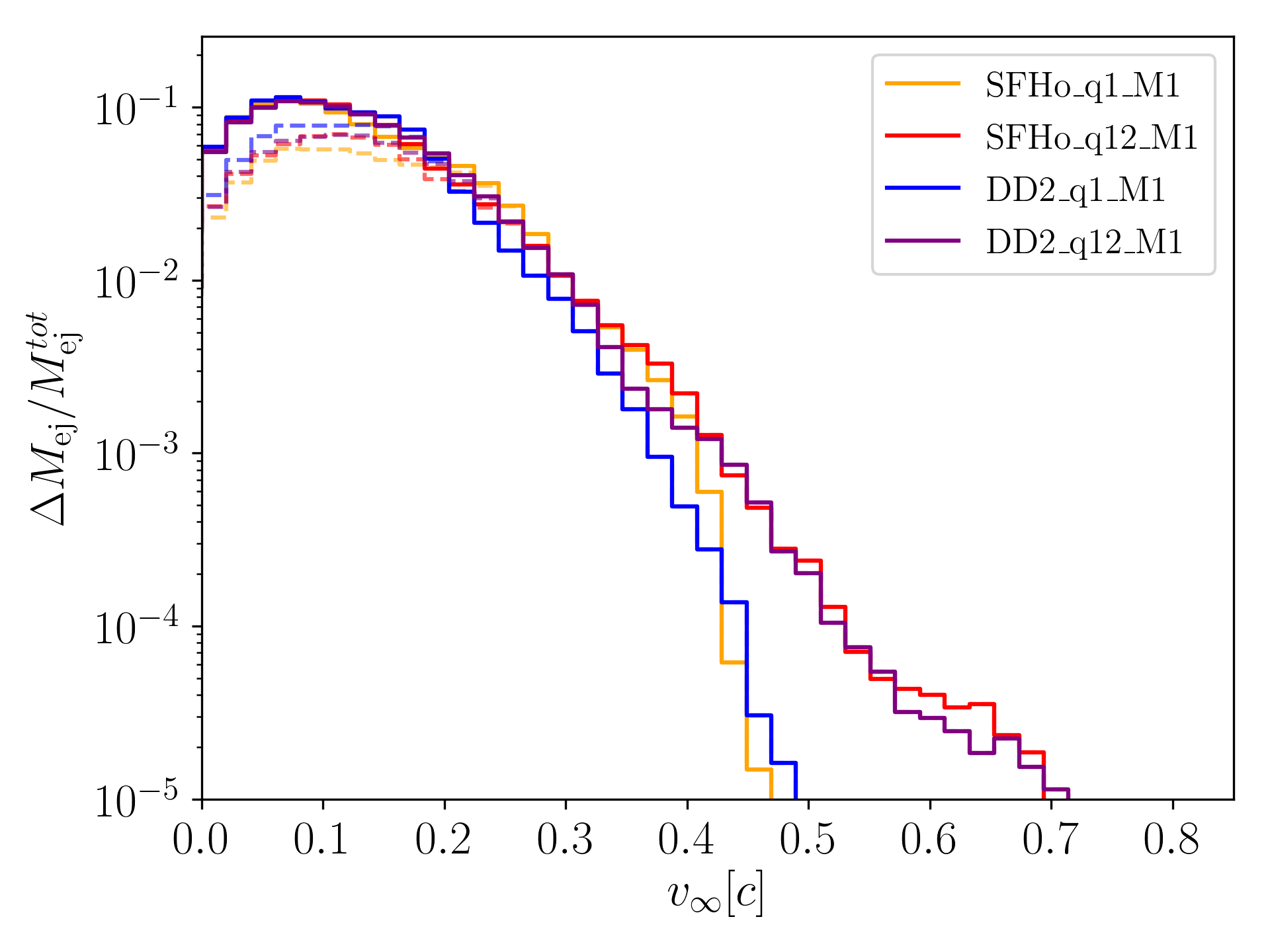}
    \caption{Histogram of the ejecta's asymptotic velocity. Dashed lines refer to the dynamical ejecta only, while the continuous one includes the neutrino wind component. Histograms have been normalized with respect to $M_{\rm ej}^{\rm tot}$.}
    \label{fig:v_hist} 
\end{figure}

All simulations show an overall monotonically increasing electron fraction since matter ejected later remains longer next to the remnant, having more time for protonizing due to neutrino absorption. In addition, most systems show a more or less pronounced plateau at about $5-15$~ms after the merger with a visible dependence on the mass ratio. This is likely due to tidal tails containing material with an almost uniform and low $\langle Y_e \rangle$ of $\simeq 0.1$. Tidal tails are then reached and partially reprocessed by the faster and more proton-rich shocked ejecta, giving rise to the plateau we observe\footnote{Note that this plateau is absent for SFHo\_q1\_M1 due to the smaller amount of tidal ejecta for this equal mass, soft-EoS configuration.}. 
$\langle v_{\infty} \rangle$ has a sharp velocity peak at early times followed by slow late-time ejecta. The fact the initial peak does not show a bimodal shape is another indicator of the fact that tidal and shock ejecta already merged together at the extraction radius. Finally, we see that asymmetric binaries present a higher velocity peak of the early ejecta. This feature is consistent with Fig.~\ref{fig:v_hist} and is responsible for the tails with $v\sim 0.5c - 0.7c$. The velocity histogram in the same figure shows no dependence on the EoS, with the mass ratio being the only feature determining the velocity profile.

\subsubsection{Angular dependence}

\begin{figure}
    \centering 
    \includegraphics[width=0.45\textwidth]{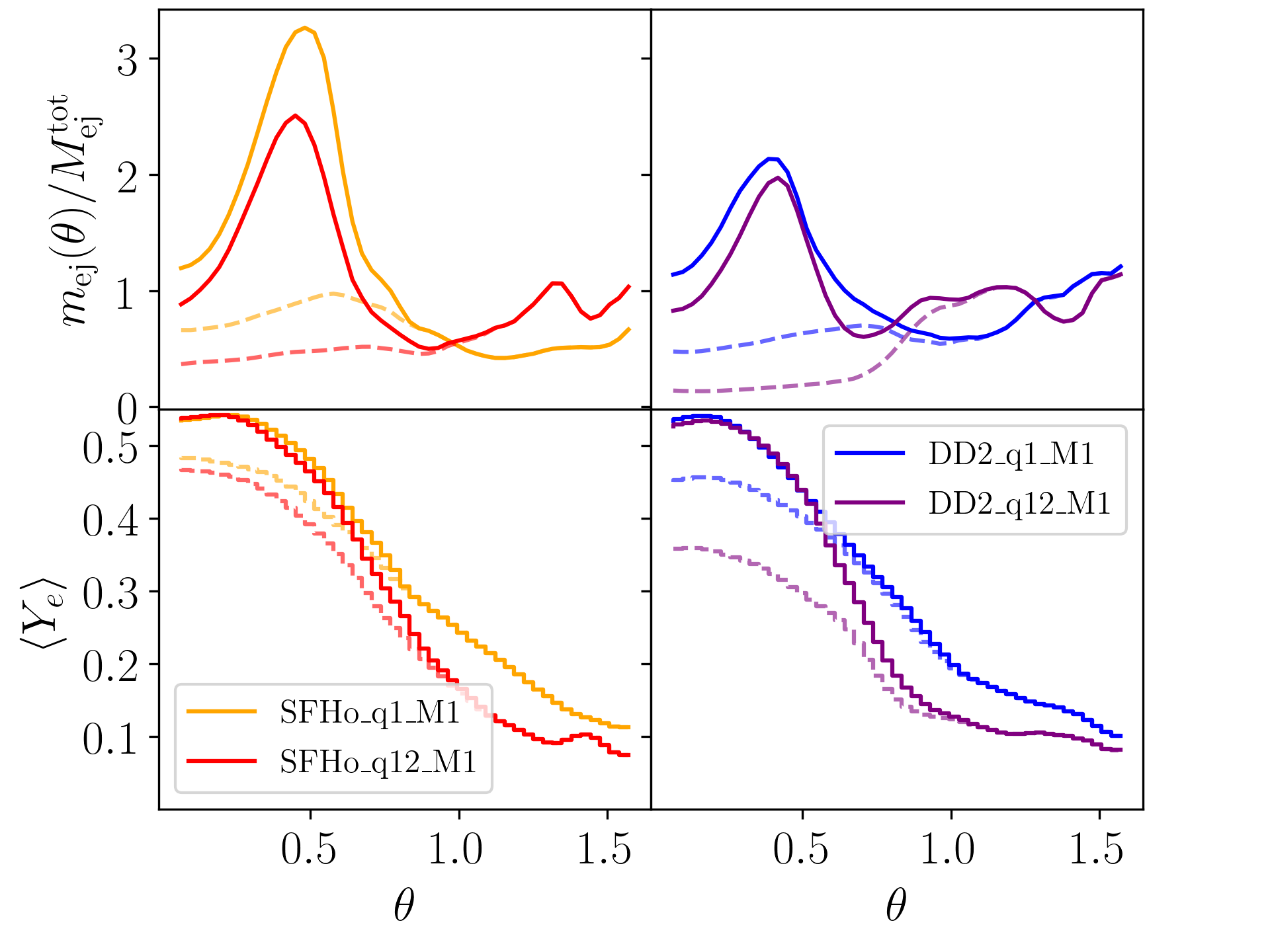}
    \caption{Mass distribution and average $\langle Y_e \rangle$ along the polar angle, with $\theta=0$ being the pole and $\theta=\pi/2$ being the equator. Mass distribution has been normalized with respect to $M_{\rm ej}^{\rm tot}$. Continuous lines refer to the total amount of ejecta flowed during the whole simulation time, the dashed ones account only for the matter being ejected before 20 ms.}
    \label{fig:Mej_vs_theta} 
\end{figure}

\begin{figure}
    \centering 
    \includegraphics[width=0.48\textwidth]{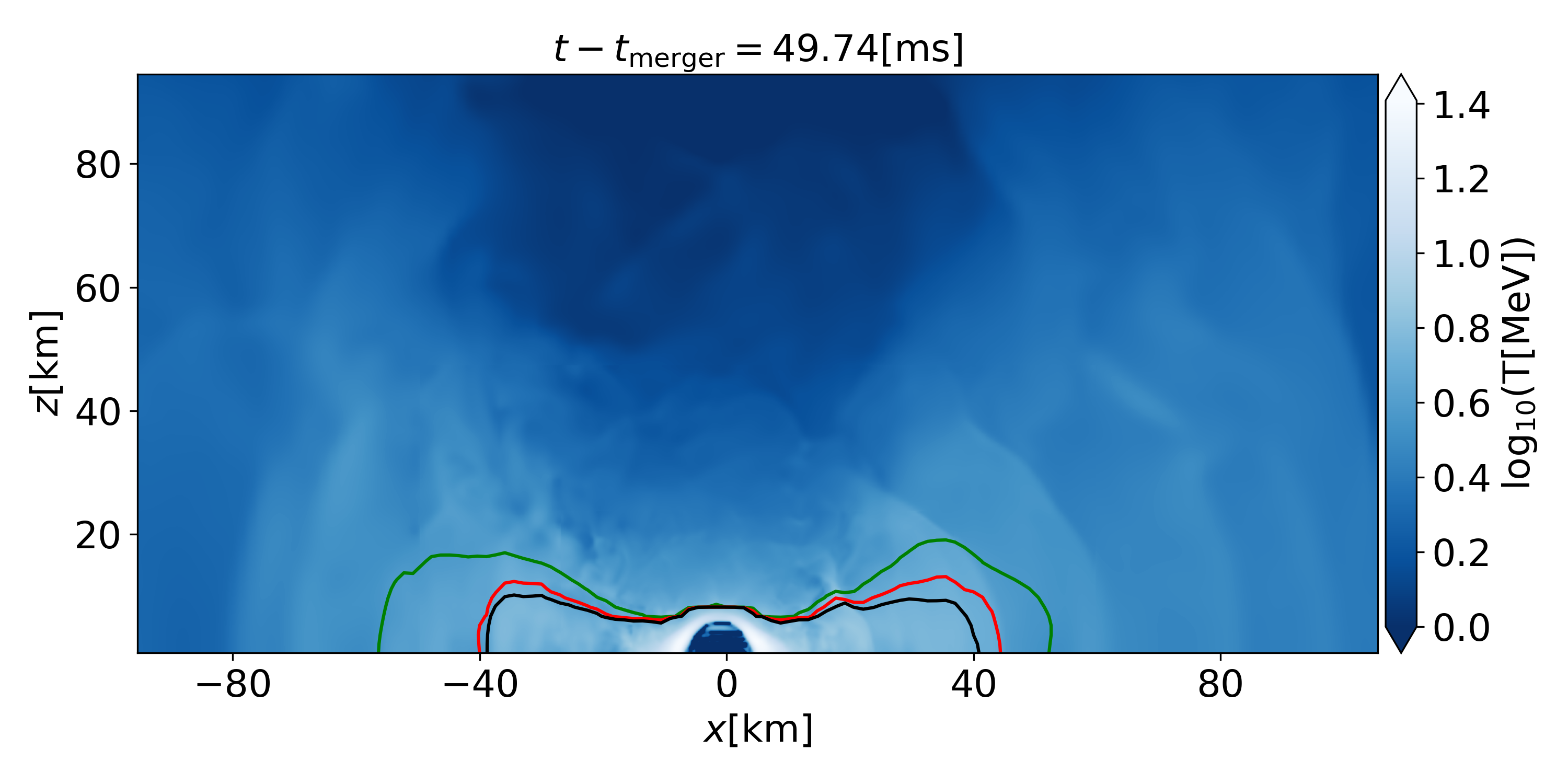}
    \caption{2D snapshot of the fluid's temperature and neutrino photosphere in the post-merger for SFHo\_q1\_M1. The latter is defined as the place where $\tau=2/3$ and is shown by a different colored contoured line for every species: green for $\nu_e$, red for $\overline{\nu}_e$ and black for $\nu_x$.}
    \label{fig:T_tau} 
\end{figure}

In the upper panel of Fig.~\ref{fig:Mej_vs_theta}, we show the normalized polar angle distribution of the ejecta defined as
\begin{equation}
    m_{\rm ej}(\theta) = r^2 \int^T_0 \int_0^{2\pi} F_{D_u}(t,\theta,\phi) dt d\phi,
\end{equation}
with $T$ being the final time of the simulation and $r$ the radius of the detection sphere (in this case $\simeq 450$~km). According to this definition $M_{\rm ej} = \int_0^\pi \sin(\theta) m_{\rm ej}(\theta) d\theta$. Then $m_{\rm ej}(\theta)$ is normalized by the total mass of the ejecta $M_{\rm ej}^{\rm tot}$.
The peak at $\theta \simeq 0.5$ due to the post-merger neutrino wind is immediately visible. At lower latitudes, the neutrino wind mechanism is indeed heavily suppressed by the disk, which is cold and optically thick and stops the neutrinos emitted by the remnant (see Fig.~\ref{fig:T_tau}). For asymmetric binaries, there is also a peak at low latitudes visible, caused by the tidal tail ejecta.
The effect of such a component on the electron fraction is visible in the lower panels of the same figure. It is responsible for the lower $\langle Y_e \rangle$ of the equatorial region, and, as expected, it is more evident for asymmetric binaries. 
In the same panel, we can also observe that when neutrino wind is included in the ejecta, the $\langle Y_e \rangle$ of the polar regions increases significantly, reaching up to $0.5$, while regions with polar angles above one radiant are unchanged by the phenomenon. This is due to the intense neutrino irradiation that this matter received, which increased its electron fraction. The fact that dynamical ejecta from equal mass binaries has an overall higher $\langle Y_e \rangle$ can be explained by the higher amount of shocked ejecta that such configurations are known to produce. Shock ejecta is indeed supposed to have a higher entropy and electron fraction with respect to tidal tail ejecta and is more isotropically distributed. The last characteristic can explain why symmetric binaries give a higher $\langle Y_e \rangle$ than their respective asymmetric counterparts at lower latitudes.

\begin{figure}
    \centering 
    \includegraphics[width=0.45\textwidth]{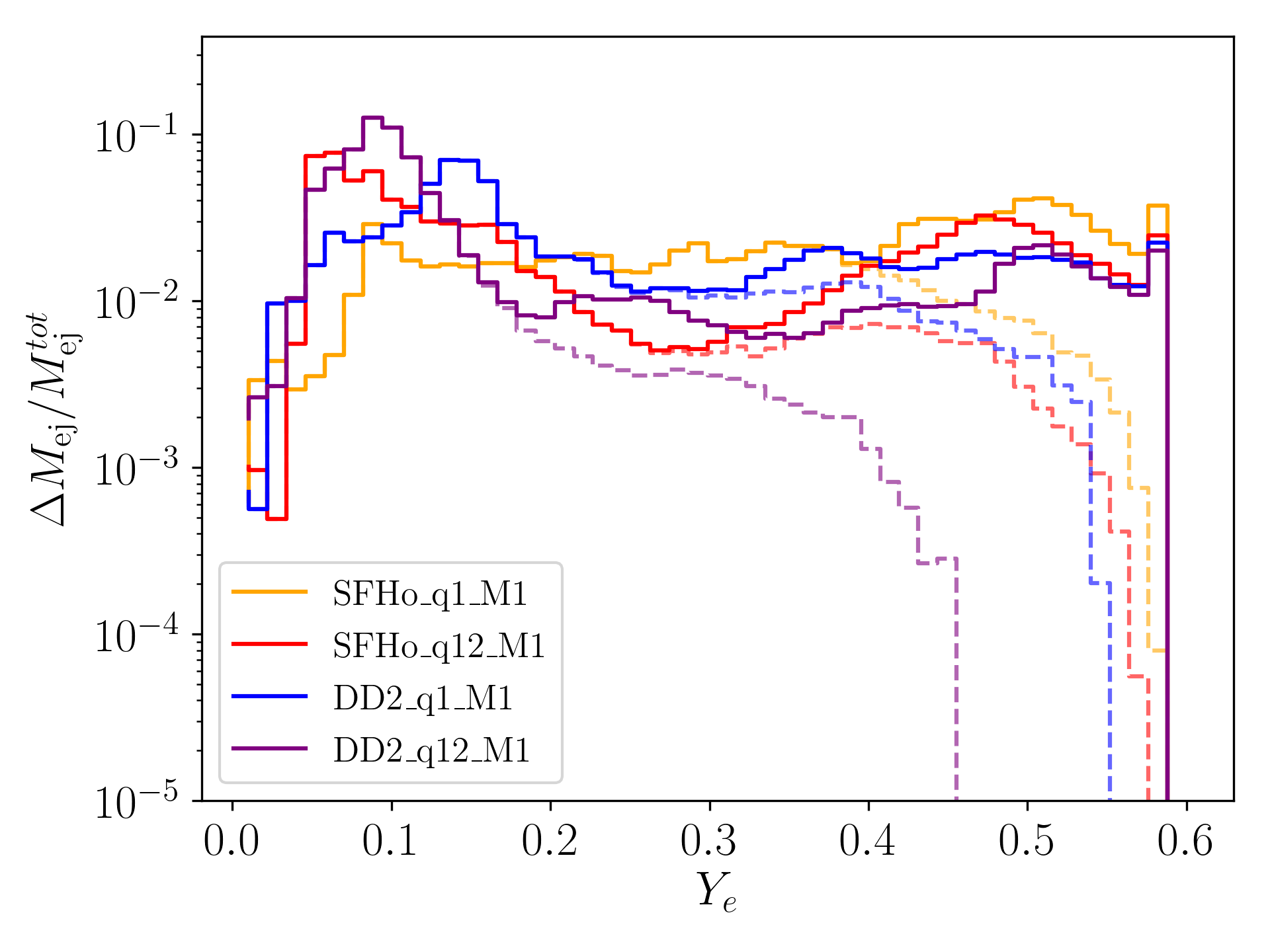}
    \caption{Histogram of the ejecta's electron fraction. Again the dashed lines refer to the ejecta before 20 ms only, while the continuous ones include the later time component.  Both dashed and solid lines have been normalized to their respective $M_{\rm ej}^{\rm tot}$.}
    \label{fig:Ye_hist} 
\end{figure}

In Fig.~\ref{fig:Ye_hist}, a histogram of the ejecta's $\langle Y_e \rangle$ is shown. Dynamical ejecta of equal mass binaries produces a fairly uniform distribution of mass with a drop for $\langle Y_e \rangle \lesssim 0.1$. In the unequal mass scenario, the situation changes. Here, we have indeed a clear peak at $\langle Y_e \rangle \simeq 0.1$ produced by tidal tails. In both cases, the inclusion of neutrino wind leads to an increase of ejecta with $0.3 \lesssim \langle Y_e \rangle \lesssim 0.6$.

\begin{figure*}
    \centering 
    \includegraphics[width=0.485\textwidth]{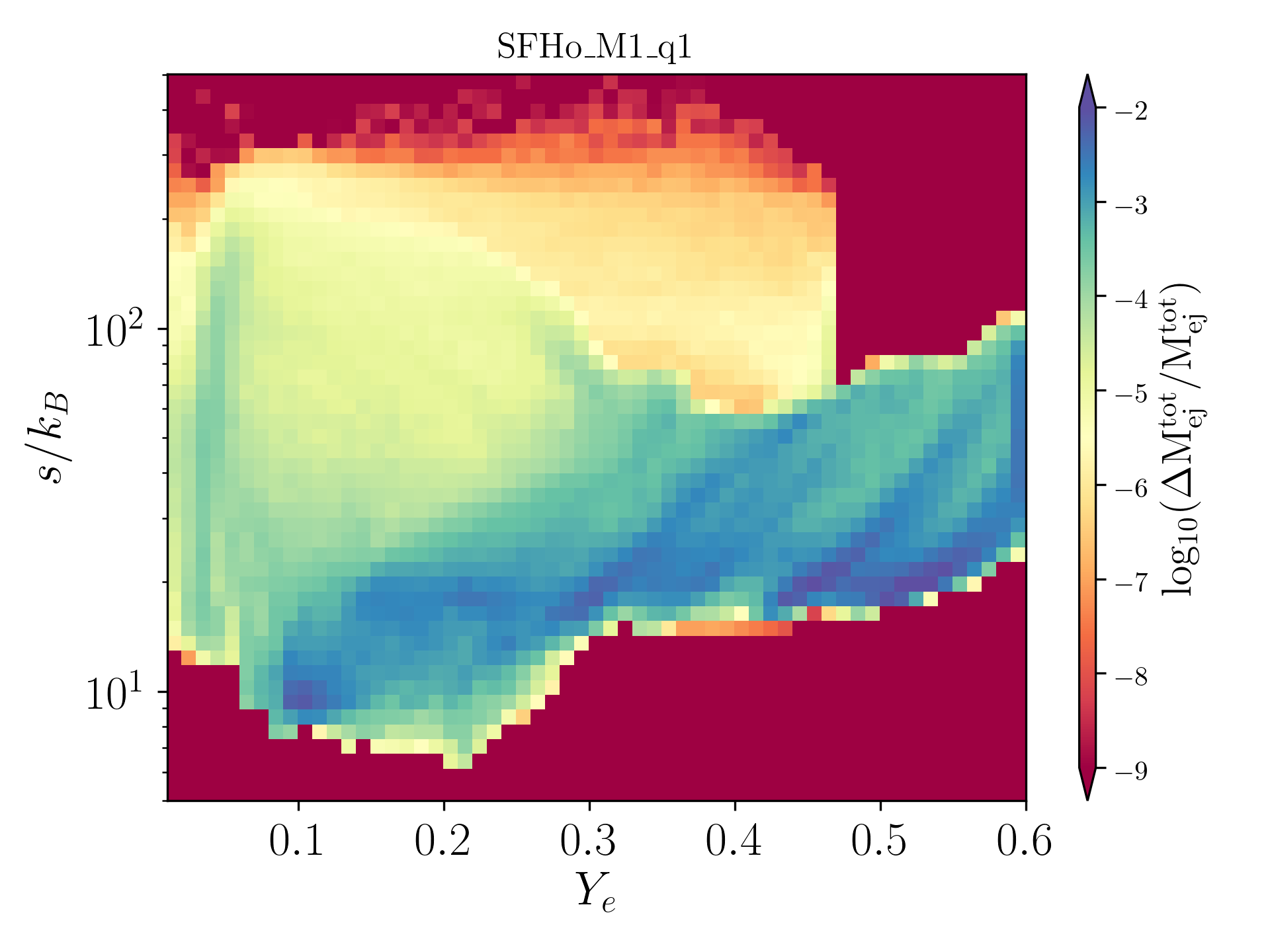}
    \includegraphics[width=0.485\textwidth]{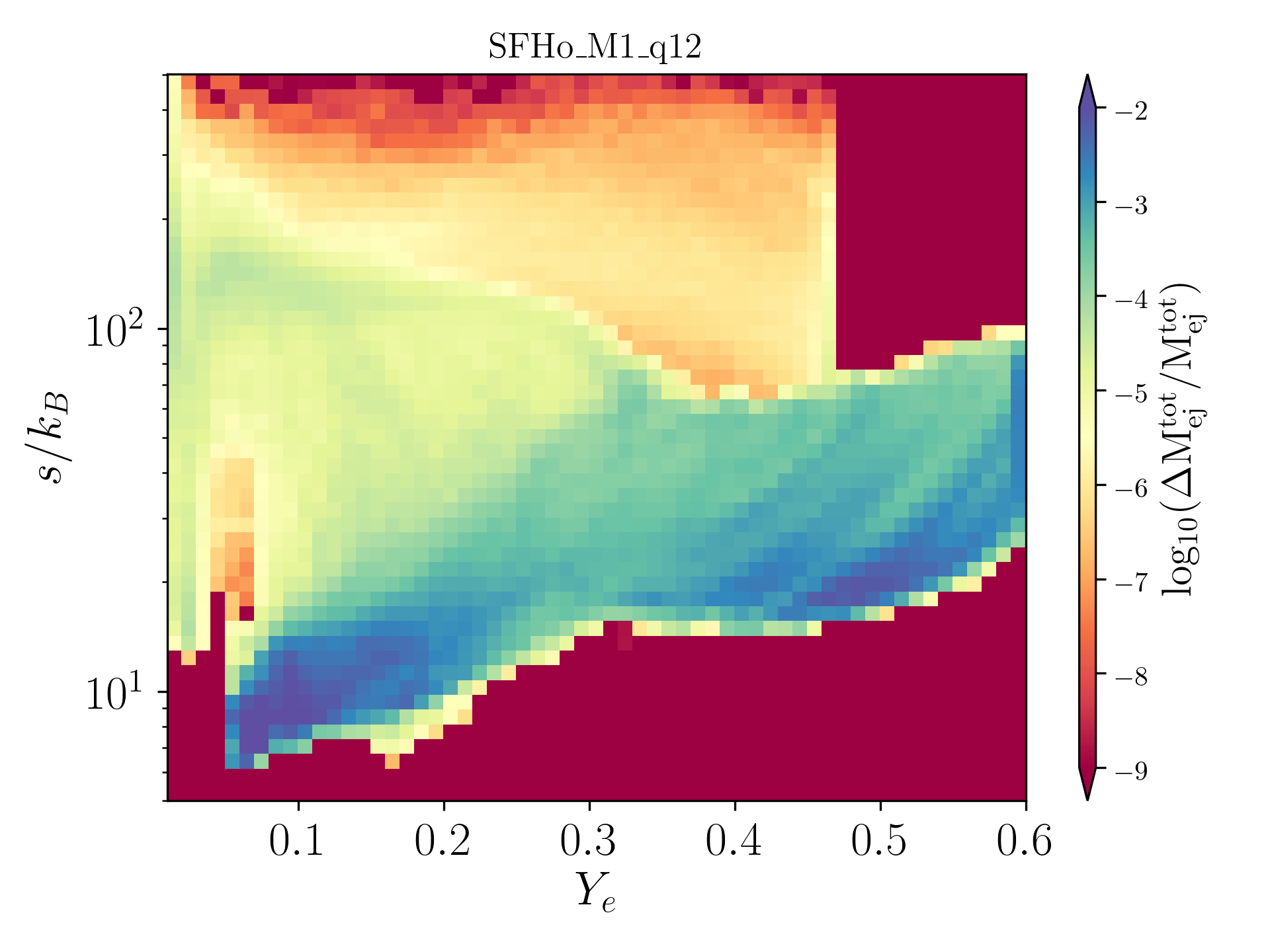}
    \includegraphics[width=0.485\textwidth]{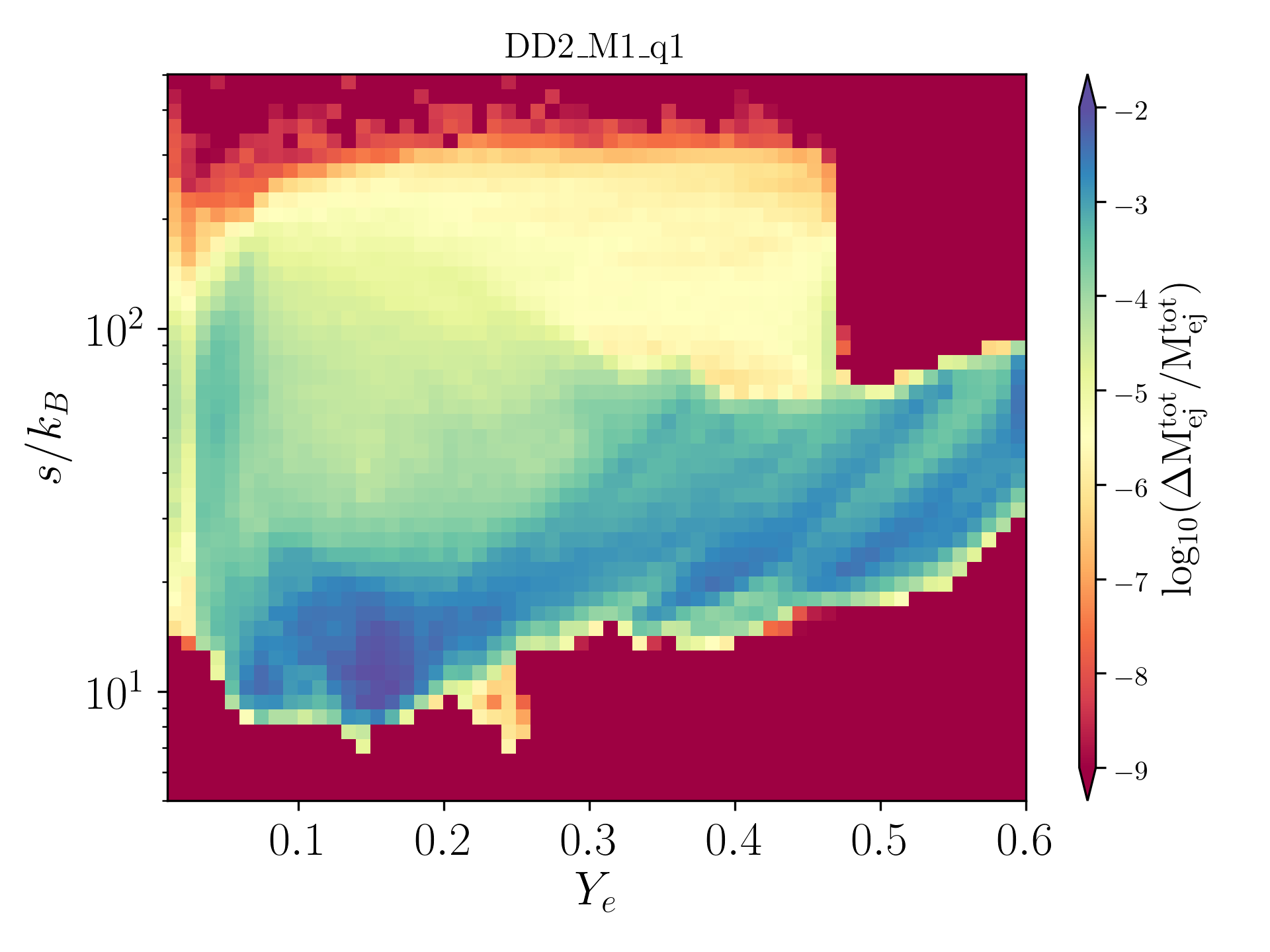}
    \includegraphics[width=0.485\textwidth]{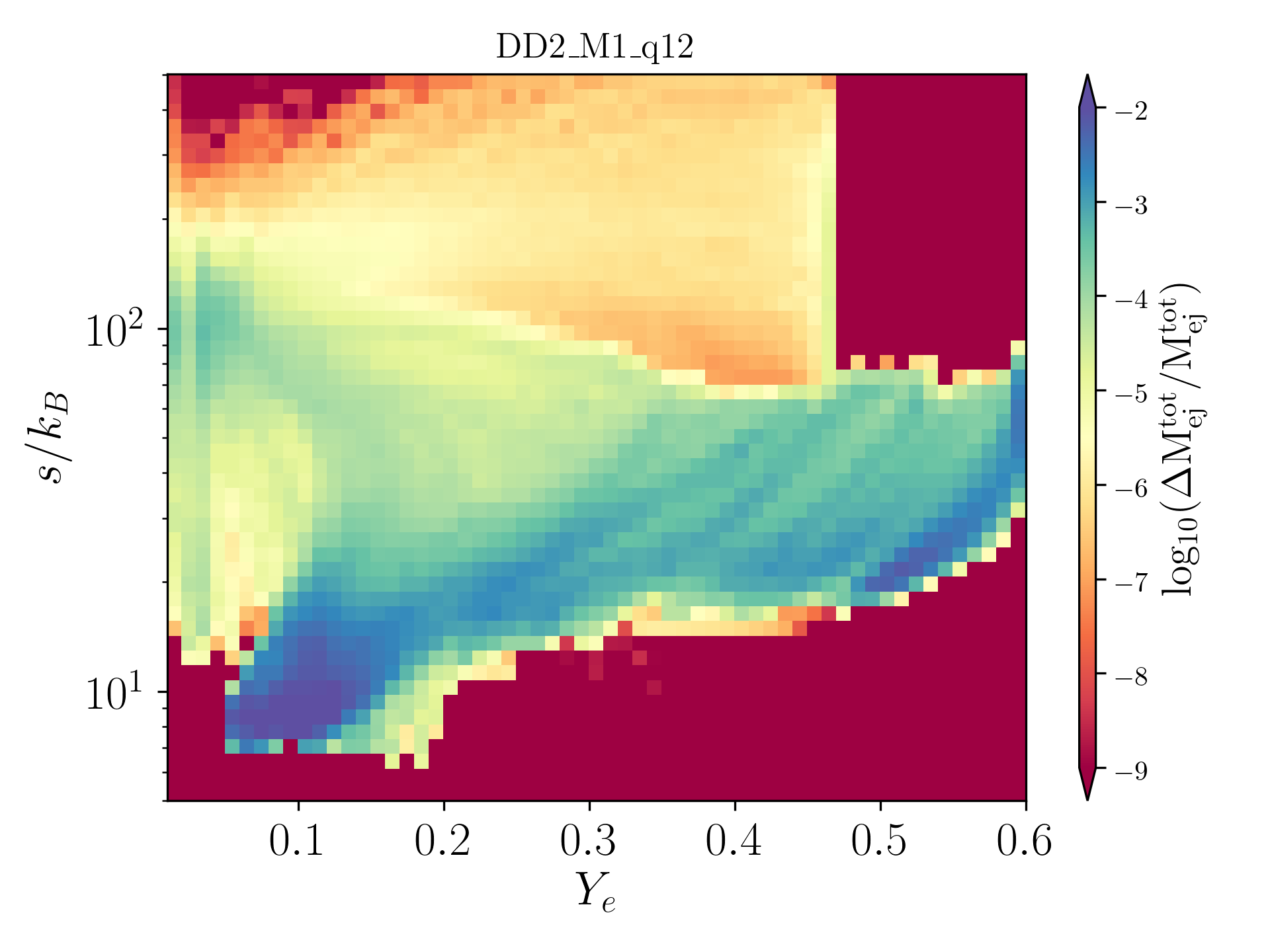}
    \caption{Two-dimensional histogram of the ejecta mass as a function of the entropy per baryon and $Y_e$ for all simulations at resolution R2.}
    \label{fig:S_Ye_hist} 
\end{figure*}

Another important feature of the ejecta that has been investigated in literature is the correlation between $Y_e$ and entropy ($s/k_B$). In Fig.~\ref{fig:S_Ye_hist}, we show a 2D histogram of the total ejecta in these two variables. Most of the ejecta mass lies within a main sequence with a positive monotonic correlation between entropy and $Y_e$. This is a consequence of the fact that fluid with a higher entropy is characterized by a more proton-rich thermodynamical equilibrium configuration. The exception to this rule is made by matter with $Y_e \lesssim 0.3$ and entropy in a very wide range going up to $s \sim 100 k_B$. This matter is present in every simulation and is believed to be a consequence of the interaction between tidal tails and shocked ejecta \cite{Fujibayashi:2020dvr}. When the latter hits the former, it generates indeed a violent shock that increases the fluid's entropy. Since this happens at low density, when the neutrino-matter interaction timescale is bigger than the dynamical one, this does not leave time for the fluid to settle to an equilibrium configuration with higher $Y_e$.

Ejecta's average properties of our simulations are summarized in Table \ref{tab:BNS_results_table}. Here we see, as expected, a strong dependence of $\langle Y_e \rangle$ on the mass ratio, with asymmetric binaries producing a more neutron-rich and an overall more massive outcome. An imprint of the tidal deformability can also be observed, with the more deformable EoS (DD2) producing less massive but neutron-rich ejecta. For SFHo simulations, the dependence of the ejecta mass and $\langle Y_e \rangle$ on the mass ratio is consistent through all resolutions. We do not observe any significant dependence of the average asymptotic velocity on the mass ratio or tidal deformability. \\

\subsection{Neutrino luminosity}
We determine the neutrino luminosity as the total flux of neutrino energy $\tilde{E}$ through the same series of spheres used for the analysis of the ejecta, i.e., $L_{\nu} = r^2 \int d\Omega (\alpha \tilde{F}^r - \tilde{E}\beta^r)$. Similarly to the ejecta detection, also here the two spheres, located at 450~km and 600~km, are able to save the flux angular direction together with its values of $J$ and $n$, enabling a more detailed study that includes the geometry of neutrino luminosity and its average energy.

\begin{figure*}
    \centering 
    \includegraphics[width=0.45\textwidth]{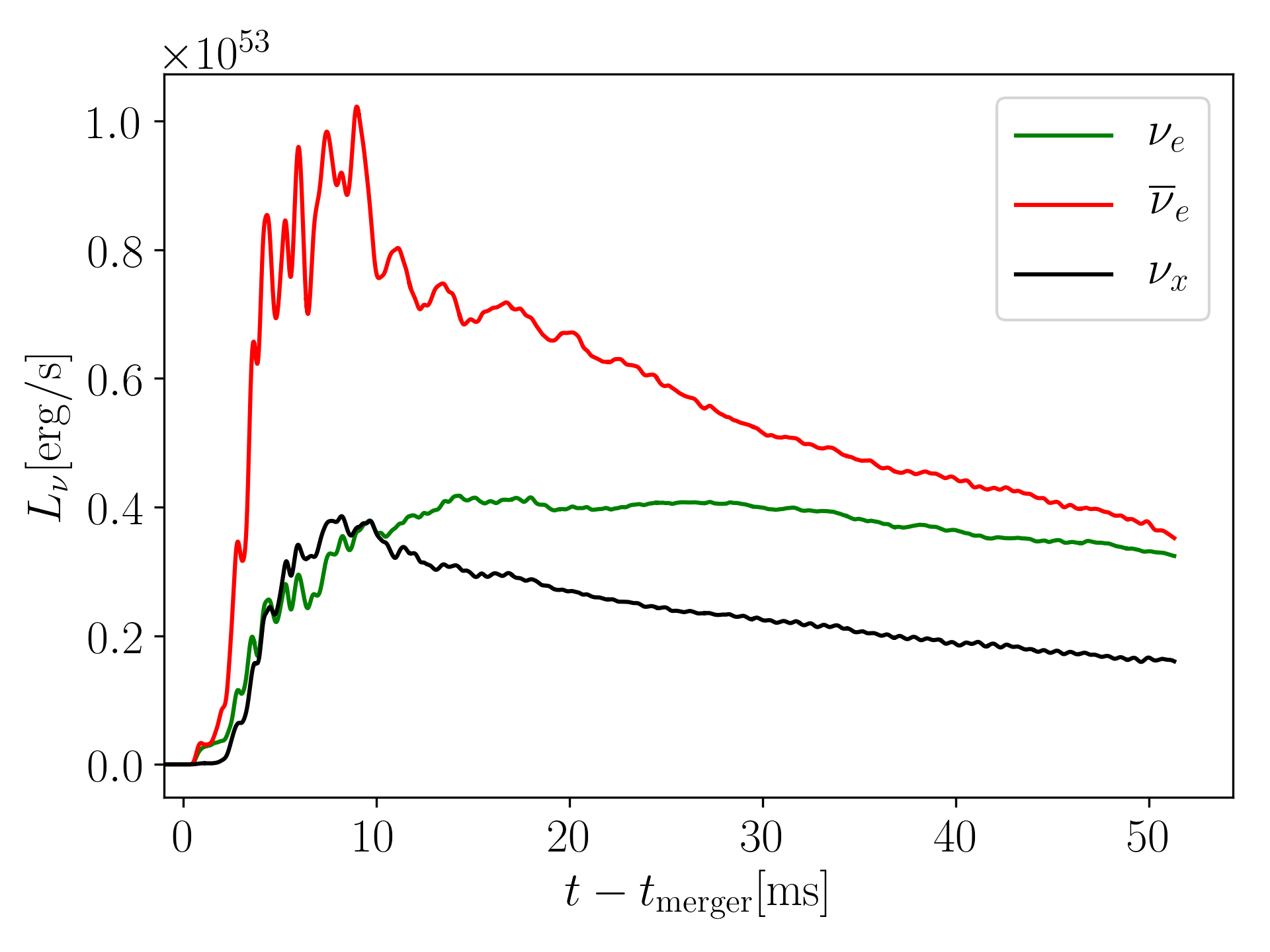}
    \includegraphics[width=0.45\textwidth]{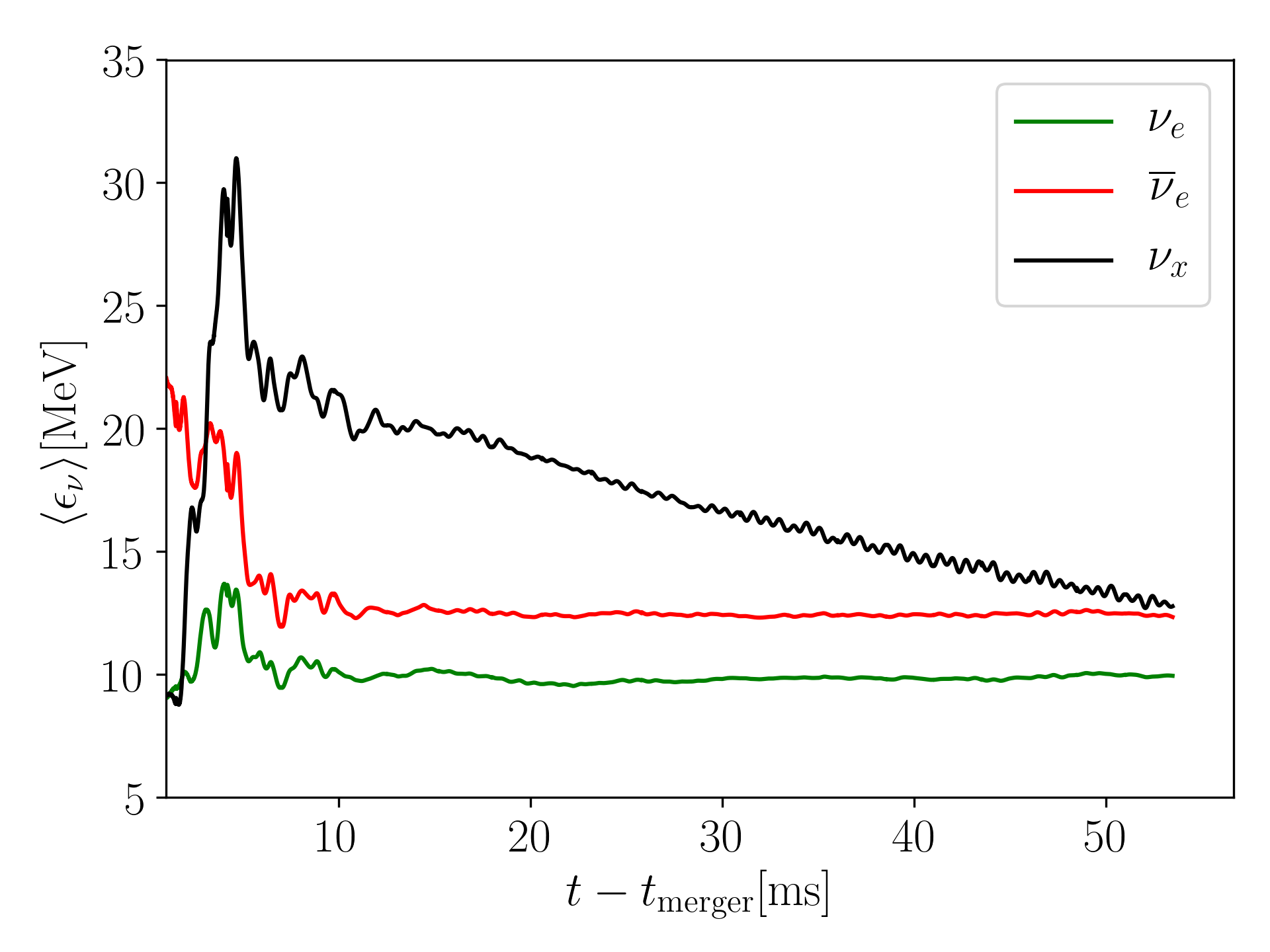}
    \caption{Neutrino luminosity and average energy as a function of time for the three different species from the simulation SFHo\_q1\_M1\_R1. The data refer to the extraction sphere located at $r \simeq 450$~km.}
    \label{fig:Lnu_vs_time} 
\end{figure*}

Looking at the total neutrino luminosity in the left panel of Fig.~\ref{fig:Lnu_vs_time}, we find that $\overline{\nu}_e$ emission is brighter in the early post-merger with respect to the other species. Its peaking luminosity of $\sim 10^{53} \rm erg/s$ is consistent with results obtained by similar simulations \cite{Sekiguchi:2015dma, Radice:2021jtw, Foucart:2015gaa, Cusinato:2021zin, Zappa:2022rpd, Radice:2023zlw}. The initial $\overline{\nu}_e$ burst is a consequence of the fast protonization that the material undergoes right after the merger, when the beta equilibrium is broken, and the system evolves toward a new meta-stable configuration characterized by a higher entropy and $Y_e$. Approximately 10~ms after the merger, the $\overline{\nu}_e$ starts decreasing and approaches the luminosity of $\nu_e$ a few tens of ms later. This is a signal that the system is approaching the weak equilibrium configuration within the late simulation time. Both $\nu_e$ and $\nu_x$ show similar behavior, with a peak at $\sim 10$~ms and roughly half of the intensity of $\overline{\nu}_e$. In the early post-merger, we have, as reported in the literature, $L_{\overline{\nu}_e} > L_{\nu_x} > L_{\nu_e}$. The brightness oscillations that appear in this phase for every neutrino species are due to the remnant oscillations, which cause shocks propagating outward and perturbing the surface of the neutrino sphere. The last inequality is inverted after the luminosity peak. $L_{\nu_x}$ drops faster because of the remnant's cooling. $\nu_x$ interactions indeed include only thermal processes that are independent of $Y_e$. This makes the heavy neutrino emission more sensitive to temperature with respect to other species. The right panel of Fig.~\ref{fig:Lnu_vs_time} shows the average energy of neutrinos flowing through the detection sphere at $r = 450$~km as a function of time. As reported in the literature $\nu_x$ have significantly higher energy with respect to the other two species in the early post-merger. This is an expected feature since heavy neutrinos are less interacting with matter and decouple at higher densities, where matter is usually also hotter. All the features described above have been already explored in more detail in, e.g., \cite{Cusinato:2021zin}.

\begin{figure}
    \centering 
    \includegraphics[width=0.45\textwidth]{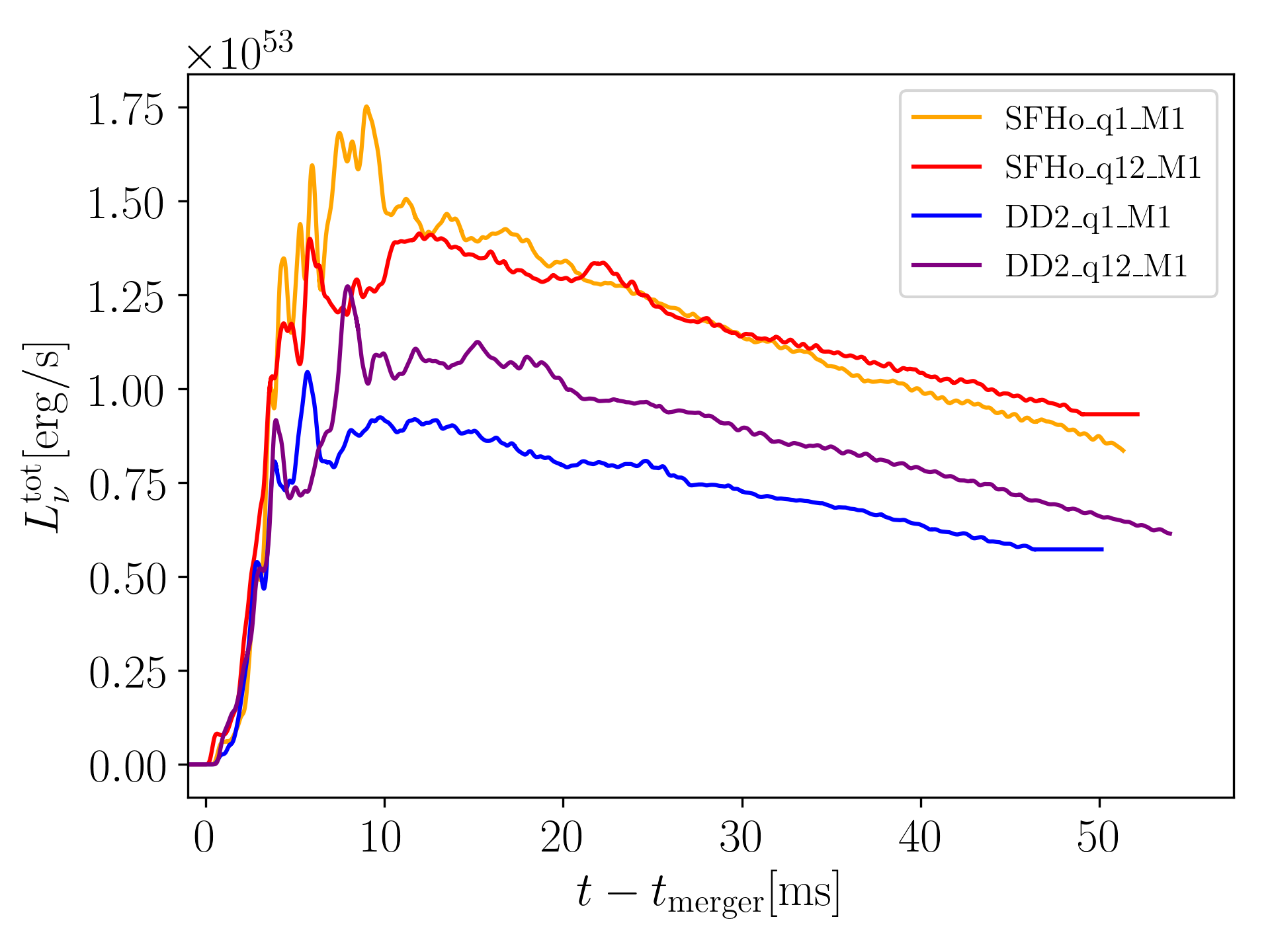}
    \caption{Total neutrino luminosity for all simulations at resolution R2.}
    \label{fig:Lnutot_vs_time} 
\end{figure}

Finally, in Fig.~\ref{fig:Lnutot_vs_time}, we show the total luminosity for all four configurations at resolution R2. The first observation is that SFHo systems emit significantly more neutrinos than their DD2 counterparts due to the temperature difference visible in Fig.~\ref{fig:rho_max}, with a difference of almost $50\%$ at the brightness peak. Such an important difference could explain, or at least contribute to, the significant difference in the neutrino wind emission between the two EoSs. 

\subsection{Remnant properties}
\begin{figure*}
    \centering 
    \includegraphics[width=0.45\textwidth]{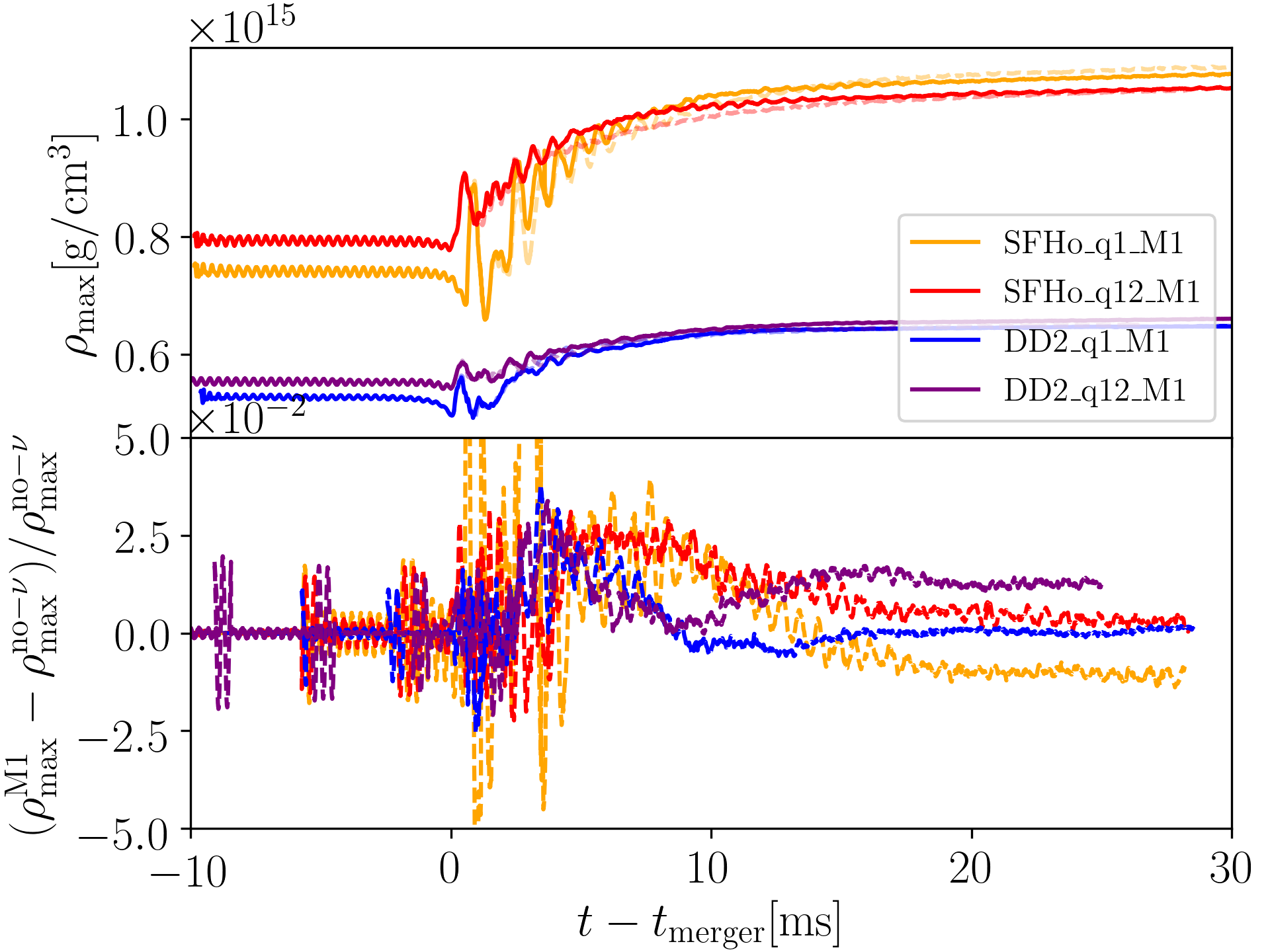}
    \includegraphics[width=0.45\textwidth]{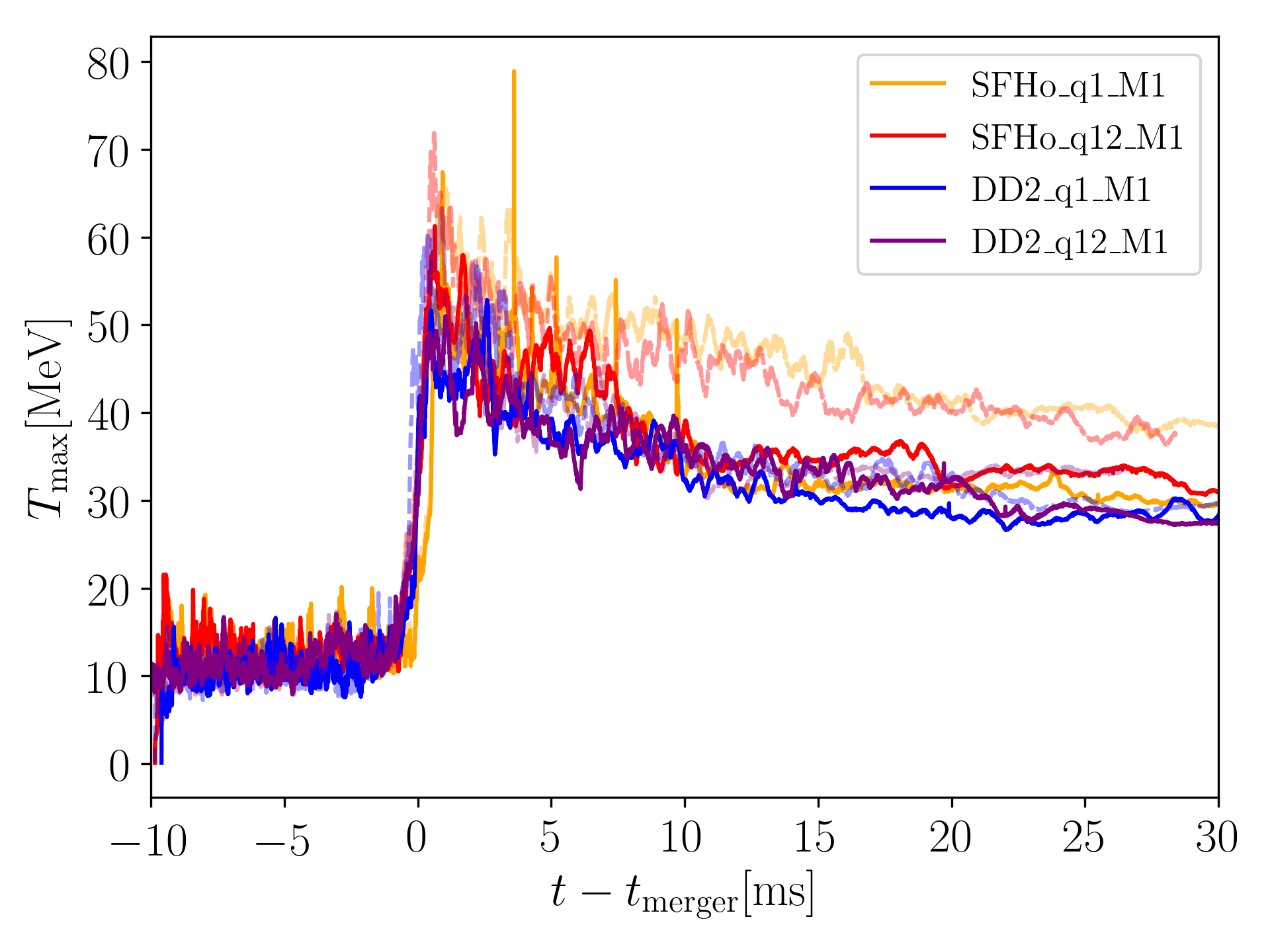}
    \caption{Maximum rest mass density and temperature. Solid lines refer to simulations with neutrino transport and dashed lines to simulations where neutrinos interactions have been neglected. In the left plot the relative difference between simulations with neutrino transport and without neutrinos is shown.}
    \label{fig:rho_max} 
\end{figure*}

We begin the analysis of the remnant by looking at Fig.~\ref{fig:rho_max}, showing the evolution of density and temperature maxima. In the left panel, we can observe that maximum density is not significantly affected by neutrino radiation, with differences rarely exceeding 5\% during the post-merger oscilation ond settling to smaller values after $\simeq 15$~ms. Considering the temperature evolution, we find that the maximum temperature for the simulations using SFHo is indeed lower for systems using M1 compared to simulations without evolving the neutrinos. Contrary, the setups employing the DD2 EOS show an almost unchanged maximum temperature. 
The difference between M1 and neutrinoless simulations is more pronounced for SFHo EoS because of the higher amount of neutrino energy involved. We explain this result as an indirect effect of neutrino cooling affecting the remnant in the early post-merger. 

Since all the binary simulations performed in this work produce a stable massive neutron star (MNS) surrounded by a disk, we decide to adopt the usual convention of defining the disk of a MNS+disk system as the region where matter is gravitationally bound and $\rho<10^{13}$~~g/cm$^3$, by contrary the MNS is defined by $\rho>10^{13}$~g/cm$^3$; \cite{Shibata:2017xdx,Radice:2018pdn,Kiuchi:2019lls,Vincent:2019kor}. This allows us to provide an estimate of the mass of the disk and the MNS.

\begin{figure}
    \centering 
    \includegraphics[width=0.45\textwidth]{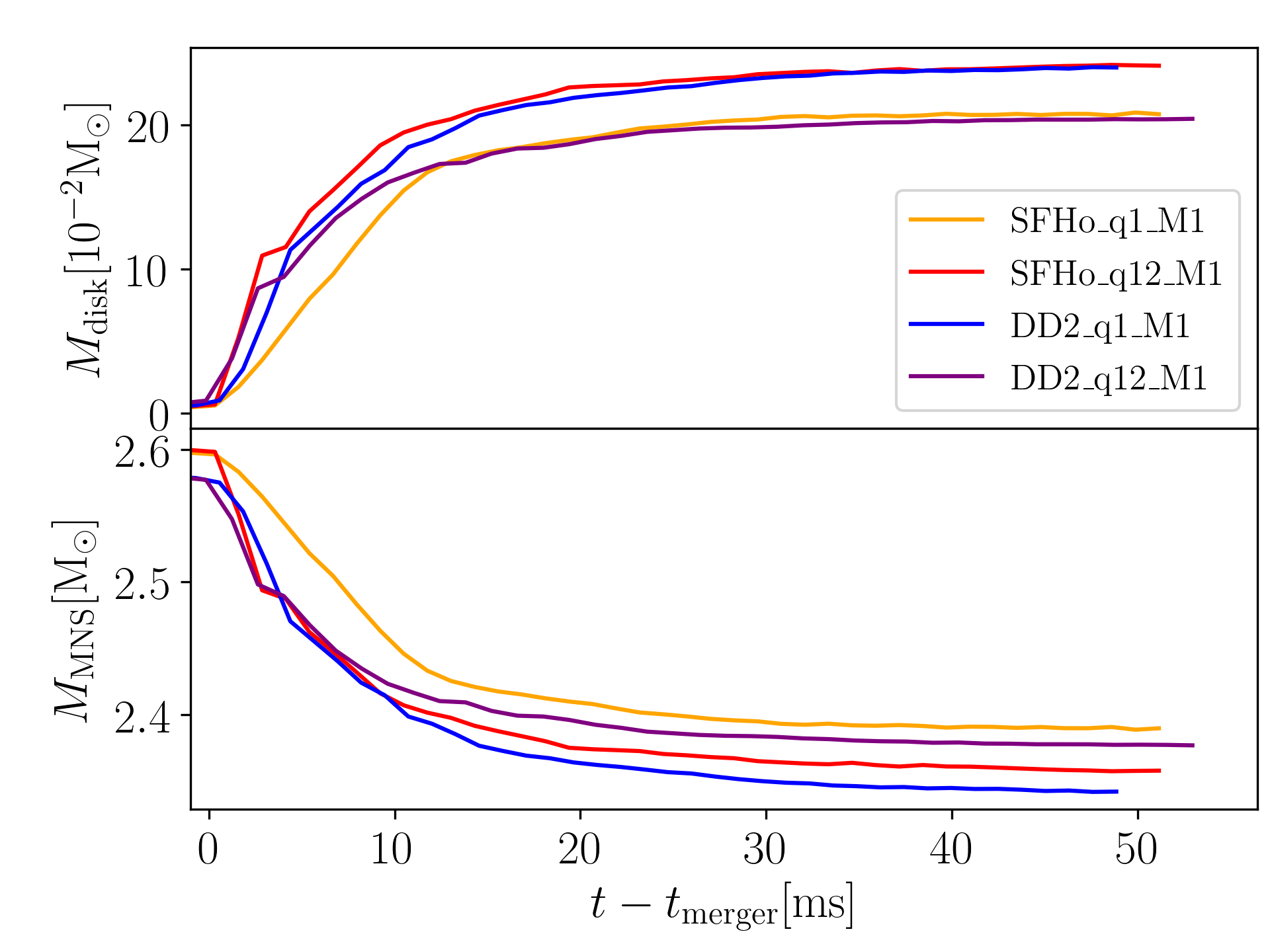}
    \caption{Mass of the disk and of MNS as a function of time. The last one is computed as the integral of bound $D$ in the regions where $\rho>10^{13}$~g/cm$^3$.}
    \label{fig:M_disk} 
\end{figure}

In Fig.~\ref{fig:M_disk}, we show the masses of the disk and the MNS as a function of post-merger time. After an initial time where the disk is growing fast, acquiring mass from the remnant, the disk mass stabilizes at $\simeq 20$~ms after the merger. Such disk accretion phenomena are usually sustained by angular momentum viscous transport and shocks generated by the $m=1$ bar mode oscillations of the central object \cite{Nedora:2021eoj, Nedora:2019jhl} and contrasted by the gravitational pull of the central object. The effect of neutrino transport on the disk's mass for SFHo simulations is negligible (see Table~\ref{tab:BNS_results_table}), and the results look robust also at lower resolutions. This is an expected result since the disk formation takes place at times when neutrino cooling is not the dominant source of energy loss.

\subsection{Nucleosynthesis}

\begin{figure}
    \centering 
    \includegraphics[width=0.45\textwidth]{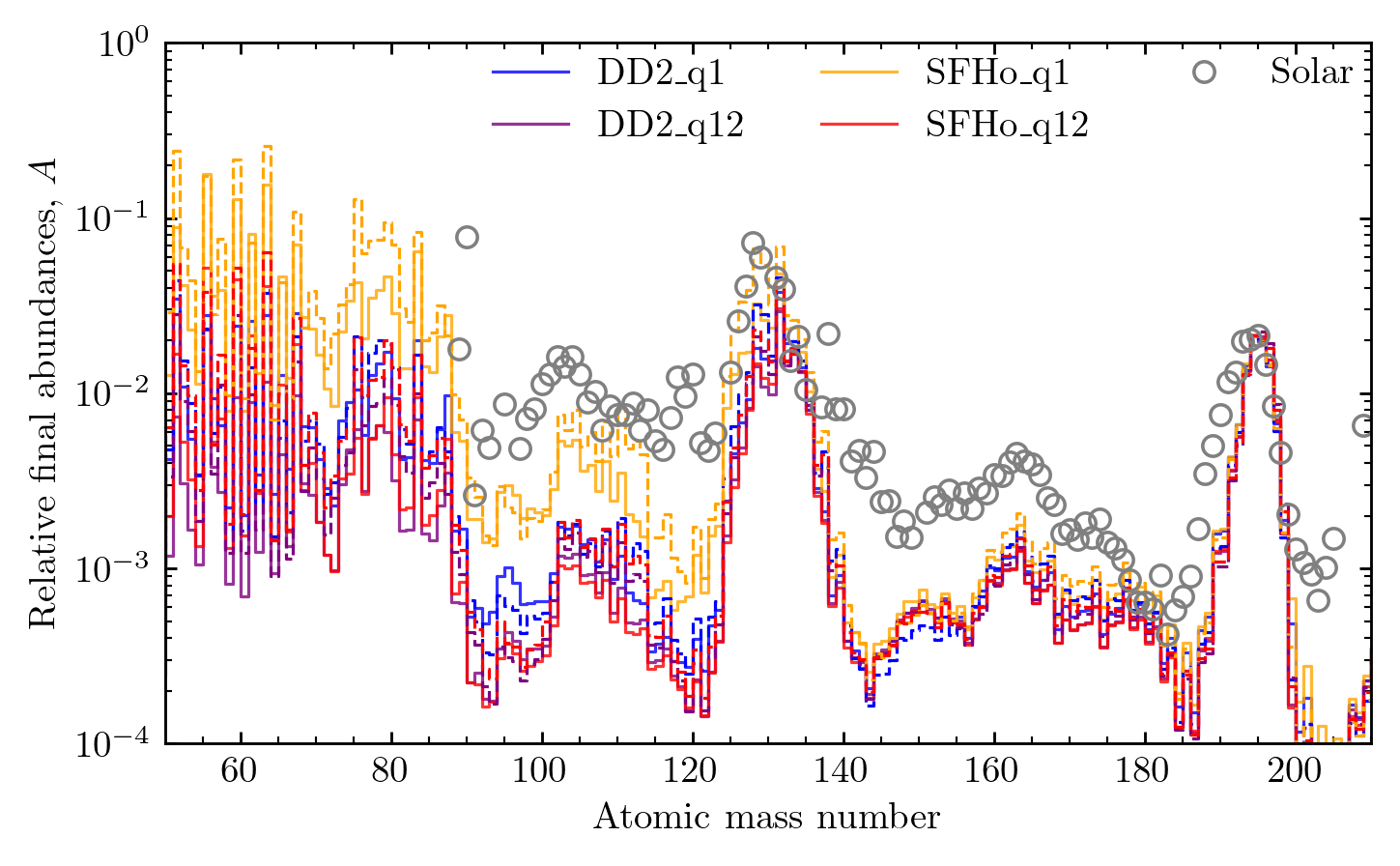}
    \caption{Nucleosynthesis yields for all simulations. 
      The nucleosynthesis is computed dynamical ejecta only ($t=20$~ms).
      (In the top panel the normalization to $A_{\rm sol}=195$ is used, while in the bottom the normalization to $A_{\rm sol}=135$ is done. 
      Solid lines indicate the dynamical ejecta, while dashed lines correspond to the dynamical ejecta extracted with the Bernoulli criterion.}
    \label{fig:yields} 
\end{figure}

The nucleosynthesis calculations are performed in postprocessing
following the same approach as in \cite{Radice:2016dwd,Radice:2018pdn}
employing the results from the nuclear reaction network \texttt{Skynet} 
of \citep{Lippuner:2015gwa}.
In Fig.~\ref{fig:yields}, we show the abundances as a function of the mass 
number $A$ of the different isotopes synthesized by the $r$-process $32$ years after the merger in ejecta. 
To compare the results for different simulations, we shift the 
abundances from all models such that they are always the same as the 
solar one for $A=195$. The solar residual $r-$process abundances are taken 
from \cite{Prantzos2020} (for a review of the solar system abundances; see
\citep{Pritychenko:2019xvf}). 
The normalization to $A_{\rm sol}=195$ is chosen as nucleosynthesis in neutron-rich ejecta from BNS mergers was shown to robustly reproduce the third $r-$process peak \cite{Lippuner:2017tyn}. We also consider normalization to $A_{\rm sol}=135$ and $A_{\rm sol}=152$ commonly considered in literature \cite{Curtis:2021guz}. The former leads to only a minor qualitative change while the latter leads to the overall overestimation of the abundances at both, second and third $r-$process peaks.

As the mass-averaged electron fraction of the dynamical ejecta from most models (except SFHo $q=1$ model) is small (see Fig.~\ref{fig:Ye_hist}), the $r$-process nucleosynthesis results in the underproduction of lighter, $1$st and $2$nd peak elements. Additionally, the elements around the rare-earth peak are underproduced. This can be also attributed to the systematic uncertainties in the simplified method we employ to compute nucleosynthesis yields. The simulation with SFHo EOS and mass-ratio $q=1$ displays a more flat electron fraction distribution in its ejecta, and relative abundances at $2$nd peak are consistent with solar. 

The Bernoulli ejecta displays on average higher electron fraction, as it undergoes strong neutrino irradiation, being ejected on a longer timescale. Higher $Y_e$ leads to a larger amount of lighter elements produced. However, the overall underproduction of $1$st $r$-process elements for all simulations but the SFHo $q=1$ model remains.

\subsection{Gravitational waves}
While neutrinos are supposed not to play any role during the inspiral, they could, in principle, be relevant in the post-merger dynamics, e.g., through the additional cooling channel of the formed remnant, which might change the compactness of the remnant and, therefore, the post-merger GW frequency and the time until black-hole formation. We investigate this possibility in the following subsection by comparing the GW signal produced by each simulation and its `neutrinoless' counterpart.
We compute the GW strain $h$ on a series of concentric spheres using the $\Psi_4$ Newman-Penrose scalar~\cite{Newman:1961qr}, following the method of \cite{Reisswig:2010di}.

\begin{figure}
    \centering 
    \includegraphics[width=0.45\textwidth]{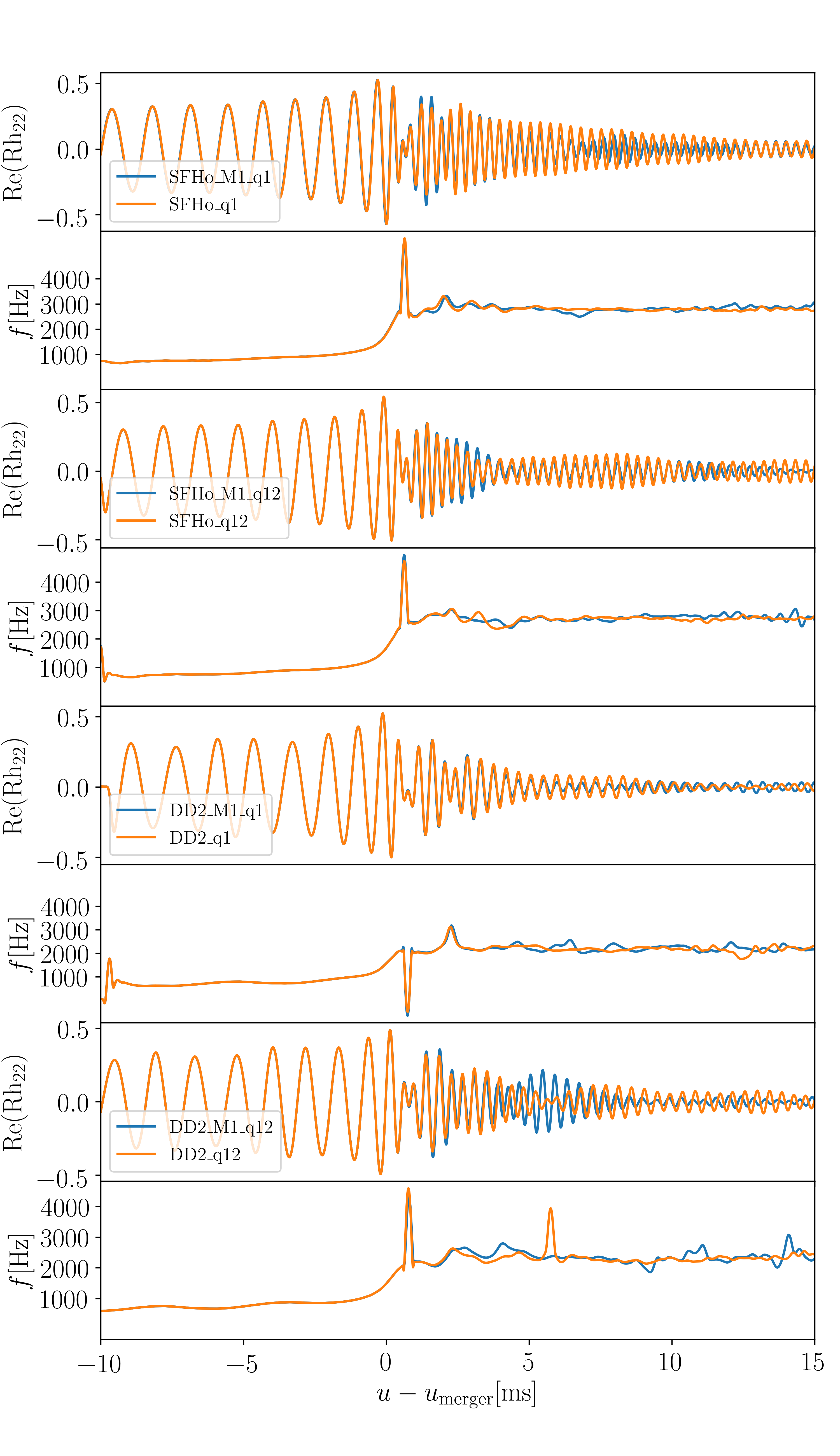}
    \caption{Amplitude and frequency of the GW's (2,2) mode as a function of the retarded time $u$. The waveforms are extracted at $r \simeq 1200$~km.}
    \label{fig:waveforms} 
\end{figure}

In Fig.~\ref{fig:waveforms}, we show the GW strain $h$ and its frequency for the dominant (2,2) mode of each simulation.
Overall, one can observe only minimal changes in the GW amplitude and frequency caused by neutrino cooling\footnote{We note that the spike at 6~ms for DD2\_q12 is due to numerical inaccuracies when computing the instantaneous GW frequency for a GW signal with almost vanishing amplitude.}. 
Given the large challenge in measuring the post-merger GW signal from future detections~\cite{Rezzolla:2016nxn,Chatziioannou:2017ixj,Soultanis:2021oia,Wijngaarden:2022sah,Breschi:2022xnc,Puecher:2022oiz} and the presumably large uncertainties regarded the extracted postmerger frequencies, 
we expect that the differences visible here are not measurable, potentially not even with the next generation of detectors. 
However, a more systematic study involving Bayesian parameter estimation is needed to verify this hypothesis.

\subsection{Lightcurves}

To compute the kilonova signal associated with the extracted ejecta profiles from the performed simulations, we use the 3D Monte Carlo radiative transfer code \texttt{POSSIS}~\cite{Bulla:2019muo,Bulla:2022mwo}. The code allows us to use the 3D simulation output of the unbound rest-mass density $D_u$ and the electron fraction $Y_e$ of the ejecta as input. The required input data represents a snapshot at a reference time $t_{0}$ and is subsequently evolved following a homologous expansion, i.e., the velocity $v^i$ of each fluid cell remains constant. 
In Appendix.~\ref{sec:appendixC}, we outline the exact procedure employed to obtain \texttt{POSSIS} input data. 

For the generation of photon packets (assigned energy, frequency, and direction) at each time step, \texttt{POSSIS} employs the heating rate libraries from \cite{Rosswog:2022tus} and computes the thermalization efficiencies as in \cite{Barnes:2016umi,Wollaeger:2017ahm}. The photon packets are then propagated through the ejecta, taking into account interactions with matter via electron scattering and bound-bound absorption. \texttt{POSSIS} uses wavelength- and time-dependent opacities from \cite{Tanaka:2019iqp} as a function of local densities, temperatures, and electron fraction within the ejecta. We perform the radiative transfer simulations with a total of $N_{\rm ph} = 10^6$ photon packets.

In contrast to previous works in which we used \texttt{POSSIS}~\cite{Neuweiler:2022eum,Markin:2023fxx}, we are now able to use the electron fraction of the material directly and do not have to approximate it through the computation of the fluid's entropy. This is an important improvement since this quantity is fundamental in determining the kilonova luminosity and spectrum. Matter with low $Y_e$ (like tidal tails) can indeed synthesize Lanthanides and Actinides, which have high absorption opacities in the blue (ultraviolet-optical) spectrum, making the EM signal redder. On the contrary, high $Y_e$ material (like shocked ejecta and winds) synthesizes lighter elements that have a smaller opacity and are more transparent to high-frequency radiation, i.e., it will produce a bluer kilonova.

\begin{figure}
    \centering 
    \includegraphics[width=0.45\textwidth]{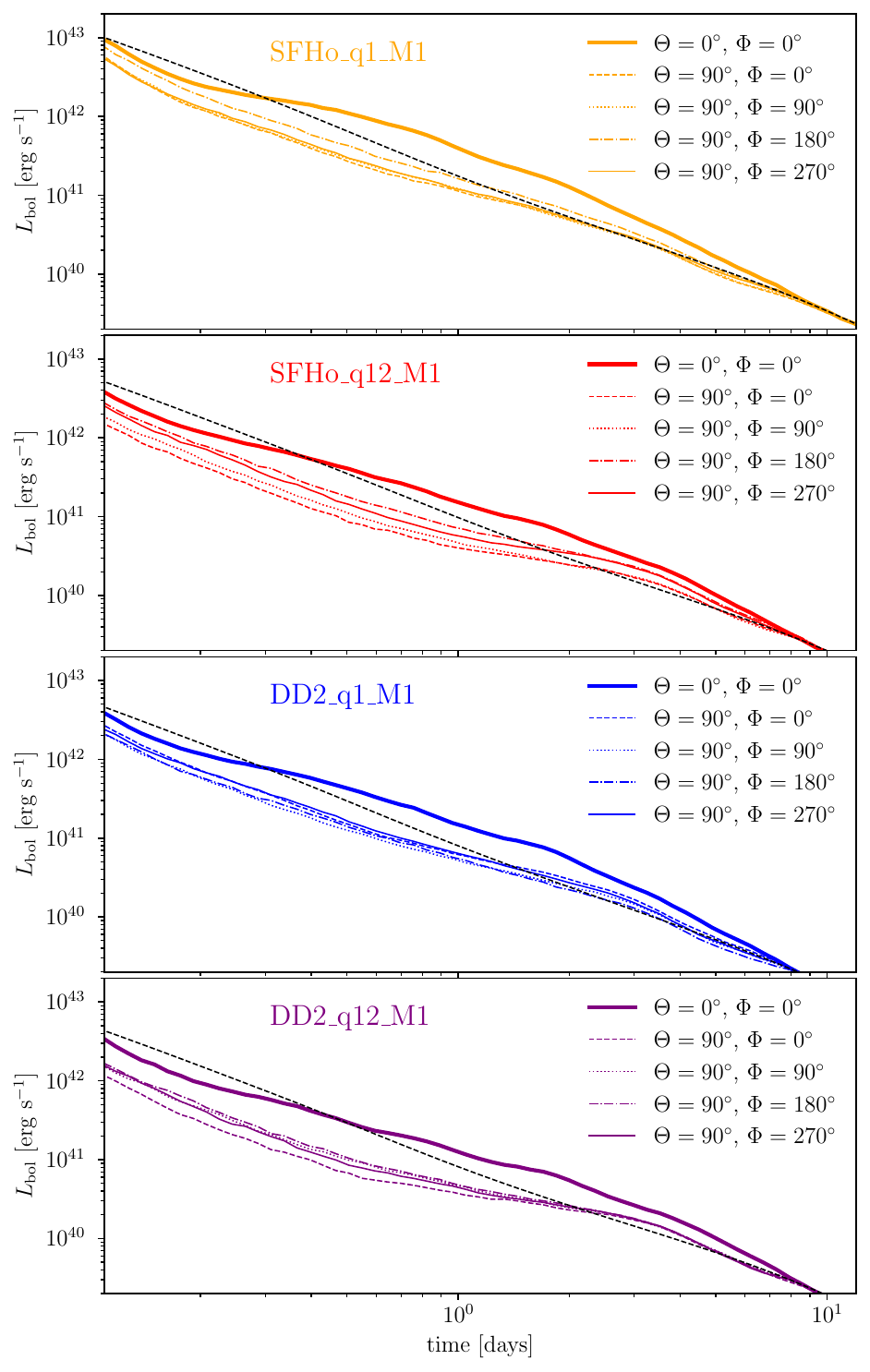}
    \caption{Bolometric luminosity for all four BNS systems. We show the luminosity for the pole with $\Theta = 0^\circ$ (thick solid line), and in the orbital plan with $\Theta = 90^\circ$ for different azimuths: $\Phi = 0^\circ$ (thin dashed line), $\Phi = 90^\circ$ (thin dotted line), $\Phi = 180^\circ$ (thin dash-dotted line), and $\Phi = 270^\circ$ (thin solid line). The deposition curve, based on the amount of energy available, is shown as black dashed line for each system.}
    \label{fig:BNS_lcbol} 
\end{figure}

In Fig.~\ref{fig:BNS_lcbol}, we show the bolometric luminosity for each simulation for five different observation angles: For the pole with $\Theta = 0^\circ$, and in the orbital plane with $\Theta = 90^\circ$ for $\Phi = 0^\circ$, $\Phi = 90^\circ$, $\Phi = 180^\circ$, and $\Phi = 270^\circ$. 
In general, we find that the luminosity at the pole is higher than in the equatorial plane, because of the smaller opacities and the higher amount of mass. At the same time, light curves obtained for the four angles in the orbital plane are rather similar in the $q=1$ simulations. For the systems with unequal mass, the differences are more prominent, but they tend to decrease in time within a timescale of a few days.
This can be explained by the fact that the ejecta input in \texttt{POSSIS} for these systems is less axisymmetric than for the systems with equal masses (see ejecta maps in Appendix~\ref{sec:appendixC}).

\begin{figure}
    \centering 
    \includegraphics[width=0.45\textwidth]{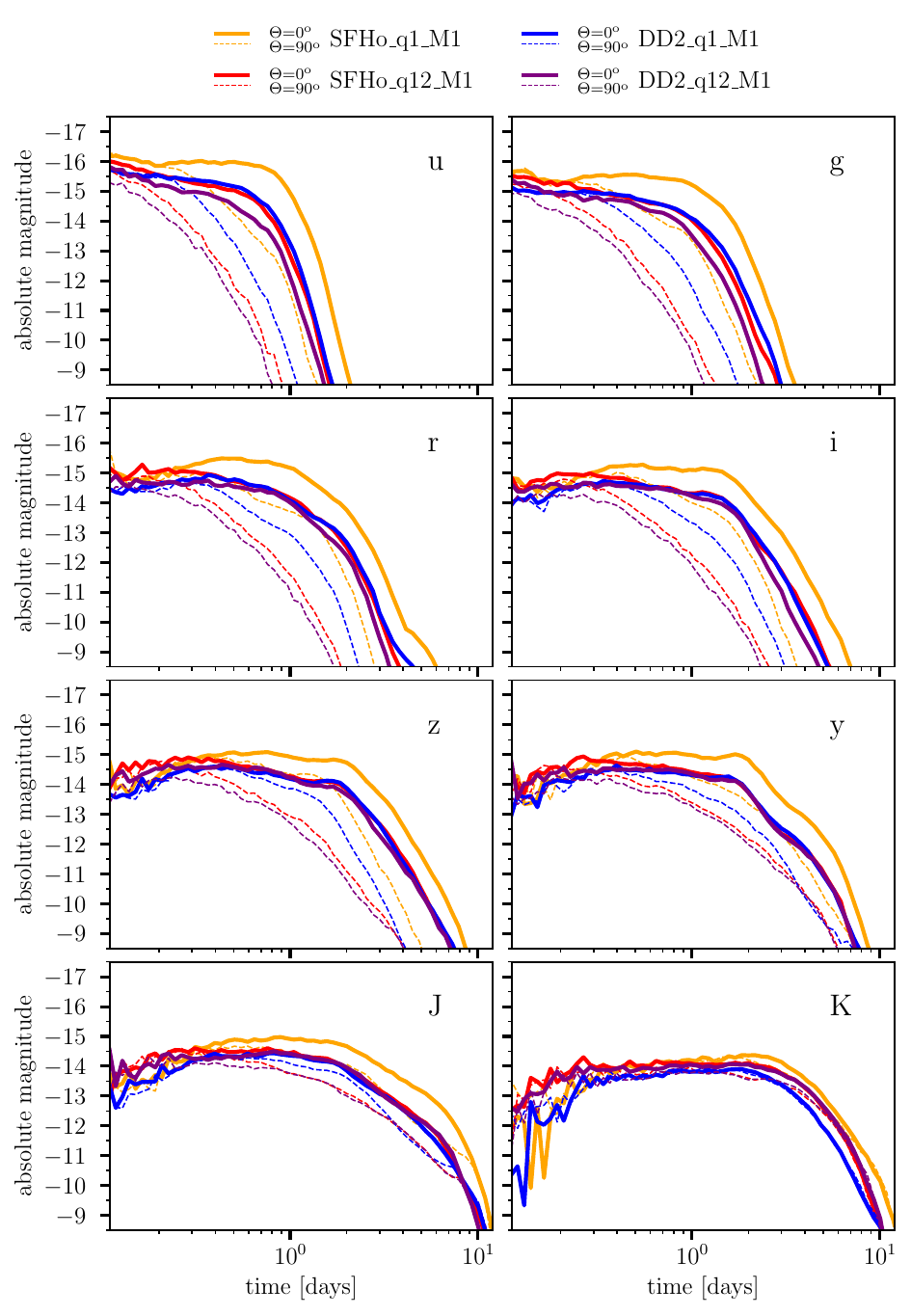}
    \caption{Light curves at the azimuth $\Phi = 0^\circ$. We show each light curve for the polar angle $\Theta = 0^\circ$ (solid line), i.e., the pole, and $\Theta = 90^\circ$ (dashed line), i.e., in the orbital plane.}
    \label{fig:BNS_lc} 
\end{figure}

Furthermore, we show in Fig.~\ref{fig:BNS_lc} the light curves for the four systems in different frequency bands, ranging from ultraviolet to optical and infrared. We focus on one $\Phi$-angle only, i.e., $\Phi=0^\circ$. Still, we want to note here that the results for other $\Phi$ angles for the systems with unequal masses differ up to about $\sim 1$~mag in the first two days after the merger.

We observe that the magnitude difference between polar angles is more pronounced in the ultraviolet and optical bands than in the infrared bands, particularly, in the J- and K-bands. 
The light curves for the systems with SFHo EoS are on average brighter due to the larger ejecta mass than systems employing the DD2 EoS (at the same mass ratio). Even more importantly, we observe that the ratio between the blue and the red component of the kilonova is strongly affected by both the EoS and mass ratio, with more deformable EoS (DD2) and asymmetric configurations giving a redder kilonovae due to the bigger amount of tidal tails with respect to shocked ejecta.

Moreover, we find that in the orbital plane ($\Theta=90^\circ$) the infrared bands are generally more dominant. This is due to the neutron-rich matter of tidal tails located at low latitude, which absorbs most of the radiation at high frequencies. In contrast, for an observer at the pole ($\Theta = 0^\circ$), the ultraviolet and optical bands are brighter in the first two days. However, these diminish rapidly, and at later times the red and infrared bands dominate the kilonova signal here as well. Accordingly, a blue kilonova will be observed in the first days, shifting to the red spectra in the following days. These observations indicate again the need for quick follow-up observations of GW signals with upcoming UV-satellites, e.g.,~\cite{Shvartzvald:2023ofi}.

\section{Conclusions}
In this article, we implemented a gray M1 multipolar radiation transport scheme following~\cite{Thorne:1981,Shibata:2011kx,Foucart:2015vpa,Foucart:2016rxm,Radice:2021jtw} in the BAM code. The main features of the implementation are summarized in Tab.~\ref{tab:numerical_methods}.

\begin{table}
\begin{tabular}{l|l|l}
                    & Method                                 & Reference                  \\ \hline \hline
Fluxes              & Composed, Eq.~(\ref{eq:fluxes})        & \cite{Radice:2021jtw}       \\ \hline
Collisional sources  & Linearized                             &  \cite{Foucart:2015vpa}     \\ \hline  
Opacities           & Modified \texttt{Nulib} tables         &  \cite{OConnor:2014sgn}     \\ \hline
Time step           & Implicit-Explicit, Eqs.~(\ref{eq:evolution_substep1}, \ref{eq:impl_sys}) & \cite{Radice:2021jtw}   
\end{tabular}
\caption{Summary of the M1 implementation and the employed methods.}
\label{tab:numerical_methods}
\end{table}

We performed a series of standard tests: transport along lightlike geodesics in vacuum, absorption by static and moving fluid, advection by a moving fluid in the scattering-dominated regime, and emission by a thick uniform sphere. 
The main difficulty was to properly account for the collisional sources implicitly and to suppress artificial dissipation in the trapped regime in order to capture the correct diffusion rate. We show that our implementation is able to correctly handle all these regimes employing linearized implicit sources of \cite{Foucart:2015vpa} and the flux reconstruction of \cite{Radice:2021jtw}.

In addition, we also performed simulations of a single, isolated, hot neutron star. In this case, both the spacetime and the fluid are dynamically evolved. Opacities are motivated by nuclear physics theory and computed using the \texttt{nulib} library. In this last test, we show that neutrinos correctly thermalize inside the star, where they form gas in thermal equilibrium with the nuclear matter and decouple at the star's surface at different temperatures according to their species (with the hierarchy $T^{\nu_e}_\textrm{eff} < T^{\overline{\nu}_e}_\textrm{eff} < T^{\nu_x}_\textrm{eff}$). 
Moreover, we showed neutrinos correctly start developing a non-zero average momentum at the neutrinosphere $\tau=2/3$. In the last part of the article, we simulated four different low-mass BNS configurations using two different EoS and two mass ratios.

Ejecta from our simulations had the following properties: masses of the order of $\sim 10^{-3}~M_{\odot}$ with $\langle V_{\infty} \rangle = 0.1c - 0.2c$ and $\langle Y_e \rangle = 0.2-0.4$, the latter with a strong dependence on the mass ratio. In general, more asymmetric systems and systems with a stiffer EoS (DD2) produce lower $\langle Y_e \rangle$ due to the larger mass of tidal tail ejecta, with the lowest $\langle Y_e \rangle$ given by the asymmetric DD2 configuration. 
We also illustrated that, on average, more asymmetric binaries produce more ejecta with respect to their symmetric counterparts for both EoSs. Softer EoS (SFHo) eject more than stiffer ones due to the more violent impact of the merger.

Overall, the mechanisms we identified in our simulations are consistent with those reported in the literature for the dynamical ejecta. 
Moreover, similar to \cite{Foucart:2015gaa}, we found a neutrino wind ejecta component in the polar region during the whole duration of the simulation, albeit with decreasing matter flux. Such a component is significantly more important for softer EOSs, in our case SFHo, due to the higher outflow of neutrino energy. It can contribute up to 50\% of the total ejecta mass and significantly increase $\langle Y_e \rangle$. This component could get even more dominant if the simulation is run for longer.

All our simulations produce a MNS remnant surrounded by a massive, neutrino-thick disk with baryonic mass $M_{\rm disk} \sim 10^{-1}~M_{\odot}$. The mass of the disk increases with the mass ratio for SFHo EoS while having the opposite behavior for the stiffer DD2. The results summarized so far are valid for all resolutions.

Finally, we used our new ejecta analysis tools to employ our NR-extracted ejecta properties as inputs for the codes \texttt{Skynet} and \texttt{POSSIS}, which we used to compute nucleosynthesis yields and kilonova lightcurves, respectively. The use of $Y_e$ obtained directly from the NR simulations produces much more realistic results with respect to the previous assumption based on fluid's entropy that was used in \texttt{POSSIS}.

We plan to use the implementation described in this article as the standard for our future BNS simulations oriented to the study of ejecta properties and post-merger dynamics and of the associated kilonova light curves and nucleosynthesis yields. 

\section*{Acknowledgements}

We thank  H.~Andresen, S.~Bernuzzi, B.~Br\"ugmann, F.~Foucart, E.~O'Connor, M.~Shibata, and W.~Tichy for helpful discussions. 
FS and TD acknowledge funding from the EU Horizon under ERC Starting Grant, no.\ SMArt-101076369. TD and AN acknowledge support from the Deutsche
Forschungsgemeinschaft, DFG, project number DI 2553/7. 
TD and VN acknowledge support through the Max Planck Society funding the Max Planck Fellow group 
`Multi-messenger Astrophysics of Compact Binaries'. HG acknowledges funding by FAPESP grant number 2019/26287-0. MU acknowledges support through the UP Reconnect Program from the Alumni Researcher Program of the University of Potsdam.
The simulations were performed on the national supercomputer HPE Apollo Hawk at the High Performance Computing (HPC) Center Stuttgart (HLRS) under the grant number GWanalysis/44189, on the GCS Supercomputer SuperMUC\_NG at the Leibniz Supercomputing Centre (LRZ) [project pn29ba], and on the HPC systems Lise/Emmy of the North German Supercomputing Alliance (HLRN) [project bbp00049]. 


\bibliography{references}


\appendix

\section{Linearized implicit timestep solution}
\label{sec:appendixA}

 The projections of tensors $A^{\alpha}$ and $B^{\alpha}_i$ of Eq.~\eqref{eq:sources_linear} perpendicular to the spacelike hypersurface $\Sigma_t$ can be written as:
\begin{equation}
    \begin{aligned}
    n^{\alpha}A_{\alpha} & = k_a A_{(J)} - (k_a+k_s) A_{(H)},
    \end{aligned}
\end{equation}
\noindent with
\begin{equation}
    \begin{aligned}
    A_{(J)} = & W[W^2 + a W^2 (v \cdot f)^2 + b \frac{W^2-1}{2 W^2+1}(3-2 W^2)], \\
    A_{(H)} = & W[-1 + W^2 + a W^2(v \cdot f)^2 + b \frac{W^2-1}{2W^2+1}(3-2W^2)],
    \end{aligned}
\end{equation}
\noindent and
\begin{equation}
    \begin{aligned}
    n^{\alpha}B^i_{\alpha} & = k_a B^i_{(J)} - (k_a+k_s) B^i_{(H)},
    \end{aligned}
\end{equation}
\noindent where we define
\begin{equation}
    \begin{aligned}
    B^i_{(J)} = &W[-2W + b\frac{W^2-1}{2W^2+1}4W^2]v^i, \\
    B^i_{(H)} = &W [1 - 2W^2 + b\frac{W^2-1}{2W^2+1}4W^2]v^i. 
    \end{aligned}
\end{equation}
\noindent While for the parallel component, we have
\begin{equation}
    \begin{aligned}
    \gamma^{\alpha}_i A_{\alpha} &= k_a A_{i,(J)} - (k_a+k_s) A_{i,(H)},
    \end{aligned}
\end{equation}
\noindent with
\begin{equation}
    \begin{aligned}
    A_{i,(J)} = &-W[W^2 + aW^2(v \cdot f)^2 + b\frac{W^2-1}{2W^2+1}(3-2W^2)]v_i, \\
    A_{i,(H)} = &-[W^3 + aW^3(v \cdot f)^2 + b W\frac{W^2}{2W^2+1}(3-2W^2)]v_i \\
             &- a W(v \cdot f) f_i,
    \end{aligned}
\end{equation}
\noindent and
\begin{equation}
    \begin{aligned}
    \gamma^{\alpha}_i B_{\alpha}^j &= k_a B^j_{i,(J)} - (k_a + k_s) B^j_{i,(H)},
    \end{aligned}
\end{equation}
\noindent with
\begin{equation}
    \begin{aligned}
    B^j_{i,(J)} =& W[2W^2 - b\frac{W^2-1}{2W^2+1}4W^2]v_iv^j, \\
    B^j_{i,(H)} =& [2W^3 - bW\frac{W^2-1}{2W^2+1}4W^2 - \\
    & -b\frac{W}{2W^2+1}(2W^2-1)]v_iv^j \\
    & + (1-bv^2)W\delta_i^j,
    \end{aligned}
\end{equation}
\noindent where $a = (3\chi-1)/2$ and $b = 1- a$ are the thin and thick closure coefficients respectively and $f^i=F^i/|F|$.

Using these projections we can write $\tilde{E}$ and $\tilde{F_i}$ at time $n+1$ as:
\begin{equation}
    \begin{aligned}
        \tilde{F}_i^{n+1} &= (M^{-1})_i^j S_j, \\
        \tilde{E}^{n+1} & = \frac{1}{1 + \alpha \Delta t n^\alpha A_{\alpha}} [ \tilde{E}^n + \Delta t (-\partial_i \mathcal{F}^i_E +G_E) \\
        &+\alpha \Delta t (\eta \sqrt{\gamma} W - n^{\alpha}B^i_{\alpha}\tilde{F}_i^{n+1}) ], 
    \end{aligned}
\end{equation}
\noindent with
\begin{equation}
    M_i^j = \delta_i^j - \alpha \Delta t \gamma_i^{\alpha} B_{\alpha}^j + \frac{\alpha^2 \Delta t ^2}{1+\alpha \Delta t n^{\alpha}A_{\alpha}} A_{\alpha}\gamma^{\alpha}_i n^{\beta}B_{\beta}^j, \\
\end{equation}
\noindent and
\begin{equation}
    \begin{aligned}
    S_i &= \tilde{F}_i^n + \Delta t ( -\partial_j \mathcal{F}^j_{F_i} + G_{F_i})+ \Delta t \Bigl[ \alpha \sqrt{\gamma} \eta W v_i \Bigr.\\ 
    & \left. +\frac{\alpha A_{\alpha}\gamma_i^{\alpha}}{1+\alpha \Delta t n^{\alpha}A_{\alpha}} \left ( \tilde{E}^n +\Delta t (-\partial_i \mathcal{F}^i_{E} + G_E) + \alpha \Delta t \sqrt{\gamma}W\eta \right ) \right ],
\end{aligned}
\end{equation}
\noindent where $G_E$ and $G_{F_i}$ represent the gravitational sources of Eq.~(\ref{eq:evolution_E}) and Eq.~(\ref{eq:evolution_F}), respectively, computed using $\tilde{E}^n$ and $\tilde{F}_i^n$.
Since $M_i^j$ and $S_i$ only depend on variables at time $n$, $\tilde{F}_i^{n+1}$ must be first computed and then plugged into the expression of $\tilde{E}^{n+1}$ to complete the solution. 

\section{Hamiltonian constraint violation}
\label{sec:appendixB}

\begin{figure}
    \centering 
    \includegraphics[width=0.45\textwidth]{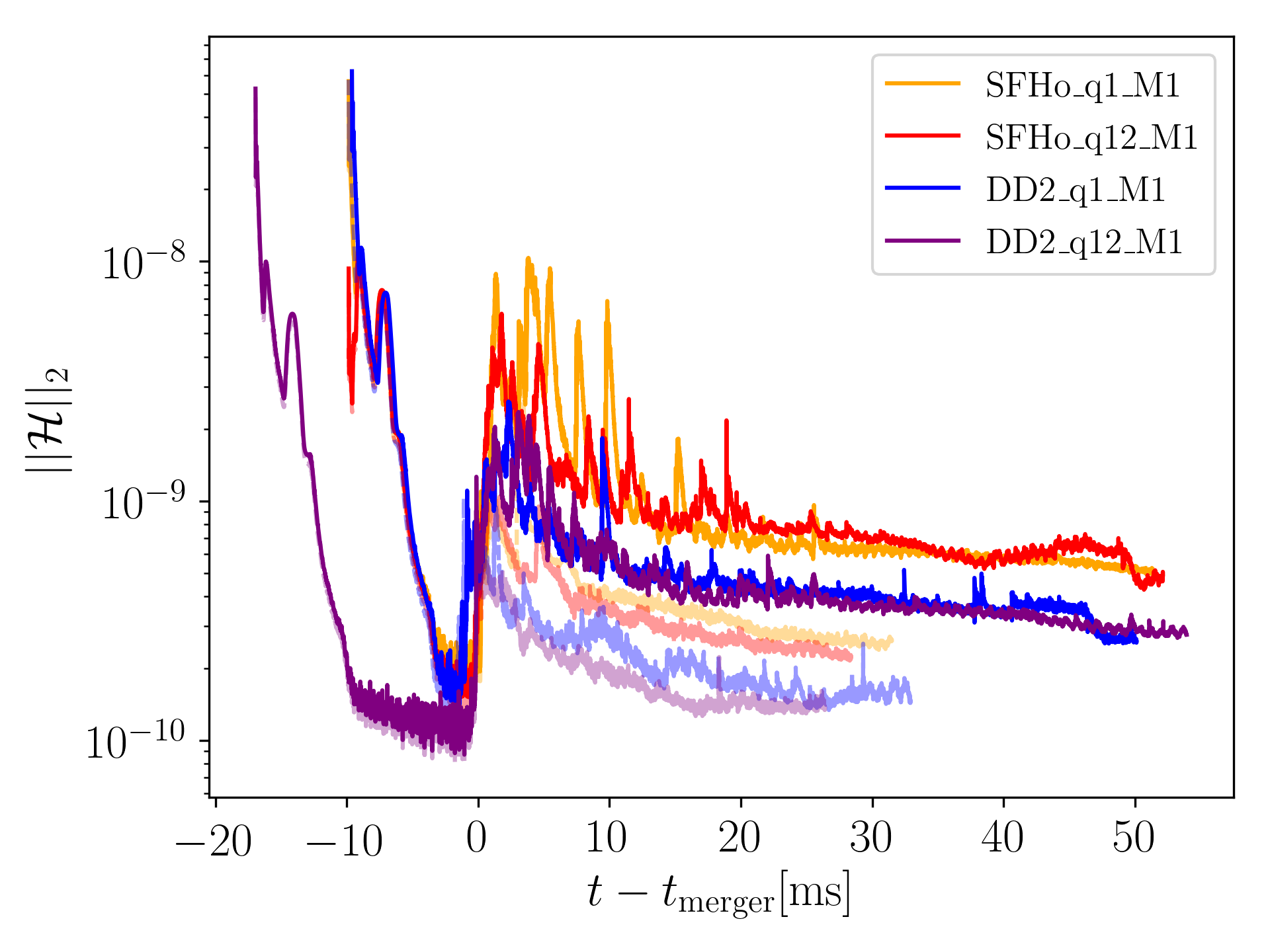}
    \caption{Hamiltonian constraint violation for R2 simulations as function of post-merger time. Shaded lines represent the correspondent neutrinoless counterparts.}
    \label{fig:hamiltonian} 
\end{figure}

In Fig.~\ref{fig:hamiltonian}, we show the L$_2$ norm of the Hamiltonian constraint as a function of the time. The latter follows the same qualitative evolution as in Ref.~\cite{Gieg:2022}. When initial data are interpolated from \textsc{sgrid}, the Hamiltonian constraint is of the order of $10^{-8}$. The evolution with the Z4c formulation reduces this value order $10^{-10}$ due to its constraint-damped properties. At merger time, the value increases, due to the formation of shocks in the hydrodynamics variables, reaching a peak shortly after. After the peak, the value decreases and stabilizes between $10^{-9}$ and $10^{-10}$. Simulations including neutrino transport systematically show a bigger violation of the Hamiltonian constraint after the merger. One of the reasons might be that, as common in the literature, the neutrino's stress-energy tensor is not included in the matter term of the spacetime evolution equations. This leads to a mathematical violation of General Relativity constraints proportional to neutrino's stress-energy tensor. However, we can observe that the value of the Hamiltonian constraint is always lower than its initial value.

\section{Ejecta Data for \texttt{POSSIS}}
\label{sec:appendixC}

\begin{figure}
    \centering 
    \includegraphics[width=\linewidth]{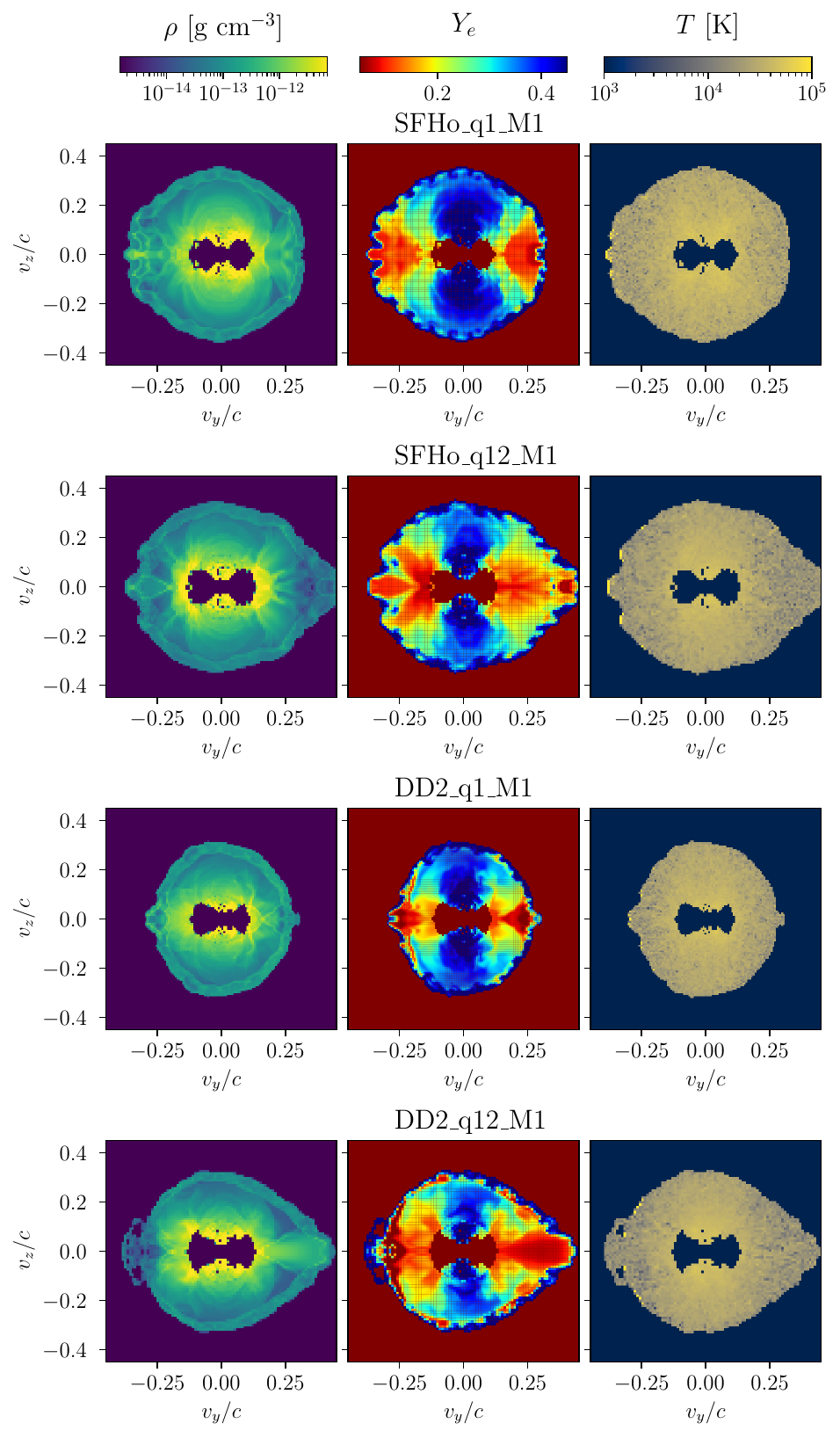}
    \caption{Maps of the matter density $\rho$, electron fraction $Y_e$, and temperature $T$ in the $v_y$-$v_z$ plane as used in \texttt{POSSIS}. We show the configurations for all four systems, i.e., SFHo\_q1\_M1, SFHo\_q12\_M1, DD2\_q1\_M1, and DD2\_q12\_M1, scaled for $1$\,day after the merger by homologous expansion.}
    \label{fig:BNS_maps} 
\end{figure}

\begin{figure*}
    \centering 
    \includegraphics[width=\linewidth]{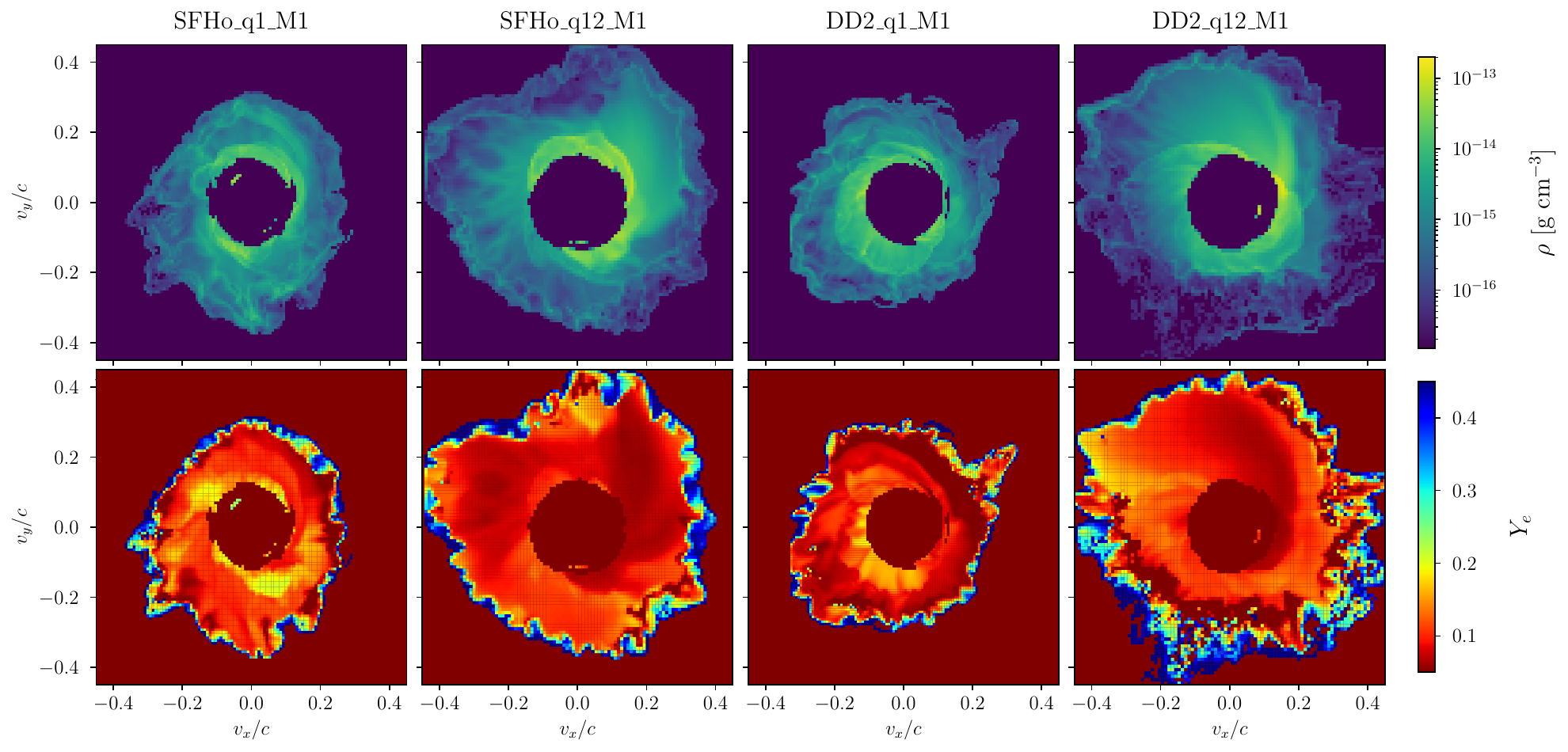}
    \caption{Maps of the matter density $\rho$ and electron fraction $Y_e$ in the $v_x$-$v_y$ plane as used in \texttt{POSSIS}. We show the configurations for all four systems, i.e., SFHo\_q1\_M1, SFHo\_q12\_M1, DD2\_q1\_M1, and DD2\_q12\_M1, scaled for $1$\,day after the merger by homologous expansion.}
    \label{fig:BNS_mapsxy} 
\end{figure*}

Given the limited length of our simulations and the issue of covering both early-time and postmerger ejecta with individual snapshots, we employ 3D snapshots together with information from the detection sphere at $r \simeq 450$~km. The detailed procedure is as follows: 
\begin{itemize}
    \item We find the latest 3D snapshot in which all the ejecta is contained within the simulation domain. We mark the time of this snapshot as $t_{\rm cut}$. From it, we cut out the matter still contained within the detection sphere. This component includes most of the ejecta mass, including the tidal tails and the shocked component.
    \item We rescale the ejecta from the previous step assuming homologous expansion the same way \texttt{POSSIS} does, i.e., assuming every fluid element moves with a constant velocity $v^i = x^i/(t-t_{\rm merger})$. This is equivalent to defining a scale factor $\alpha(t) = (t - t_{\rm merger})/ (t_{\rm cut} - t_{\rm merger})$ and rescaling coordinates and mass density as $x^i \rightarrow \alpha(T) x^i$, $\rho \rightarrow \rho / \alpha^3(T)$, where $T$ is the final time of the simulation. After this step, the radius of the inner cut (initially corresponding to the detection sphere) moved outwards, leaving a gap between the ejecta and the detection sphere that we are going to fill using data from the sphere itself. \\
    \item From the sphere we select data with $t \in [t_{\rm cut}, T]$. Assuming homologous expansion like for the 3D data, we can map the time into a radius by $R(t) = r (T-t_{\rm merger})/(t-t_{\rm merger})=r \alpha(T)/\alpha(t)$ where $r$ is the fixed coordinate radius of the detection sphere. At the same time, we rescale the mass density by $\rho(t,\theta,\phi) \rightarrow \rho(t,\theta,\phi) (\alpha(t)/\alpha(T))^3$. After this procedure, we will have the $\rho(R,\theta,\phi)$, and we interpolate it into the Cartesian grid, where the ejecta from 3D data is defined. This way, we fill the gap between the ejecta and the detection sphere left by the previous rescaling step.
\end{itemize}

It is important to point out that the ejecta at the detection sphere is not fully homologous, and assuming a constant velocity with $v^i = x^i/(t-t_{\rm merger})$ might introduce biases. This is due to the different velocities of components ejected at different times, with shock ejecta that is faster than tidal tails, although it is ejected later. Although deviations from homologous expansion are shown to be present even at $\mathcal{O}(100~{\rm ms})$ after the merger \cite{Rosswog:2013kqa,Kawaguchi:2020vbf}, it has been shown that their influence for the light curve computation using \texttt{POSSIS} is negligible, i.e., within the range of Monte Carlo noise, if the ejecta is extracted at $t > 80$~ms after the merger \cite{Neuweiler:2022eum}. (In \cite{Neuweiler:2022eum}, only the dynamical ejecta was included, and GRHD simulations were performed without the evolution of the electron fraction. The inclusion of other ejecta components or neutrino radiation probably leads to a delay in reaching the homologous phase.) \par 

Because of this reason, we let the ejecta evolve as long as possible out of the detection sphere before assuming homologous expansion and starting the procedure described above. In order to alleviate the issue, an even longer evolution would be required to produce accurate lightcurves. 

The resulting input data for the radiative transfer simulations are shown in Fig.~\ref{fig:BNS_maps} and Fig.~\ref{fig:BNS_mapsxy}. 
In Fig.~\ref{fig:BNS_maps}, we present maps in the $v_y$-$v_z$ plane of the matter density $\rho$, electron fraction $Y_e$, and temperature $T$ used in \texttt{POSSIS} and computed at $1$~day after the merger for all four BNS systems using the M1 scheme. In addition, we show in Fig.~\ref{fig:BNS_mapsxy} the distribution of density and electron fraction in the $v_x$-$v_y$ plane to show how the configurations deviate from axisymmetry.

\end{document}